%% file: main.tex
\documentclass[9pt,sigconf]{acmart}

\renewcommand\footnotetextcopyrightpermission[1]{} 

\usepackage{booktabs} 
\usepackage[normalem]{ulem}
\usepackage{hyperref}
\usepackage[export]{adjustbox}

\usepackage{datetime}
\usepackage{url}
\usepackage{hyperref}
\usepackage[ruled, noend]{algorithm2e}
\usepackage{float}
\usepackage{color}
\usepackage{multirow}
\usepackage{soul}
\usepackage{array}
\usepackage{subcaption}
\usepackage{soul}
\usepackage{balance}
\usepackage{enumitem}
\usepackage{graphicx}
\usepackage{amsmath}
\usepackage{kotex}
\usepackage{mathtools}
\usepackage{physics}

\usepackage{stmaryrd}
\usepackage{trimclip}
\usepackage{xcolor}
\usepackage{fontawesome}

\makeatletter
\DeclareRobustCommand{\shortto}{%
  \mathrel{\mathpalette\short@to\relax}%
}
\DeclareMathOperator*{\E}{\mathbb{E}}
\DeclareMathOperator*{\Var}{\mathbb{V}ar}
\DeclareMathOperator*{\Cov}{\mathbb{C}ov}

\DeclareMathOperator*{\argmin}{arg\,min}

\newcommand{\short@to}[2]{%
  \mkern2mu
  \clipbox{{.3\width} 0 0 0}{$\m@th#1\vphantom{+}{\shortrightarrow}$}%
  }
\makeatother

\newcolumntype{L}[1]{>{\raggedright\let\newline\\\arraybackslash\hspace{0pt}}m{#1}}

\newcommand{\eat}[1]{}

\usepackage{tikz}
\usetikzlibrary{calc}

\usepackage{algorithmicx}
\usepackage{algpseudocode}
\algdef{SE}[DOWHILE]{Do}{doWhile}{\algorithmicdo}[1]{\algorithmicwhile\ #1}%

\newtheorem{theorem}{\bf Theorem}

\newtheorem{definition2}{\bf Definition}
\newtheorem{lemma2}{\bf Proposition}
\newtheorem{lemma3}{\bf Lemma}

\newtheorem{p-rule}{\bf Rule}

\newtheorem{example}{\bf Example}

\begin{document}
\fancyhead{}

\input{preamble}

\title{Combining Sampling and Synopses with Worst-Case Optimal Runtime and Quality Guarantees for Graph Pattern Cardinality Estimation}

\author{~~~~~~~~~~~~~~~~~~~~~~~Kyoungmin Kim}\authornotemark[2]
\affiliation{\vspace*{-0.2cm}}
\author{~~~~~~~~~~~~~~~~~~~~~~~Hyeonji Kim}\authornotemark[2]
\affiliation{\vspace*{-0.2cm}}
\author{~~~~~~~~~~~~~~~~~~~~~~~George Fletcher}\authornotemark[3]
\affiliation{}
\author{~~~~~~~~~~~~~~~~~~~~~~~Wook-Shin Han}\authornotemark[2]\authornotemark[1]
\affiliation{}

\affiliation{ 
  \institution{\mbox{\hspace*{-9.2cm}Pohang University of Science and Technology (POSTECH), Korea\footnotemark[2], Eindhoven University of Technology (TU/e), Netherlands\footnotemark[3]}}
}
\affiliation{ 
  \institution{\mbox{\hspace*{-8.5cm}\{kmkim, hjkim, wshan\}@dblab.postech.ac.kr\footnotemark[2], \{g.h.l.fletcher\}@tue.nl\footnotemark[3]}}
}

\begin{abstract}
Graph pattern cardinality estimation is the problem of estimating the number of embeddings $\QueryNum$ of a query graph in a data graph. This fundamental problem arises, for example, during query planning in subgraph matching algorithms.
There are two major approaches to solving the problem: sampling and synopsis. Synopsis (or summary)-based methods are fast and accurate if synopses capture information of graphs well. However, these methods suffer from large errors due to loss of information during summarization and inherent assumptions. Sampling-based methods are unbiased but suffer from large estimation variance due to large sample space.

To address these limitations, we propose \Alley, a hybrid method that combines both sampling and synopses. \Alley employs 1) a novel sampling strategy, \emph{random walk with intersection}, which effectively reduces the sample space, 2) \emph{branching} to further reduce variance, and 3) a novel mining approach that extracts and indexes \emph{tangled} patterns as synopses which are inherently difficult to estimate by sampling. \redtext{By using them in the online estimation phase, we can effectively reduce the sample space while still ensuring unbiasedness.}
We establish that \Alley has {worst-case optimal runtime and approximation quality guarantees \release{for any given error bound $\epsilon$ and required confidence $\mu$}}. 
In addition to the theoretical aspect of \Alley, our extensive experiments show that \Alley outperforms the state-of-the-art methods by up to orders of magnitude higher accuracy with similar efficiency. 
\end{abstract}


\maketitle

\setcounter{definition2}{0}
\setcounter{figure}{0}
\setcounter{section}{0}
\setcounter{page}{1}

\footnotetext[1]{corresponding author}



\input{01.introduction}

\input{02.background}

\input{03.ht}

\input{04.overview}

\input{05.sampling}

\input{06.index}

\input{07.proof}

\input{08.experiments}

\input{09.related}
\input{10.conclusion}

\ifFullVersion
\else
\vspace*{-0.4cm} 
\section{Acknowledgment}
\release{This work was supported by the National Research Foundation of Korea (NRF) grant funded by the Korea government (MSIT) (No. NRF-2021R1A2B5B03001551).}
\fi



\bibliographystyle{ACM-Reference-Format}
\bibliography{main}

\end{document}

%% file: preamble.tex
\let\oldnl\nl
\newcommand{\nonl}{\renewcommand{\nl}{\let\nl\oldnl}}

\DeclarePairedDelimiter{\ceil}{\lceil}{\rceil}

\newcommand{\FuncName}[1]{\textsc{{#1}}}
\newcommand{\QueryNum}{\abs{\mathcal{M}}}
\newcommand{\CondQueryNum}{c(q, g|M)}
\newcommand{\AGM}{AGM(q)}
\newcommand{\SampleSpace}{\mathcal{P}}
\newcommand{\Matches}{\mathcal{M}}
\newcommand{\Branch}{b \cdot w(i-1)}

\useunder{\uline}{\ul}{}
\newcommand{\veryshortarrow}[1][3pt]{\mathrel{%
   \hbox{\rule[\dimexpr\fontdimen22\textfont2-.2pt\relax]{#1}{.4pt}}%
   \mkern-4mu\hbox{\usefont{U}{lasy}{m}{n}\symbol{41}}}}

\newcommand{\DataGraph}{g}
\newcommand{\DataVertexSet}{V_G}
\newcommand{\DataEdgeSet}{E_G}
\newcommand{\QueryGraph}{q}
\newcommand{\QueryVertex}{u}
\newcommand{\DataVertex}{v}
\newcommand{\Degree}[1]{Deg({#1})}
\newcommand{\Neighbor}[1]{N({#1})}
\newcommand{\Edge}[2]{\left({#1},{#2}\right)}
\newcommand{\VSet}[1]{V({#1})}
\newcommand{\ESet}[1]{E({#1})}
\newcommand{\GESet}[1]{GE({#1})}
\newcommand{\ELabel}[1]{L({#1})}

\newif\ifFullVersion

\def\FullVersion{\let\ifFullVersion=\iftrue}
\def\ShortVersion{\let\ifFullVersion=\iffalse}
\FullVersion

\newcommand{\Catalog}{\FuncName{Catalog}\xspace}
\newcommand{\CSET}{\FuncName{C-SET}\xspace}
\newcommand{\CS}{\FuncName{CS}\xspace}
\newcommand{\GCARE}{\FuncName{G-CARE}\xspace}
\newcommand{\Alley}{\FuncName{Alley}\xspace}
\newcommand{\AlleyPM}{\FuncName{Alley+TPI}\xspace}
\newcommand{\IBJS}{\FuncName{IBJS}\xspace}
\newcommand{\JSUB}{\FuncName{JSUB}\xspace}
\newcommand{\BSK}{\FuncName{BSK}\xspace}
\newcommand{\AlleyD}{\FuncName{AlleyD}\xspace}
\newcommand{\AlleyTPI}{\FuncName{Alley+TPI}\xspace}
\newcommand{\AlleyNaive}{\FuncName{Alley+Naive}\xspace}
\newcommand{\GenericJoin}{Generic-Join\xspace}
\newcommand{\WanderJoin}{\FuncName{WanderJoin}\xspace}
\newcommand{\WJ}{\FuncName{WJ}\xspace}
\newcommand{\ST}{\FuncName{SSTE}\xspace}
\newcommand{\Estimator}{\FuncName{OurMethod}\xspace}
\newcommand{\LogicBlox}{\FuncName{LogicBlox}\xspace}
\newcommand{\EmptyHeaded}{\FuncName{EmptyHeaded}\xspace}
\newcommand{\Graphflow}{\FuncName{Graphflow}\xspace}
\newcommand{\Recursive}{\FuncName{Recur}}
\newcommand{\ChooseRandomVertices}{\FuncName{ChooseRandomVerticesWOR}}
\newcommand{\ChooseRandomVertex}{\FuncName{ChooseRandomVertex}}
\newcommand{\TurboHOM}{\FuncName{TurboHom++}\xspace}
\newcommand{\TurboH}{\FuncName{TH}\xspace}
\newcommand{\ChooseSamplingOrder}{\FuncName{ChooseSamplingOrder}}
\newcommand{\SamplePotentialEmbedding}{\FuncName{SamplePotentialEmbedding}}
\newcommand{\SamplePotentialEmbeddings}{\FuncName{RandomWalkWithIntersect}}
\newcommand{\SamplePotentialEmbeddingsAndEstimate}{\FuncName{SampleAndEstimate}}
\newcommand{\SampleOnePotentialEmbedding}{\FuncName{SampleOnePotentialEmbedding}}
\newcommand{\SampleOneEdge}{\FuncName{SampleOneEdge}}
\newcommand{\SampleOneVertex}{\FuncName{SampleOneVertex}}
\newcommand{\TangledPatternMining}{\FuncName{TangledPatternMining}}
\newcommand{\ExtendPattern}{\FuncName{ExtendPattern}}
\newcommand{\CalculateDomains}{\FuncName{CalculateDomains}}
\newcommand{\HasUnindexedSubgraph}{\FuncName{HasSubgraphWithEmptyDomain}}
\newcommand{\GetMininumDomainsFromSubgraphs}{\FuncName{GetMininumDomainsFromSubgraphs}}
\newcommand{\GetFailureRate}{\FuncName{GetFailureRate}}
\newcommand{\SearchDomainsRecursive}{\FuncName{SearchDomainsRecursive}}
\newcommand{\SearchDomains}{\FuncName{SearchDomains}}
\newcommand{\PriorityFirstSearchFrom}{\FuncName{PriorityFirstSearchFrom}}
\newcommand{\Intersect}{\FuncName{Intersect}}

\newcommand{\GraphGrep}{\FuncName{GraphGrep}\xspace}
\newcommand{\GRAMI}{\FuncName{GraMi}\xspace}
\newcommand{\VF}{\FuncName{VF2}\xspace}
\newcommand{\QuickSI}{\FuncName{QuickSI}\xspace}
\newcommand{\GraphQL}{\FuncName{GraphQL}\xspace}
\newcommand{\gIndex}{\FuncName{gIndex}\xspace}
\newcommand{\SPath}{\FuncName{SPath}\xspace}
\newcommand{\TurboISO}{\FuncName{TurboIso}\xspace}
\newcommand{\CFL}{\FuncName{CFL-Match}\xspace}
\newcommand{\DAF}{\FuncName{DAF}\xspace}
\newcommand{\RDFX}{\FuncName{RDF-3X}\xspace}
\newcommand{\Trinity}{\FuncName{Trinity.RDF}\xspace}
\newcommand{\gStore}{\textsf{gStore}\xspace}
\newcommand{\gMark}{\textsf{gMark}\xspace}
\newcommand{\Chau}{Chaudhuri's Method}
\newcommand{\SumRDF}{\FuncName{SumRDF}\xspace}
\newcommand{\Bernoulli}{\FuncName{BernoulliSampling}\xspace}
\newcommand{\EndBiased}{\FuncName{End-biasedSampling}\xspace}
\newcommand{\Correlated}{\FuncName{CorrelatedSampling}\xspace}
\newcommand{\IMPR}{\FuncName{IMPR}\xspace}

\newcommand{\AIDS}{\textsf{AIDS}\xspace}
\newcommand{\HPRD}{\textsf{HPRD}\xspace}
\newcommand{\Youtube}{\textsf{Youtube}\xspace}
\newcommand{\LUBM}{\textsf{LUBM}\xspace}
\newcommand{\DBLP}{\textsf{DBLP}\xspace}
\newcommand{\Human}{\textsf{Human}\xspace}
\newcommand{\YAGO}{\textsf{YAGO}\xspace}
\newcommand{\Epinions}{\textsf{Epinions}\xspace}
\newcommand{\Amazon}{\textsf{Amazon}\xspace}
\newcommand{\Google}{\textsf{Google}\xspace}
\newcommand{\WatDiv}{\textsf{WatDiv}\xspace}
\newcommand{\UniProt}{\textsf{UniProt}\xspace}

\newcommand{\Lfun}{\mathrm{L}}
\newcommand{\Vset}{\mathrm{V}}

\newcommand{\vertex}{v}
\newcommand{\none}{-}
\newcommand{\QueryTree}{q'}
\newcommand{\convert}{convert}
\newcommand{\precondition}{precondition}
\SetKw{Continue}{continue}

\newcommand{\GMO}{mo_{g}}
\newcommand{\PartialOrders}{PO}

\newcommand{\SWITCH}[1]{\STATE \textbf{switch} (#1)}
\newcommand{\ENDSWITCH}{\STATE \textbf{end switch}}
\newcommand{\CASE}[1]{\STATE \textbf{case} #1\textbf{:} \begin{ALC@g}}
\newcommand{\ENDCASE}{\end{ALC@g}}
\newcommand{\CASELINE}[1]{\STATE \textbf{case} #1\textbf{:} }
\newcommand{\DEFAULT}{\STATE \textbf{default:} \begin{ALC@g}}
\newcommand{\ENDDEFAULT}{\end{ALC@g}}
\newcommand{\DEFAULTLINE}[1]{\STATE \textbf{default:} }

\def\QEDmark{\ensuremath{\square}}
\def\endproof{\hfill\QEDmark}

\newcommand{\bluecomment}[1]{}
\newcommand{\redcomment}[1]{\tcc{#1}}

\newcommand{\spellcheck}[1]{#1}
\newcommand{\newredtext}[1]{#1}
\newcommand{\newbluetext}[1]{#1}

\newcommand{\profwshan}[1]{{#1}}
\newcommand{\profwshangreen}[1]{{#1}}
\newcommand{\profwshangr}[1]{#1}

\ifFullVersion
\newcommand{\revision}[1]{#1}
\newcommand{\todelete}[1]{#1}
\newcommand{\release}[1]{#1}
\else
\newcommand{\revision}[1]{#1}
\newcommand{\todelete}[1]{\textcolor{red}{#1}}
\newcommand{\release}[1]{#1}
\fi
\newcommand{\redtext}[1]{#1}
\newcommand{\bluetext}[1]{#1}

\LinesNumbered
\SetAlgoCaptionSeparator{.}
\SetKwProg{Fn}{Function}{}{end}
\SetKwFor{uForEach}{foreach}{do}{}
\SetStartEndCondition{ (}{) }{)}
\SetNlSty{texttt}{}{:}
\SetArgSty{}

\def\checkmark{\tikz\fill[scale=0.4](0,.35) -- (.25,0) -- (1,.7) -- (.25,.15) -- cycle;}

%% file: 01.introduction.tex
\vspace*{-0.2cm}\section{Introduction}\label{sec:introduction}

\begin{figure*}[htb] 
\centering
\includegraphics[width=6.3in]{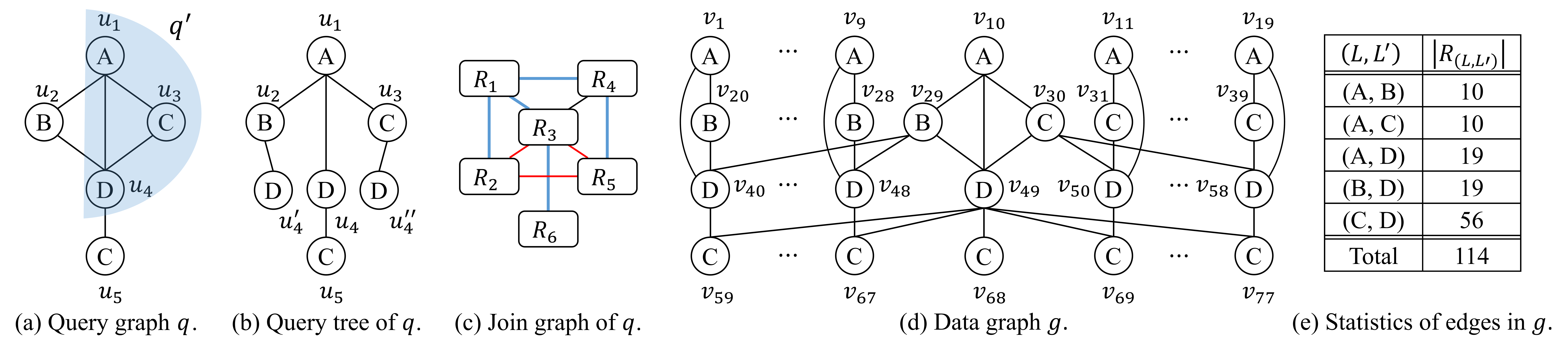}
\vspace*{-0.3cm}
\caption{\revision{A running example.}}\label{fig:running-example} %
\vspace*{-0.4cm}
\end{figure*}

\ifFullVersion
Subgraph matching is one of the most fundamental and heavily researched types of graph querying \cite{sahu2017ubiquity}. Given a query graph $\QueryGraph$ and a data graph $\DataGraph$, subgraph matching is the problem of finding the set $\Matches$ of all \newredtext{(isomorphic or homomorphic)} embeddings of $q$ in $g$. Figure \ref{fig:running-example} shows an example of subgraph matching with 20 embeddings $\{u_1 \veryshortarrow v_{10}, u_2 \veryshortarrow v_{29}, u_3 \veryshortarrow v_{30}, u_4 \veryshortarrow v_{49}, u_5 \veryshortarrow v_{30}\}$, $\{u_1 \veryshortarrow v_{10}, u_2 \veryshortarrow v_{29}, u_3 \veryshortarrow v_{30}, u_4 \veryshortarrow v_{49}, u_5 \veryshortarrow v_{59}\}$, ..., $\{u_1 \veryshortarrow v_{10}, u_2 \veryshortarrow v_{29}, u_3 \veryshortarrow v_{30}, u_4 \veryshortarrow v_{49}, u_5 \veryshortarrow v_{77}\}$. Here, $u \veryshortarrow v$ denotes the mapping from a query vertex $u$ to a data vertex $v$.
Subgraph matching has many well-known applications, including chemical compound search \cite{zhang2009gaddi}, protein interaction analysis \cite{zhao2010graph}, social analysis \cite{fan2012graph}, and \mbox{knowledge base queries \cite{hu2018answering}.}

As well as subgraph matching, graph pattern cardinality estimation (i.e., estimating the number of embeddings, $\QueryNum$) is an important problem. For example, estimates are used to determine the cost of query plans in subgraph matching systems \cite{moerkotte2009preventing}. Accurate estimation  enhances the quality of plans while efficient estimation minimizes the query optimization overhead.
Estimates can also be used in fast approximate query answering when calculating the exact answer takes too long, {and approximate results are sufficient for data analytics}. An online aggregation method can even progress the estimation towards a more accurate answer \cite{li2016wander}.

\else

\revision{Given a data graph $g$ and a query graph $q$, the graph pattern cardinality estimation problem is to estimate the number of all (isomorphic or homomorphic) embeddings of $q$ in $g$.} Figure \ref{fig:running-example} shows an example with 20 embeddings $\{u_1 \veryshortarrow v_{10}, u_2 \veryshortarrow v_{29}, u_3 \veryshortarrow v_{30}, u_4 \veryshortarrow v_{49}, u_5 \veryshortarrow v_{30}\}$, $\{u_1 \veryshortarrow v_{10}, u_2 \veryshortarrow v_{29}, u_3 \veryshortarrow v_{30}, u_4 \veryshortarrow v_{49}, u_5 \veryshortarrow v_{59}\}$, ..., $\{u_1 \veryshortarrow v_{10}, u_2 \veryshortarrow v_{29}, u_3 \veryshortarrow v_{30}, u_4 \veryshortarrow v_{49}, u_5 \veryshortarrow v_{77}\}$.  Here, $u \veryshortarrow v$ denotes the mapping from a query vertex $u$ to a data vertex $v$. \revision{This problem is important since} estimates are used to determine the cost of query plans in subgraph matching systems \cite{moerkotte2009preventing}. Accurate estimation enhances the quality of plans while efficient estimation minimizes the query optimization overhead.
Estimates can also be used in fast approximate query answering when calculating the exact answer takes too long, \redtext{and approximate results are sufficient for data analytics}. An online aggregation method can even progress the estimation towards a more accurate answer \cite{li2016wander}.

\fi

\revision{In this paper, we focus on combining two major approaches -- sampling and synopses -- to solve the graph pattern cardinality estimation problem.}
\revision{
Synopsis-based methods pre-build summary structures in the offline phase and use the structures to perform cardinality estimation in the online phase. Sampling-based methods perform sampling, calculate weights of samples, and aggregate these weights to estimate the cardinality in the online phase.}

\revision{
Synopsis-based methods have been mainly developed for RDF graphs.}
\CSET \cite{neumann2011characteristic} summarizes the data graph into a set of star-shaped structures, while \SumRDF summarizes it into a smaller graph.
\revision{
These methods are fast and accurate if synopses capture information of data graphs well.}
However, it is a well-known problem that the information loss in summarization and the ad-hoc assumptions (e.g., uniformity and independence) in computing the estimates may produce large errors \cite{park2020g, leis2017cardinality, markl2005consistently}.
On various real and synthetic datasets, {a very recent work} by Park et al.\ \cite{park2020g} shows that these methods actually suffer from serious under-estimation problems due to these limitations.

\revision{Sampling-based methods have been mainly developed for relational data, but show good performance on graph data.} 
Park and colleagues further show that, surprisingly, an online aggregation method designed for relational data, \WanderJoin, significantly outperforms all techniques designed for graph data \cite{park2020g}.
\revision{However, sampling-based methods suffer from large \emph{sample space} that leads to large estimation variance and a high probability of sampling failures (i.e., samples have zero-weights). The failures lead to significant under-estimation in sampling-based methods, including \WanderJoin, which we will show in \redtext{Section~\ref{sec:ht}}. Here, the sample space is the set of all possible random walks following a particular sampling strategy and order. {The set of candidates for each vertex/edge in order is called the \emph{local} sample space.}
We illustrate these problems using an example in Figure \ref{fig:running-example} and \WanderJoin.}

\revision{\WanderJoin for relational data can be translated into an estimator for graph data which performs \emph{edge-at-a-time} sampling, by regarding query edges as relations and overlapping query vertices as join keys. For example, the query graph in Figure \ref{fig:running-example}a is equivalent to a join graph in Figure \ref{fig:running-example}c.} \WanderJoin first determines the order of query edges, e.g., $\langle(u_1, u_2), (u_2, u_4),$ $(u_1, u_4), (u_1, u_3), (u_3, u_4),$ $(u_4,$ $u_5)\rangle$. 
Following the order, it breaks cycles and transforms the query graph into a tree with split vertices, as in Figure \ref{fig:running-example}b. For example, $u_4$ is first split into $u_4$ and $u'_4$ since the first three edges form a cycle and $u_4$ is the latest visited. $u_4$ is then split into $u_4$ and $u''_4$ again due to $(u_3, u_4)$. The sampling order is defined on the edges of the tree as {$\langle(u_1, u_2), (u_2, u'_4), (u_1, u_4),$ $(u_1, u_3), (u_3, u''_4),$ $(u_4, u_5)\rangle$}.
\revision{Blue edges in Figure \ref{fig:running-example}c represent the corresponding join path.}

\WanderJoin then randomly walks on the data graph following the sampling order. For instance, $(v_1, v_{20})$ is first uniformly sampled for $(u_1, u_2)$ out of ten data edges that match $(u_1, u_2)$, i.e., $\{(v_1, v_{20})$, $(v_2, v_{21})$, ..., $(v_{10}, v_{29})\}$. 
For the second edge $(u_2, u'_4)$, the walk continues from \redtext{$v_{20}$}. While the walk continues to sample \redtext{$(v_{20}, v_{40})$ and $(v_{1}, v_{40})$} for $(u_2, u'_4)$ and $(u_1, u_4)$, respectively, it cannot continue to sample for \redtext{$(u_1, u_3)$} since $v_1$ has no candidate that matches $(u_1, u_3)$ in its adjacency list. We say that the random walk (or sampling) failed at $(u_1, u_3)$. 
\revision{If \WanderJoin starts the walk from $(v_{10}, v_{29})$ and succeeds to sample a data edge for each query edge, it finally checks the join conditions between the split query vertices \redtext{$u_4$, $u'_4$, and $u''_4$} (equivalent to red edges in Figure \ref{fig:running-example}c)}.
For example, if $(v_{10}, v_{49})$ and $(v_{29}, v_{40})$ are sampled for $(u_1, u_4)$ and $(u_2, u'_4)$, respectively, the join fails between the split vertices $u_4$ and $u'_4$ \redtext{since $v_{49}$ is sampled for $u_4$ while $v_{40}$ is for $u'_4$}. Similarly, the condition between $u_4$ and $u''_4$ is checked. 
If any of the join conditions fails, we also say that the sampling has failed.
\revision{Sampling succeeds if and only if the sampled edges form an embedding of $q$ in $g$.}

\WanderJoin suffers from a large sample space for two main reasons. First, in performing random walks, it considers only one query edge at a time, which blindly leads to zero candidates for later edges in the sampling order. \bluetext{Second, it breaks cycles in $q$ and samples multiple data vertices for the same query vertex (e.g., $\{u_4, u'_4, u''_4\}$) which must eventually be joined. This can lead to a significant number of failures for cyclic queries.}

\revision{Since performing a successful sampling (i.e., sampling an embedding) is important in sampling-based estimators, a naive approach to combine sampling and synopses would pre-compute embeddings by subgraph matching and use them as sample space. However, this is an NP-hard problem, and since we do not know which patterns will be queried online, we might have to calculate embeddings for all possible patterns that appear in the data graph. This is, of course, an infeasible approach.}

\revision{A second naive approach would to pre-compute \emph{domains} of small patterns (e.g., up to five edges), inspired by work for frequent pattern mining in a single large graph \cite{elseidy2014grami}.} Here, the domain is defined as the set of data vertices that participate in an \revision{embedding} of a pattern. For example, the domain of $u_1$ in \redtext{$q^{\prime}$} in Figure \ref{fig:running-example}a is $\{v_{10}\}$, since only $v_{10}$ participates in an embedding of $q^{\prime}$ in $g$.
\redtext{By indexing the domains in the offline phase and using them as the local sample spaces of vertices in $q$ in the online phase, we can safely prune out the candidates that lead to sampling failures.}

However, frequent pattern mining methods \revision{(aka \emph{fpms})} \redtext{suffers from scalability for large heterogeneous graphs~\cite{abdelhamid2016scalemine, teixeira2015arabesque, jiang2013survey}.} Enumerating all small-size patterns and calculating their domains using NP-hard subgraph matching incurs significant overhead. Especially in our scenario, the scalability issue would be even more onerous since \revision{1) we do not differentiate between frequent or infrequent patterns so the number of patterns to index can be much larger, 2) we have to materialize all the domains to use them online, rather than repeatedly discarding the searched ones as in \emph{fpms}, and 3) at least a superset of each domain should be indexed (no false-negative candidates), where \emph{fpms} are optimized to probe a subset of domain and determines whether a pattern is frequent or not.}

\noindent \underline{\textbf{Contributions.}}
\revision{To address the above limitations, we present \Alley, a hybrid method that carefully combines both sampling and synopses. We first design a novel sampling strategy that reduces the sample space. Instead of using sampling only in the online phase, we interleave sampling and mining domains in the offline phase, and index only hard patterns that are inherently difficult to handle by sampling. We call} such patterns \emph{tangled}, if it is hard to reduce sample space, and the rate of sampling failures exceeds some threshold. By indexing only tangled patterns as important synopses, we can greatly improve scalability of the offline phase. \revision{Again, in the online phase, we use the domains of tangled subqueries to prune local sample spaces of queries. In short, we improve the efficiency of building synopses by combining with sampling, also the accuracy of sampling-based estimation by combining with synopses.}

\revision{
We briefly illustrate our ideas using examples, starting from a novel sampling strategy, called \emph{random walk with intersection}.}
Instead of considering only one edge at a time as in \WanderJoin, \Alley samples a vertex at a time from a reduced number of candidates by considering all its incident edges. For example, if the sampling starts from $u_1$ in Figure \ref{fig:running-example}a, \Alley intersects the candidates for $(u_1, u_2)$, $(u_1, u_3)$, and $(u_1, u_4)$ on $u_1$, i.e., $\{v_1, v_2, ..., v_{10}\} \cap \{v_{10}, v_{11}, ..., v_{19}\} \cap \{v_1, v_2, ...,$ $v_{19}\} = \{v_{10}\}$. We readily obtain the candidates that \redtext{are more likely to} lead to a successful sampling.
Along with intersections, \Alley uses \emph{branching} to bound variance and increase efficiency. Branching is to sample multiple data vertices \emph{without replacement} for a query vertex \redtext{$u$} if the number of candidates \redtext{for $u$} is large.
\revision{These enable a significant accuracy increase with a reasonable trade-off of efficiency.} 

\begin{figure}[h!]
\centering
\vspace*{-0.3cm}
\includegraphics[width=0.75\columnwidth]{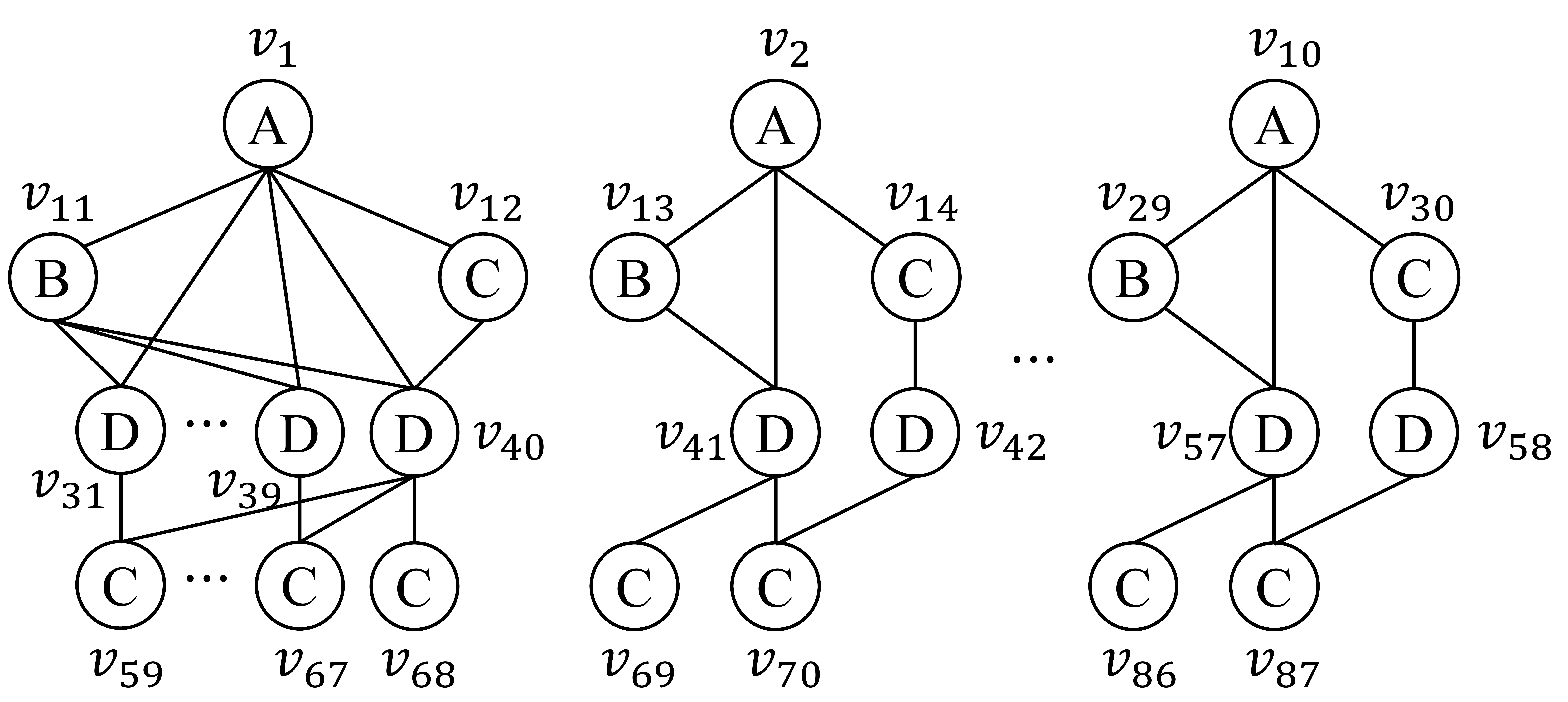}
\vspace*{-0.3cm}
\caption{Another data graph $g_2$ with same edge statistics.}
\ifFullVersion
\vspace*{-0.2cm}
\else
\vspace*{-0.5cm}
\fi
\label{fig:index-example}
\end{figure}

\revision{For indexing tangled patterns, we use a novel mining approach, called \emph{walk-fail-then-calculate}.} \Alley performs random walks using the domains of smaller patterns and calculates the domains if the rate of sampling failures exceeds some threshold. For a data graph $g_2$ in Figure \ref{fig:index-example} and a subquery \redtext{$q^{\prime}$} in Figure \ref{fig:running-example}a, the number of embeddings of $q^{\prime}$ is just one, i.e., $\{u_1 \veryshortarrow v_1, u_3 \veryshortarrow v_{12}, u_4 \veryshortarrow v_{40}\}$. However, the number of candidates for any vertex $u$ in \redtext{$q^{\prime}$}\redtext{, i.e., the local sample space for $u$,} is at least \redtext{10, even if} we consider the intersections. \revision{The domains of smaller subqueries, e.g., $\{(u_1, u_3), (u_1, u_4)\}$, have no effect on pruning local sample spaces of $q^{\prime}$.}
Therefore, a random walk will fail with a probability near $0.9$. If we set the threshold for sampling failures less than $0.9$, say $0.8$, such patterns can be considered tangled. Then, we index the domains for $u_1$, $u_3$, and $u_4$ of \redtext{$q^{\prime}$}, which are \redtext{$\{v_1\}$, $\{v_{12}\}$, and $\{v_{40}\}$}, respectively. 
\ifFullVersion
\redtext{Along with our novel mining approach, we propose \revision{three} optimization techniques based on the fact that we only need to obtain reduced local sample spaces, not necessarily the exact domains of each tangled pattern, \revision{and we can determine tangled patterns without performing too many random walks.}}
\fi



In addition to its practicality, we establish that \Alley has formal guarantees on estimation error, confidence, and time complexity as follows: given an error bound $\epsilon$ and confidence $\mu$, \Alley guarantees $\Pr(\abs{Z-\QueryNum} < \epsilon \cdot \QueryNum)$ $> \mu$ always in $O(\AGM)$ time, where $Z$ is the random variable for the cardinality estimate and $\AGM$ is the worst-case optimal bound of $\QueryNum$ \cite{ngo2012worst}.
Achieving this guarantee is meaningful since it bounds both estimation error and runtime. 
In addition, an emerging class of subgraph matching algorithms that run in $O(\AGM)$ time has recently been studied actively \cite{ngo2012worst, DBLP:journals/pvldb/MhedhbiS19}.
\ifFullVersion
\fi
In summary, in this paper
\begin{itemize}[topsep=0pt]
  \item We show that small sample space leads to small variance and fewer sampling failures, resulting in high accuracy in sampling-based estimators (Section \ref{sec:ht}).
  \item We present \Alley, an accurate and efficient graph pattern cardinality estimator based on a novel sampling strategy (Sections \ref{sec:overview}-\ref{sec:sampling}).
  \item \bluetext{We propose a novel mining method to increase effectiveness for \redtext{tangled patterns} and to make \Alley a hybrid method that combines both sampling and synopses (Section \ref{sec:index}).}
  \item We prove the probabilistic theoretical guarantees of \Alley in worst-case optimal time (Section \ref{sec:proof}). 
  \item With extensive experiments, we establish that \redtext{\Alley consistently and significantly outperforms all the state-of-the-art estimators (Section \ref{sec:exp}). Specifically, \Alley outperforms \WanderJoin by up to two orders of magnitude in terms of accuracy with similar efficiency.}
\end{itemize}

%% file: 02.background.tex
\section{Background} \label{sec:prelim}


\subsection{Problem Definition} \label{subsec:problem}

An undirected labeled graph is a triple, ($V$, $E$, $L$), such that $V$ is a set of vertices, $E$ is a set of edges, and $L$ is a label function that maps a vertex to a set of labels. Given two labeled graphs $\QueryGraph$ and $\DataGraph$, we can define a match or an embedding between $q$ and $g$ as follows \cite{kim2015taming}.

\begin{definition2}
\label{def:match}
A graph $q$ = $(V_q, E_q, L_q)$ is homomorphic to a subgraph of a data graph $g$ = $(V_g, E_g, L_g)$ if there is a mapping (or an embedding) $m: V_q \rightarrow V_g$ such that 1) $\forall u \in V_q$, $L_q(u) \subseteq L_g(m(u))$, and 2) $\forall (u, u') \in E_q$, $(m(u), m(u')) \in E_g$.
\end{definition2}

Here, $(u, u')$ represents an undirected edge between vertex $u$ and $u'$.
Given a query graph $\QueryGraph$ and a data graph $\DataGraph$, let $\Matches$ denote the set of all embeddings of $\QueryGraph$ in $\DataGraph$. Finding $\Matches$ is called the \emph{subgraph matching} problem, and estimating $\abs{\Matches}$ is called the \emph{graph pattern cardinality estimation} problem.

We assume that $\QueryGraph$ is connected, otherwise we can estimate the cardinalities of disjoint subqueries and multiply them.
For ease of explanation, we assume \redtext{edges are not labeled}, and the query and the data graph are not multi-graphs. 
The actual implementation of \Alley supports directed, edge-labeled, and multi-graphs as well.

\vspace{-0.2cm}
\subsection{Notation} \label{subsec:notations}

We next present notation used throughout the paper (Table \ref{table:notations}). 
Set \redtext{operations} are often used, for example, \release{we regard $e = \{u, u'\}$ for an edge $e = (u, u')$.} 
An embedding $m$ is also a set of mappings $u \veryshortarrow v$ where $u \in V_q$ and $v \in V_g$. If $u \veryshortarrow v \in m$, we write $m(u) = v$. 

We also use notation from relational algebra. Given a query edge $e$, $R_e$ denotes the set of data edges that match $e$. Hence, $R_e = R_{(u, u')} := \{(v, v') \in E_g \,|\, L_q(u) \subseteq L_g(v) \wedge L_q(u') \subseteq L_g(v')\}$. 
\bluetext{Since $R_{(u, u')}$ is determined by labels of $u$ and $u'$, we also use $R_{(L, L')}$ ($L = L_q(u)$, $L' = L_q(u')$).}
This corresponds to a binary relation in relational algebra, regarding $u$ and $u'$ as two attributes and each $(v, v') \in R_{(u, u')}$ as a tuple. \revision{Note that, a subgraph matching query can be translated into a natural join query $\bowtie_{e \in E_q}{R_e}$ as in \mbox{Figure \ref{fig:running-example}c.}}

We use \profwshangreen{a} set projection of $R_{(u, u')}$ onto $u'$ as $\pi_{u'}(R_{(u, u')}) := \{v' \in V_g \,|\, \exists v \in V_g: (v, v') \in R_{(u, u')}\}$ \revision{which we simplify as
$V^{u'}_{u}$, representing the set of data vertices that match $u'$ and have an incident edge in $R_{(u, u')}$.}
\revision{We also use $adj^{u'}_{u}(v) := \{v' \in V^{u'}_{u} \,|\, (v, v') \in E_g\}$, which is the set of adjacent vertices of $v$ in $V^{u'}_{u}$. For example, $V_{u_2}^{u_1} = \{v_1, v_2, ..., v_{10}\}$ and $adj_{u_2}^{u_1}(v_{20}) = \{v_1\}$ in Figure \ref{fig:running-example}.}


\begin{table}[htp]
\scriptsize
\vspace*{-0.2cm}
\caption{\revision{Notation used in the paper.}}
\vspace{-0.3cm}
\small
\begin{tabular}{ |c|l| }
 \hline
 
 $\Matches$ & set of \revision{embeddings} of $\QueryGraph$ in $\DataGraph$ \\
 \hline
 $\SampleSpace$, $P_{u}$ & sample space and local sanple space for $u$ \\
 \hline
 $p$, $w(p)$ & potential embedding in $\SampleSpace$ and weight of $p$ \\
 \hline
 $\mathbb{I}(p)$ & indicator variable for $p$ whether $p \in \Matches$ or not \\
 \hline
 
 $R_e$ & set of data edges that match a query edge $e$ \\
 \hline
 \revision{$V^{u'}_{u}$} & \revision{set of data vertices that match $u'$} \\
 & \revision{and have an incident edge in $R_{(u, u')}$} \\
 \hline
 
 \revision{$adj^{u'}_{u}(v)$} & \revision{set of adjacent vertices of $v$ in $V^{u'}_{u}$} \\
 \hline
 $o$ & sampling order of query vertices (or edges) \\
 \hline
 $D_q(u)$ & domain of a vertex $u$ in a pattern $q$ \\
 \hline
\end{tabular}\label{table:notations}
\ifFullVersion
\else
\vspace*{-0.4cm}
\fi
\end{table}

\subsection{AGM Bound} \label{subsec:agm}

\bluetext{The AGM bound \cite{atserias2008size} is defined for a given query graph $q$ and a set of data graphs $G_q$, denoted as $\AGM$. Here, $G_q$ consists of all possible data graphs with \redtext{the same} $\abs{R_e}$ for each $e \in E_q$ \release{(e.g., Figures~\ref{fig:running-example}d and~\ref{fig:index-example} have the same edge statistics in Figure \ref{fig:running-example}e.)}}

The AGM bound is worst-case optimal in the sense that it is a \emph{tight} upper bound of $\QueryNum$, i.e., there exists a data graph $g^* \in G_q$ that satisfies $\abs{\Matches} = \AGM$.
If a subgraph matching algorithm runs in  $O(\AGM)$ time, the algorithm is also called worst-case optimal. \release{Recent systems \cite{aref2015design, kankanamge2017graphflow, aberger2016emptyheaded} implement such algorithms.}
\ifFullVersion
As in typical worst-case optimal algorithms \cite{assadi2018simple, ngo2014skew}, we omit $|V_q|$ and $\log|E_g|$ term in $O$ notation.
\fi



%% file: 03.ht.tex
\section{HT Estimator}\label{sec:ht}

Horvits-Thompson (HT) estimators are a class of unbiased sampling-based estimators \cite{sirken1999population}.
In our problem, an HT estimator samples data vertices/edges for query vertices/edges and checks whether the sampled vertices/edges form an \revision{embedding} of $q$ in $g$. Here, sampled vertices/edges may not form an \revision{embedding} but any (partial) mapping, so we call each sample a \emph{potential} embedding of $q$ as \cite{assadi2018simple}.


After sampling a potential embedding $p$, the estimator assigns a \emph{weight}, $w(p)$, as $w(p) = 1/\Pr(p) \cdot \mathbb{I}(p)$. Here, $\Pr(p)$ represents the probability of sampling $p$ dependent on a sampling method, and $\mathbb{I}(p)$ is an indicator variable for sampling failure of $p$, i.e.,  $\mathbb{I}(p) = 1$ if $p \in \Matches$ ($p$ is an \revision{embedding} of $q$) and $\mathbb{I}(p) = 0$ otherwise. This weight is used as the estimated cardinality. \bluetext{For instance, using \WanderJoin as an HT estimator in our running example in Figure \ref{fig:running-example}, $\Pr(m) = \frac{1}{10} \cdot \frac{1}{10}  \cdot \frac{1}{1} \cdot \frac{1}{1} \cdot \frac{1}{1} \cdot \frac{1}{20}$ for any $m \in \Matches$ \redtext{whose weight $w(m)$ is $1/\Pr(m) \cdot \mathbb{I}(p) = 2,000 \cdot 1 = 2,000$. The sampling is repeated $s$ times, and the final estimate is calculated as the average of weights, e.g., if $s=1,000$ and the sampling failed for 999 times out of 1,000 repetitions, the estimate becomes $\frac{2,000}{1,000} = 2$ in the above example. Here, $s$ is called the \emph{sample size}. The whole process is given in Algorithm \ref{alg:ht} with an additional step of determining the \emph{sampling order} of query vertices/edges.}}


\vspace{-2ex}
\setlength{\textfloatsep}{4pt}
\begin{algorithm} [htb]
\caption{{SimpleHTEstimator}($\QueryGraph, \DataGraph, s$)} \label{alg:ht}
\small{
    \KwIn{\mbox{A query graph $q$, a data graph $g$, and a sample size $s$}}
    $o \leftarrow \ChooseSamplingOrder(\QueryGraph, \DataGraph)$ \\
    $sum \leftarrow 0$ \\
    \ForEach{$i = 1$ to $s$}
    {
        $p \leftarrow \SamplePotentialEmbedding(o, \QueryGraph, \DataGraph)$ \\
        $sum \leftarrow sum + w(p)$ \\
    }
    
    \Return $sum$/$s$ \\
}
\end{algorithm}

\vspace*{-0.5cm}
\subsection{Characteristics of HT Estimators}\label{subsec:char}

HT estimators have two important characteristics: they are unbiased and there is a fundamental trade-off between estimation accuracy and efficiency.
\ifFullVersion

The unbiasedness can be explained as follows. Let $Z_i$ denote the random variable for $w(p_i)$ where $p_i$ is the $i$'th potential embedding sampled ($1 \leq i \leq s$). Then, let $Z = \sum^{s}_{i=1}{Z_i}/s$ denote the random variable for the estimated cardinality. 

\vspace{-0.2cm}
\begin{equation}\label{eq:unbiased1}
\begin{split}
\E[Z_i] = \sum_{p}{\Pr(p) \cdot \frac{1}{\Pr(p)} \cdot \mathbb{I}(p)} = \sum_{p}{\mathbb{I}(p)} = \sum_{p \in \Matches}{1} = \QueryNum
\end{split}
\end{equation}

\vspace{-0.2cm}
\begin{equation}\label{eq:unbiased2}
\E[Z] = \E\bigg[ \frac{\sum_{i=1}^{s}{Z_i}}{s} \bigg] = \sum_{i=1}^{s}{ \frac{\E[Z_i]}{s} } = \QueryNum
\end{equation}

(\ref{eq:unbiased2}) holds since \profwshangreen{the }$Z_i$'s are identical and independent. The accuracy-efficiency trade-off can be explained as follows\profwshangreen{:} 

\vspace{-0.2cm}
\begin{equation}\label{eq:tradeoff}
\Var[Z] = \Var\bigg[ \frac{\sum_{i=1}^{s}{Z_i}}{s} \bigg] = \sum_{i=1}^{s}{ \frac{\Var[Z_i]}{s^2} } = \frac{\Var[Z_i]}{s}
\end{equation}

Thus, increasing the sample size $s$ increases the runtime (i.e., decreases the efficiency) of an HT estimator but reduces the estimation variance (i.e., increases the accuracy).

\else
Let $Z_i$ denote the random variable for $w(p_i)$ where $p_i$ is the $i$'th potential embedding sampled ($1 \leq i \leq s$), and let $Z = \sum^{s}_{i=1}{Z_i}/s$ denote the random variable for the estimated cardinality. \revision{Then, the following equations hold: $\E[Z] = \E[Z_i] = \QueryNum, \,\, \Var[Z] = \Var[Z_i]/{s}$. The proofs for all equations are explained in \cite{AAA}.} 
Thus, increasing the sample size $s$ increases the runtime (i.e., decreases the efficiency) of an HT estimator but reduces the estimation variance (i.e., increases the accuracy).

\fi


\vspace*{-0.1cm}
\subsection{Sample Space}\label{subsec:samplespace}

Each HT estimator has a different sampling strategy that determines the set of potential embeddings to be sampled. For example, an HT estimator $T_1$ samples any data edge from $E_g$ for each query edge, while another HT estimator $T_2$ samples a data edge from $R_e$ for each query edge $e$. 
\redtext{Obviously, $T_1$ can sample (useless) potential embeddings which $T_2$ does not.}
If $\SampleSpace_{T}$ denotes the set of all potential embeddings that can be sampled by an estimator $T$, then $\SampleSpace_{T_2} \subset \SampleSpace_{T_1}$. Here, $\SampleSpace$ is called the \emph{sample space}. If $\Matches \subseteq \SampleSpace_{T}$, then we say that $T$ is \emph{consistent}. So far, we assumed that the estimator is consistent.

Not surprisingly, the sample space plays a critical role in estimation accuracy.
\ifFullVersion
Since $\Var[Z_i] = \E[Z_i^2] - \E[Z_i]^2$, the following equation holds for a consistent HT estimator.

\vspace{-2ex}
\begin{equation}\label{eq:variance}
\begin{split}
\Var[Z_i] = \bigg(\sum_{p}{ \Pr(p) \cdot \Big({\frac{1}{\Pr(p)} \cdot \mathbb{I}(p)}\Big)^2 }\bigg) - \E[Z_i]^2 \\ 
 = \sum_{p}{\frac{1}{\Pr(p)} \cdot \mathbb{I}(p)} - \QueryNum^2 
 = \sum_{p \in \Matches}{ \frac{1}{\Pr(p)} } - \QueryNum^2.
\end{split}
\end{equation}
\vspace{-2ex}
\else
The following equation holds for a consistent HT estimator: \revision{$\Var[Z_i] = \sum_{p \in \Matches} 1/\Pr(p) - \QueryNum^2$.}
\fi

That is, the estimation variance depends on the probabilities of sampling \revision{embedding}s of $q$. If the probabilities are low, then the variance will be large. To increase such probabilities, \bluetext{we} \profwshangreen{must} \bluetext{reduce the sample space size $\abs{\SampleSpace}$ to decrease the probabilities of sampling $p \not\in \Matches$ and increase the probabilities of sampling $p \in \Matches$.} 
\ifFullVersion
If we assume uniform sampling, i.e., $\Pr(p) = 1/\abs{\SampleSpace}$, (\ref{eq:variance}) can be approximated as follows\profwshangreen{:}

\vspace{-2ex}
\begin{equation}\label{eq:space}
\revision{\Var[Z_i] \approx \QueryNum \cdot \abs{\SampleSpace} - \QueryNum^2.}
\end{equation}
\vspace{-2ex}

\else
If we assume uniform sampling, i.e., $\Pr(p) = 1/\abs{\SampleSpace}$, $\Var[Z_i]$ can be approximated as follows\profwshangreen{:} \revision{$\Var[Z_i] \approx \QueryNum \cdot \abs{\SampleSpace} - \QueryNum^2$.}

\fi

Thus, the variance is roughly proportional to the size of the sample space.
If $\abs{\SampleSpace}$ is significantly larger than $\abs{\Matches}$, it means that most of the potential embeddings sampled are not \revision{embeddings}. In Line 4 of Algorithm \ref{alg:ht}, it is highly likely that $p \not\in \Matches$, resulting in sampling failures. Since the weight is zero for such samples, this can cause a serious under-estimation problem as shown in \cite{park2020g}. For example, $\abs{\SampleSpace_{T_1}}$ is $114^{6} \approx 2 \cdot 10^{12}$ since there are 114 data edges and six query edges in Figure \ref{fig:running-example}. For $T_2$, $\abs{\SampleSpace_{T_2}}$ is $\prod_{e \in E_q}{\abs{R_e}} \approx \redtext{10^8}$. Note that both $T_1$ and $T_2$ are consistent, but have significantly large sample spaces compared to $\abs{\Matches} = \redtext{20}$. In the \redtext{following sections}, we show how we reduce the sample space. 

%% file: 04.overview.tex
\vspace{-0.1cm}

\section{Overview of Solution} \label{sec:overview}
\revision{As explained in Section \ref{sec:introduction}, a naive combination approach would mine embeddings or domains of all small-size patterns that appear in the data graph. However, this might require a tremendous amount of computation. Instead, our approach interleaves sampling during the mining to improve mining efficiency, and 
utilizes the mined domains to improve online estimation accuracy.}

\begin{figure}[h!]
\centering
\vspace*{-0.3cm}
\includegraphics[width=0.7\columnwidth]{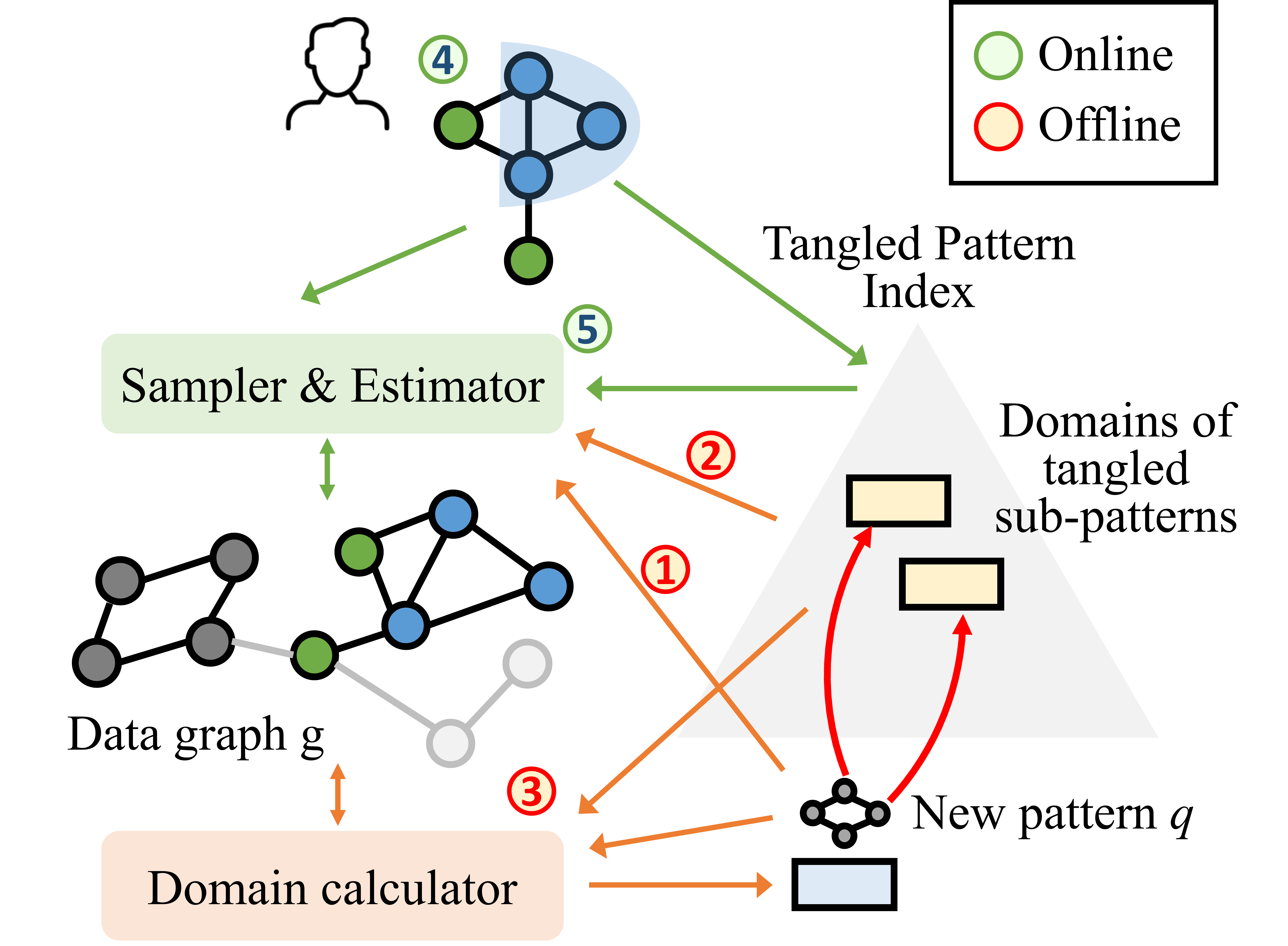}
\vspace*{-0.3cm}
\caption{\revision{Architecture of \Alley.}}
\ifFullVersion
\else
\vspace*{-0.3cm}
\fi
\label{fig:architecture}
\end{figure}

\revision{Figure~\ref{fig:architecture} shows the overall architecture of \Alley. 
In the offline phase, the \emph{tangled pattern index} is built based on the data graph $g$ only. The goal is to automatically find all tangled patterns, which are expected to result in low accuracy in the online phase.
The patterns are investigated from smaller to larger ones, that is, each pattern is extended from its sub-patterns. A new pattern $q$ is input to our sampling-based estimator (\textcircled{\small{1}}) along with the indexed domains of its sub-patterns (\textcircled{\small{2}}), which are used to prune the sample space for $q$. If the ratio of sampling failures for $q$ exceeds some threshold, i.e., the pruning effect is poor, the domains of $q$ are calculated (\textcircled{\small{3}}) and indexed as synopses. Thus, Alley interleaves sampling with mining, that is, sampling is performed to judge the tangledness of a pattern. In the online phase, given a query graph specified by a user (\textcircled{\small{4}}), we use the built index to prune the sample space in estimating the cardinality of the query (\textcircled{\small{5}}). 
We first explain online sampling-based estimation and move to building synopses offline.} 

%% file: 05.sampling.tex
\ifFullVersion
\else
\vspace*{-0.2cm}
\fi
\revision{\section{Sampling-based Estimation} \label{sec:sampling}}

\revision{This section explains the main algorithm of online estimation (Algorithm \ref{alg:main}). The algorithm consists of three main parts, 1) choosing sampling order, 2) searching domains, and 3) random walk with intersection.}




\ifFullVersion
\else
\vspace*{-0.2cm}
\fi
\setlength{\textfloatsep}{4pt}
\begin{algorithm} [htb]

\caption{{\Alley}($\QueryGraph, \DataGraph, s, I$)} \label{alg:main}
\small{
    \KwIn{A query graph $q$, a data graph $g$, a sample size $s$, and {a tangled pattern index $I$ of $g$}}
    $o \leftarrow \ChooseSamplingOrder(\QueryGraph, \DataGraph)$ \label{Algo1:samplingorder}  \leavevmode\\
    {$\{D_{q_j}\} \leftarrow \SearchDomainsRecursive(\QueryGraph, o, I)$} \label{Algo1:searchdomains} \leavevmode\\
    $sum, count \leftarrow 0, 0$ \redcomment{sum of weights and \# of calls} 
    $p \leftarrow \emptyset$ \redcomment{initialize potential embedding} 
    \While{$s > 0$}
    {\label{Algo1:whilestart}
        $w_{1}, s_{1} \leftarrow \SamplePotentialEmbeddings(q, g, \redtext{\{D_{q_j}\}}, o, p, 1)$ \label{Algo1:Sample}  \leavevmode\\
        $sum \leftarrow sum + w_{1}$  \leavevmode\\
        $s \leftarrow s - s_{1}$; $count \leftarrow count + 1$  \leavevmode\\
    } \label{Algo1:whileend}
    \Return {$sum/count$} \label{Algo1:returnt}
}
\end{algorithm}

\ifFullVersion
\else
\vspace*{-0.3cm}
\fi
\noindent \underline{\textbf{\ChooseSamplingOrder.}}
We choose the sampling order $o$ (Line \ref{Algo1:samplingorder}) by adopting a simple greedy approach considering the selectivity of each vertex. Specifically, we define the rank for each vertex $u$ as $rank(u) := \min_{(u, u') \in E_q}$\revision{${\abs{V^{u}_{u'}}}$}. This means the set of candidates that match $u$ considering the query edge $e$ only. Then, we select the minimum-rank vertex as the starting query vertex. 
\ifFullVersion
For a tie, we consider two additional ranks, one is $\min_{e \in E_q, e \ni u}{\abs{R_e}}$ and the other one is the degree of $u$.
\fi
For remaining vertices, we repeatedly select the vertex who has the largest number of neighboring selected vertices. 
\ifFullVersion
For a tie, we choose one with a smaller rank.
\fi
While selecting a good sampling order is an important research problem as noted in typical subgraph matching papers~\cite{zhao2018random, han2019efficient}, we leave it for future investigation.




\ifFullVersion
\begin{example}
In Figure \ref{fig:running-example}, $rank(u_1) =$ \revision{$\min_{(u_1, u)}{|{V^{u_1}_{u}}|}$ = $\min \{|V^{u_1}_{u_2}|$, $|V^{u_1}_{u_3}|, |V^{u_1}_{u_4}|\}$} = $\min\{10, 10, 19\}=10$. Similarly, $rank(u_2) = rank(u_3)$ $= rank(u_4) = 10$, and $rank(u_5) = \redtext{29}$. \redtext{We then compare $\min_{e \ni u}{\abs{R_e}}$ for the four tied vertices except $u_5$. Again, there's a tie between $u_1$, $u_2$, and $u_3$. We finally choose $u_1$ as the starting query vertex since it has the largest degree (=3) among $u_1$, $u_2$, and $u_3$.} In subsequent examples, we use \redtext{$o = \langle u_1, u_2, u_4, u_3, u_5 \rangle$} as the sampling order of \Alley. Since using the data graph in Figure~\ref{fig:index-example} results in the same rank values, the same sampling order $o$ is chosen. 
\end{example}
\else
\begin{example}
\vspace*{-0.2cm}
In Figure \ref{fig:running-example}, $rank(u_1) =$ \revision{$\min_{(u_1, u)}{|{V^{u_1}_{u}}|}$ = $\min \{|V^{u_1}_{u_2}|$, $|V^{u_1}_{u_3}|, |V^{u_1}_{u_4}|\}$} = $\min\{10, 10, 19\}=10$. Similarly, $rank(u_2) = rank(u_3)$ $= rank(u_4) = 10$, and $rank(u_5) = \redtext{29}$. We choose $u_1$ as the starting query vertex. In subsequent examples, we use \redtext{$o = \langle u_1, u_2, u_4, u_3, u_5 \rangle$} as the sampling order of \Alley. Since using the data graph in Figure~\ref{fig:index-example} results in the same rank values, the same sampling order $o$ is chosen.
\end{example}
\fi

\noindent \underline{\textbf{\redtext{\SearchDomainsRecursive.}}} \redtext{Next, for each vertex $u$, we find a domain in the index by searching for subqueries $q_j$ of $q$ that contain $u$ (Line~\ref{Algo1:searchdomains}). The searched domain is used for pruning a local sample space for $u$. The detailed algorithm will be explained in Section~\ref{subsec:online}.}

\begin{example}
\vspace*{-0.2cm}
\label{example:index}
\redtext{Assume that the index of the graph in Figure~\ref{fig:index-example} stores only $q'$ in Figure \ref{fig:running-example}a and its domains for simplicity. Since $q'$ is a subquery of $q$, we search for $q'$ in the index. Then, we can use the searched domains $D_{q'}$ of $q'$ which are $\{v_1\}$, $\{v_{12}\}$, and $\{v_{40}\}$, as the candidates of $u_1$, $u_3$, and $u_4$, respectively.}
\end{example}

\noindent \underline{\textbf{\SamplePotentialEmbeddings.}} \redtext{We then iteratively sample potential embeddings. As explained in Algorithm \ref{alg:ht}, } we calculate the weight \redtext{$w_{1}$} until we run out of sample size (Lines \redtext{\ref{Algo1:whilestart}-\ref{Algo1:whileend}}), and finally return the average of $w_{1}$ as the estimated cardinality (Line \redtext{\ref{Algo1:returnt}}).
However, the difference with Algorithm \ref{alg:ht} is that each (recursive) call to \SamplePotentialEmbeddings\xspace can sample $s_{i}$ $(\geq 1)$ potential embeddings~\redtext{(we call this \textit{branching}), and $w_{i}$ is the \emph{aggregated weight} of $s_{i}$ potential embeddings at each recursion depth $i$ ($1 \leq i \leq |V_q|$). This variation is essential to guarantee the worst-case optimal time complexity and approximation quality of the algorithm, which will be explained in Section~\ref{sec:proof}.}
\bluetext{Note that \Alley is a stack of HT estimators and thus unbiased, also proved in \mbox{Section \ref{sec:proof}.}}

\subsection{Random Walk with Intersection} \label{subsec:multiway}

We now explain how \SamplePotentialEmbeddings\xspace works in Algorithm \ref{alg:alley_sample}. At recursion depth $i$, the function first calculates the set \profwshan{$P_{o_i}$} of candidates for $o_i$ given the current potential embedding ($\{o_j \veryshortarrow v_j\}_{j < i}$). It then randomly walks to candidate vertices, extends the potential embedding, and continues to the next recursion. 
\revision{For simplicity, we omit in Algorithm \ref{alg:alley_sample} that, if the total number of random walks exceeds $s$, we stop further random walks and calculate the estimate from the collected random walks so far.}
\ifFullVersion
\else
\vspace*{-0.3cm}
\fi
\setlength{\textfloatsep}{4pt}
\begin{algorithm} [htb]
\caption{\mbox{{\SamplePotentialEmbeddings}($\QueryGraph, \DataGraph, \{D_{q'}\}, o, p, i$)}} \label{alg:alley_sample}
\small{
    \KwIn{\mbox{A query graph $q$, a data graph $g$, a sampling order $o$}, \mbox{current potential embedding $p$, and current recursion depth $i$}}
    \redtext{$P_{o_i} \leftarrow \Intersect(g, p, \{D_{q'}\}, o_i)$ \redcomment{multi-way intersection}}\label{alg:intersect} 
    \If{$P_{o_{i}} = \emptyset$} 
    {\label{alg:emptystart}
       \Return $0, 1$ \label{alg:emptyend} \\
    }
    \If{$i = \abs{V_q}$}
    {\label{alg:foundstart}
       \Return $\abs{P_{o_i}}, 1$\label{alg:foundend} \\
    }
    
    \redcomment{branching, sample without replacement}
    $C \leftarrow \ChooseRandomVertices\big(P_{o_{i}}, \big\lceil b \cdot \abs{P_{o_i}} \big\rceil\big)$ \label{alg:branching} \leavevmode \\
    
    $sum, num \leftarrow 0$ \redcomment{sum of weights and \# of recursion} 
    \ForEach{$v \in C$}
    {\label{alg:randomwalk_start}
        $p \leftarrow p \cup \{o_i \veryshortarrow v_i\}$ \redcomment{$p(o_i) = v_i$} 
        $w_{i+1}, s_{i+1} \leftarrow \SamplePotentialEmbeddings(\QueryGraph, \DataGraph, \redtext{\{D_{q'}\}}, o, p, i+1)$ \leavevmode \\
        $num \leftarrow num + s_{i+1}$; $sum \leftarrow sum + w_{i+1}$ \leavevmode \\
        $p \leftarrow p \setminus \{o_i \veryshortarrow v\}$ \redcomment{$p(o_i) = NIL$} 
    }\label{alg:randomwalk_end}
    \Return $\abs{P_{o_i}} \cdot sum/|C|, num$ \label{alg:return_sample}
}
\end{algorithm}
\ifFullVersion
\else
\vspace*{-0.3cm}
\fi

\redtext{\noindent \underline{\textbf{\Intersect.}} First, we calculate $P_{o_i}$ which is the local sample space for $o_i$ using a \profwshan{multi-way} intersection (Line \ref{alg:intersect}). Given the current potential embedding $p$, we perform an intersection \revision{$\big(\cap_{(o_i, o_j) \in E_q, j < i}{adj^{o_i}_{o_j}(p(o_j))}\big) \cap \big(\cap_{(o_i, o_j) \in E_q j \geq i}{V^{o_i}_{o_j}}\big)$}.} \revision{If $D_{q'}(o_{i})$ has been searched, \revision{since we assure $D_{q'}(o_{i}) \subseteq \big(\cap_{(o_i, u) \in E_{q'}}{V^{o_i}_{u}}\big)$, we can remove such $V^{o_i}_{u}$ terms (already considered in calculating $D_{q'}(o_i)$) from our initial equation and intersect with $D_{q'}(o_i)$ instead.}}


If $P_{o_i}$ is empty, it indicates that the sampling has failed at $o_i$. We directly return $w_i = 0$ and $s_i = 1$ (Lines \ref{alg:emptystart}-\ref{alg:emptyend}).
If $i = \abs{V_q}$, the sampling has succeeded with $\abs{P_{o_i}}$ \revision{embeddings} found (Lines \ref{alg:foundstart}-\ref{alg:foundend}). In Section \ref{sec:proof}, we prove that for each of $v \in P_{o_{|{V_{q}}|}}$, $p \cup \{o_{|{V_q}|} \veryshortarrow v\}$ forms an \revision{embedding} of $q$. 

\begin{example}
\vspace*{-0.2cm}
\label{example:Fig1-intersect}
\redtext{
For the graph in Figure~\ref{fig:running-example} and $i = 1$ where $o_1 = u_1$, \revision{we compute $P_{o_1}$ by intersecting three sets, i.e., $V^{u_1}_{u_2}$ $\cap$ $V^{u_1}_{u_3}$ $\cap$ $V^{u_1}_{u_4}$} = $\{v_{1}, v_{2}, ..., v_{10}\}$ $\cap$ $\{v_{10}, v_{11}, ...,$ $v_{19}\}$ $\cap$ $\{v_{1}, v_{2}, ..., v_{19}\} = \{v_{10}\}$. As in this example, we can effectively reduce the local sample space through \profwshan{3-way} intersection.
}
\end{example}
\begin{example}
\vspace*{-0.2cm}
\label{example:Fig2-noindex}
\redtext{
We now consider $g_{2}$ in Figure~\ref{example:index} and the query graph in Figure~\ref{fig:running-example}a. We compute an intersection of three sets for $P_{o_1}$, i.e., \revision{$V^{u_1}_{u_2}$ $\cap$ $V^{u_1}_{u_3}$ $\cap$ $V^{u_1}_{u_4}$} = $\{v_{1}, v_{2}, ...,$ $v_{10}\}$ $\cap$ $\{v_{1}, v_{2}, ..., v_{10}\}$ $\cap$ $\{v_{1},$ $v_{2},$ $...,$ $v_{10}\} = \{v_{1}, v_{2},$ $...,$ $v_{10}\}$. 
} \profwshan{Unfortunately, the intersection does not prune candidates in contrast to Example~\ref{example:Fig1-intersect}.}
\end{example}
\begin{example}
\vspace*{-0.2cm}
\redtext{
Unlike Example~\ref{example:Fig2-noindex}, if $D_{q'}(u_{1})=\{v_{10}\}$ is available, we only need to intersect $D_{q'}(u_{1})$ with \revision{$V^{u_1}_{u_2}$} to find the local sample space $P_{o_{1}}$ for $u_{1}$, since the other two edges $(u_{1}, u_{3})$ and $(u_{1}, u_{4})$ are already considered in $D_{q'}(u_{1})$. Then, we calculate $D_{q'}(u_{1})$ $\cap$ \revision{$V^{u_1}_{u_2}$} $=$ $\{v_{10}\} \cap \{v_{1}, v_{2}, ... , v_{10}\}=\{v_{10}\}$. This example shows that using domains for tangled patterns can further reduce local sample space.
}
\end{example}

We then sample a vertex from $P_{o_i}$ and walk to the vertex, extending $p$ and calling \SamplePotentialEmbeddings\xspace recursively (Lines \ref{alg:randomwalk_start}-\ref{alg:randomwalk_end}).
Each call returns \redtext{$w_{i+1}$ and $s_{i+1}$} where $w_{i+1}$ denotes the aggregated weight of $s_{i+1}$ potential embeddings sampled in the call. This $w_{i+1}$ is averaged and multiplied with $|{P_{o_i}}|$, then returned with the summation of $s_{i+1}$ (Line \ref{alg:return_sample}). 

\noindent \underline{\textbf{Edge/Vertex-at-a-Time Sampling.}}
\revision{
We define edge-at-a-time sampling as sampling a data edge for each query edge, and vertex-at-a-time sampling as sampling a data vertex for each query vertex considering all its incident edges. To sample an embedding that matches the query, each strategy requires sampling $|E_q|$ and $|V_q|$ times, respectively. In general, $|E_q|$ is larger than $|V_q|$ (e.g., $|E_q|$ = 6, $|V_q|$ = 5 in Figure \ref{fig:running-example}a). By considering all incident edges of each query vertex, the number of candidates is smaller in vertex-at-a-time sampling, although it’s bounded to the number of outgoing edges ($|adj_{o_i}^{o_j}|$) as in edge-at-a-time sampling except for $|P_{o_1}|$, which is bounded to $|V_{o_1}^{u}|$. In Section \ref{subsec:exp_sampling_failure}, we report the number of candidates in \Alley (vertex-at-a-time) compared to \WanderJoin's (edge-at-a-time).}

\noindent \underline{\textbf{Time and Space Complexity.}} 
In terms of efficiency, the intersection is the main bottleneck in Algorithm \ref{alg:alley_sample}. Therefore, the implementation of an intersection requires special care in order to obtain an efficient estimator. We implement the intersection by referencing the implementation of \EmptyHeaded \cite{aberger2016emptyheaded} without using its data compression techniques for a fair comparison with others. Specifically, for a multi-way intersection, we start from the smallest set and intersect it with the second smallest set and repeat this pairwise intersection. \revision{Given $k$ sets of size $N_1, N_2, ..., N_k$ (sorted in ascending order), the intersection takes $O(k{N_1}\log{N_k})$ time and requires $O(N_1)$ space.}
Meanwhile, the intersection itself has an early stopping effect; that is, it will catch sampling failures at an early stage and avoid further random walks that would eventually fail. As query graphs become larger, the effectiveness of early stopping becomes be greater. We show this in our experiments.

\vspace*{-0.2cm}
\subsection{Branching} \label{subsec:branching}

Branching is to sample data vertices in $P_{o_i}$ for $o_i$ multiple times (i.e., $\big\lceil b \cdot \abs{P_{o_i}} \big\rceil$ times) \emph{without} replacement (Line \ref{alg:branching}). The generated potential embeddings share $p(o_1), p(o_2), ...,$ and $p(o_{i-1})$ but have different $p(o_{i})$. The constant $b \in (0, 1)$ is called the \emph{branching factor} which is a hyperparameter. \release{Note that \ST \cite{assadi2018simple} uses a different branching technique for cycle (sub)queries only with branching factor $1/\sqrt{|E_g|}$ and sampling \emph{with} replacement.}

\newredtext{Making branches bounds the estimation variance of $w_i$ regardless of the local sample space $P_{o_i}$} (Proposition~\ref{lemma:2} in Section~\ref{sec:proof}). Here we explain the advantage of branching at a high level. 

\begin{figure}[h!]
\centering
\includegraphics[width=0.9\columnwidth]{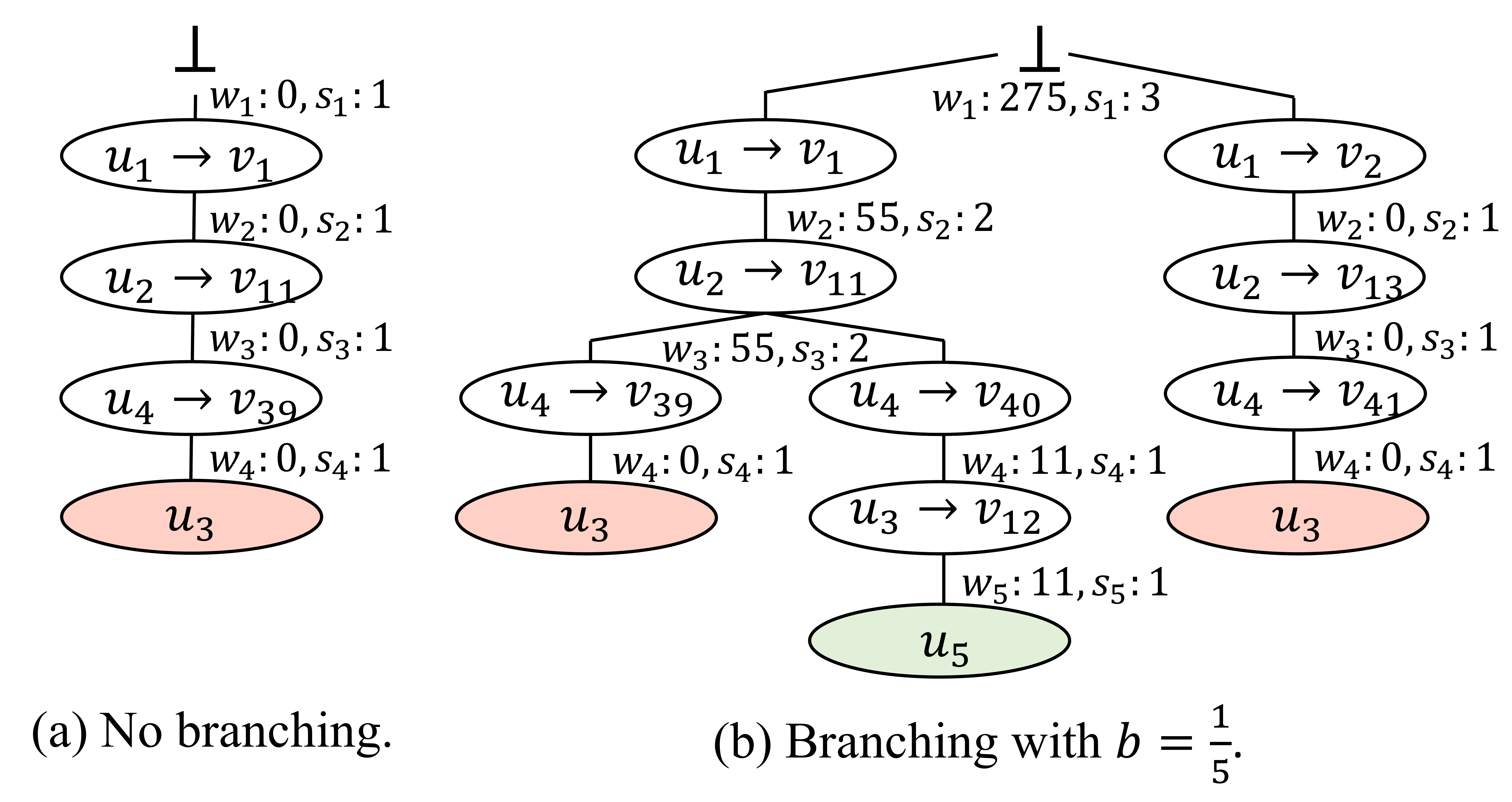}
\vspace*{-0.3cm}
\caption{Recursive calls using $g_2$ in Figure \ref{fig:index-example}.}\label{fig:alley}
\end{figure}

\begin{example}
\vspace*{-0.2cm}\redtext{
Figure~\ref{fig:alley} shows recursion trees of \SamplePotentialEmbeddings. Here, we consider the case in Example~\ref{example:Fig2-noindex}. A path from \spellcheck{the} root to each node shows a potential embedding $p_i$, and $w_i, s_i$ values to be returned are attached at right or center (aggregated). Succeeded walks are colored in green, while failed walks are colored in red.}

\redtext{If we do not make any branch (i.e., $b=\frac{1}{\inf}$), the estimator will call an independent recursion path, such as the one in Figure~\ref{fig:alley}a. This can result in multiple identical random walks.}

\redtext{In contrast, with branching, we can force a different recursion path to be performed for each recursion level. 
Assume that $b$ is $\frac{1}{5}$ (Figure~\ref{fig:alley}b). We now sample twice for $o_1$ from $P_{o_1} = \{v_1, v_2, ..., v_{10}\}$, since $\big\lceil b \cdot \abs{P_{o_1}} \big\rceil = 2$. Examples of call paths and return values for sampling $v_1$ and $v_2$ for $o_1$, are given in Figure \ref{fig:alley}b.
Here, sampling twice for $o_1$ (and $o_3$) increases the probability of sampling $v_{1}$ (and $v_{40}$) which leads to an \revision{embedding}. 
Note that in Example~\ref{example:Fig1-intersect}, we always obtain the exact cardinality  $\QueryNum$ with one possible random walk since $\abs{P_{o_1}} = \abs{P_{o_2}} = \abs{P_{o_3}} = \abs{P_{o_4}} = 1$. Thus, the variance of $w_1$ to $w_5$ is zero. For a larger $P_{o_i}$, we must sample multiple times in order to bound each $w_i$, which is the key to proving Proposition \ref{lemma:2}.}
\end{example}

\ifFullVersion
\else
\vspace*{-0.3cm}
\fi

\ifFullVersion
\else
\vspace*{-0.3cm}
\fi




Branching can also mitigate the overhead of intersections. \redtext{For example in Figure~\ref{fig:alley}b, we can reuse the intersection result of $P_{o_3}$ for sampling $v_{39}$ and $v_{40}$ given $p=\{u_1 \veryshortarrow v_1, u_2 \veryshortarrow v_{11}\}$ (the left subtree). Given the same sample size $s$, this has the effect of reducing the number of intersections. This gain results in bounded time complexity \revision{below}. Note that, in general, a larger branching factor results in a smaller bound of variance and intersection overhead, but it would require a sufficiently large sample size.}

\noindent \underline{\textbf{Time Complexity.}} \revision{
If we 1) replace Lines~\ref{alg:randomwalk_start}-\ref{alg:randomwalk_end} in Algorithm \ref{alg:alley_sample} by an exhaustive search (i.e., iterate over all $v \in P_{o_i}$ instead of sampling) and 2) replace Lines~\ref{alg:foundstart}-\ref{alg:foundend} by an exhaustive search for all $v \in P_{o_n}$ and adding $\{o_n \veryshortarrow v\}$ to $p$, then \SamplePotentialEmbeddings\xspace starting from $i = 1$ becomes a specialization of \GenericJoin \cite{ngo2012worst}; each potential embedding that can be obtained at $i = n$ with $P_{o_n} \neq \emptyset$ is an embedding in $\Matches$. 
Then it is trivial that sampling without replacement makes Algorithm \ref{alg:alley_sample} a \emph{part of} a worst-case optimal algorithm, without searching for the same embedding multiple times.
In other words, sampling \emph{with} replacement cannot guarantee this; a data vertex that leads to many embeddings in the exhaustive search can be sampled multiple times, resulting in more than $O(\AGM)$ potential embeddings sampled.
}

\ifFullVersion
\subsection{Implementation details} \label{subsec:multiway-origin}
\fi

\ifFullVersion

In order to increase efficiency, we applied the following optimization techniques. First, $P_{o_1}$ is the same for each call to \SamplePotentialEmbeddings\xspace at Line \ref{Algo1:Sample} of Algorithm \ref{alg:main}. Thus we compute $P_{o_1}$ only once. Second, we remove the sets in intersections that do not affect the result. In Figure \ref{fig:index-example}, if we start \spellcheck{a} random walk from $u_2$ (assuming that a data graph different from $g_1$ in Figure~\ref{fig:running-example} is given), we have to intersect three sets, i.e., \revision{$V^{u_2}_{u_1}$, $V^{u_2}_{u_3}$, and $V^{u_2}_{u_4}$.} However, $L_q(u_3) = L_q(u_4)$ so \revision{$V^{u_2}_{u_3} = V^{u_2}_{u_4}$} by the definition of \revision{$V^{u'}_{u}$}. Thus, we only intersect the first two sets and remove the last to avoid the redundant computation. Even when we start from $u_3$ and walk to $u_2$, calculating the candidates for $u_2$, we can ignore intersecting with \revision{$V^{u_2}_{u_4} = V^{u_2}_{u_3}$} since it is a superset of \revision{$adj^{u_2}_{u_3}(p(u_3))$} for any $p$.

To further increase efficiency, one can apply the following techniques which would be an interesting future work: 1) base and state, or a hybrid representation to compress the sets into bitmaps and apply bit intersections \cite{DBLP:conf/sigmod/Han0Y18, aberger2016emptyheaded}, 2) SIMD instructions to reduce the number of operations \cite{DBLP:conf/sigmod/Han0Y18, aberger2016emptyheaded} and 3) vertex signatures and multi-core GPUs to highly parallelize the intersections \cite{DBLP:conf/icde/ZengZOHZ20}.

\noindent \underline{\textbf{Branching.}} 
We apply the following techniques to increase the effectiveness of branching. First, we adaptively choose $b$ depending on $\abs{P_{o_i}}$. If $\abs{P_{o_i}}$ is too small so is $\big\lceil b \cdot \abs{P_{o_i}} \big\rceil$, the actual confidence of the estimated cardinality might be too low. Therefore, we use larger $b$ for small $\abs{P_{o_i}}$ as shown in Figure \ref{fig:branch}. Second, we skip unnecessary branches for a query vertex if all its adjacent vertices are sampled. For example, if the sampling order is given as $\langle u_1, u_2, u_4, u_3, u_5 \rangle$ in Figure \ref{fig:running-example}, it is unnecessary to make branches for $u_3$ and $u_5$ for any data graph, since sampling any candidate from their local sample spaces does not affect the final estimate. Otherwise\spellcheck{,} if a query vertex $u$ has an unsampled adjacent vertex, different candidates can result in different estimates. Such classification of query vertices has been studied in a subgraph enumeration paper \cite{kim2016dualsim}.

\begin{figure}[h!]                                    
\centering
\includegraphics[width=0.8\columnwidth]{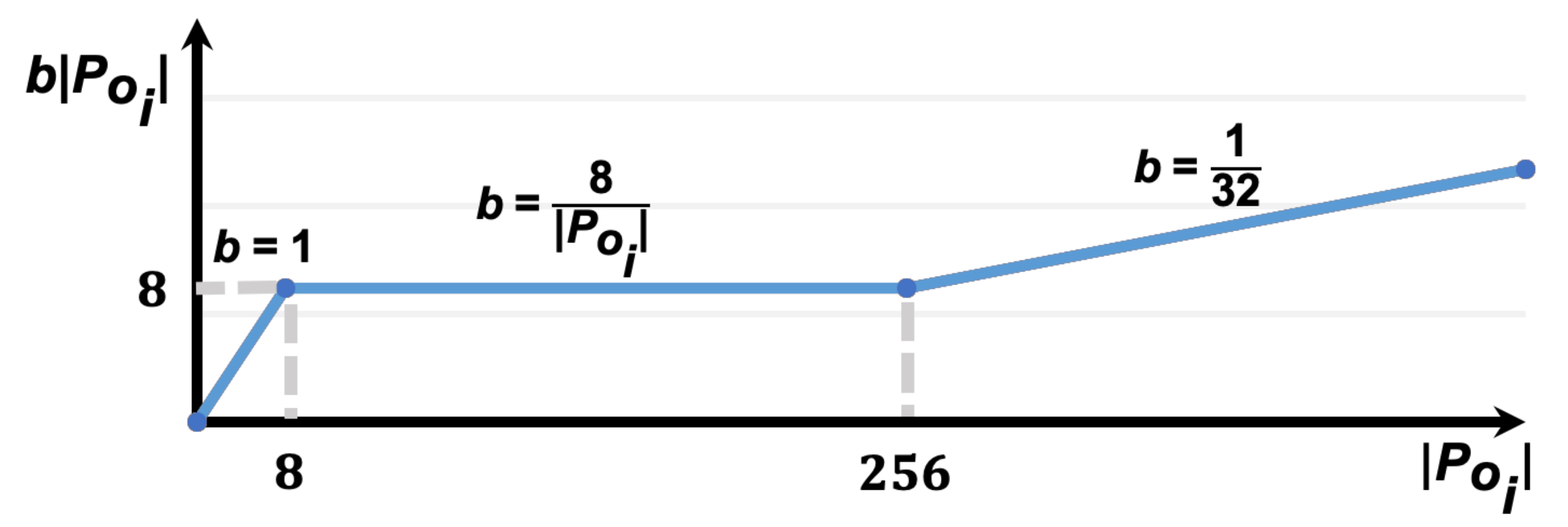}
\caption{Adaptive branch size.}
\label{fig:branch}
\end{figure}
\fi



%% file: 06.index.tex
\section{Tangled Pattern Index}\label{sec:index}

\bluetext{We now explain how to build and search in a \emph{tangled pattern index}. We have shown in Section \ref{sec:introduction} that \profwshan{tangled patterns in queries} can increase the difficulty of sampling and are, in general, inherently difficult to handle by any sampling-based methods.
To resolve the problem, we propose an index for such tangled patterns.}

\begin{definition2}
\vspace*{-0.2cm}
\revision{
Given a data graph $g$, a tangled pattern index for $g$ is a DAG $:= (N,E)$ where 1) each node $n_q \in N$ corresponds to a (tangled or untangled) pattern $q$ appearing in $g$, and 2) each edge $(n_q, n_{q^{\prime}}, [u, u^{\prime}]) \in E$ with label $[u, u^{\prime}]$ represents a \emph{PointingTo} relationship between $u \in V_q$ and $u^{\prime} \in V_{q^{\prime}}$.}
\vspace*{-0.2cm}
\end{definition2}

\revision{Here, if $D_q(u)$ denotes the domain of $u$ in $q$, then \emph{PointingTo} relationship from $u$ to $u^{\prime}$ represents that i) $D_{q^{\prime}}(u^{\prime}) \supseteq D_q(u)$ from the anti-monotone property (Lemma \ref{lemma:anti-monotone}) \cite{elseidy2014grami}, ii) $D_{q^{\prime}}(u^{\prime})$ is used to prune the local sample space of $u$ in $q$ (i.e., $P_{u}$) during the random walk, and iii) the rate of sampling failures of $q$ does not exceed a predefined threshold (i.e., $q$ is not tangled). If a node $n_{q}$ has no outgoing \emph{PointingTo} edge, $q$ is \emph{tangled} and the domain of each vertex in $q$ is materialized. 
Figure \ref{fig:offline} shows an example index, where $q_4$ is not tangled and $q_7$ is tangled.}


\begin{lemma3}
\vspace*{-0.2cm}
\label{lemma:anti-monotone}
Given a data graph $g$ and a pattern $q'$ that matches a subgraph of $q$ by a mapping $m$, $D_{q'}(u') \supseteq D_{q}(m(u'))$ for $u' \in V_{q'}$. 
\vspace*{-0.2cm}
\end{lemma3}





\subsection{Building Index}\label{subsec:offline}
In the offline phase, we start from mining since we do not know which patterns are \profwshan{tangled}. 
\newredtext{However, existing frequent pattern mining work~\cite{elseidy2014grami, abdelhamid2016scalemine} cannot be directly applied to our problem. Their purpose is to mine patterns whose \profwshan{support values are} larger than a given threshold, so they adopt optimization techniques to prune infrequent patterns earlier. \redtext{\cite{abdelhamid2016scalemine} estimates the frequency of a pattern and prunes if the estimated frequency is low. 
\cite{elseidy2014grami} models frequency evaluation as a constraint satisfaction problem (CSP), and computes the minimal set of domains that are sufficient for CSP to determine whether a pattern is frequent. This avoids finding the whole domains.}
%
In contrast, we need to enumerate tangled patterns whose frequencies would be small. Moreover, we must calculate the whole domains in order to prevent false-negatives in the sampling.} 
\bluetext{Those incur extra computational overheads to pattern mining which already has significant time complexity.}

\profwshan{To tackle this challenging problem,} we present a novel, efficient mining approach, ``walk-fail-then-calculate,'' that interleaves sampling with mining, which allows \Alley to autonomously determine and index the tangled patterns, avoiding enormous computation for frequent patterns. \revision{Like standard frequent subgraph mining algorithms, \Alley mines from smaller to larger patterns. However, instead of mining all patterns, \Alley first performs random walks (Algorithm \ref{alg:main}) regarding each pattern as a query $q$ and the domains of sub-patterns as $D_{q_{j}}$. If the random walks fail with a high chance (above a threshold), \Alley determines that this pattern $q$ is hard to estimate its cardinality from the current index thus extends the index by mining $q$.}



\profwshan{\revision{Algorithm~\ref{alg:offline} shows this procedure.} It grows patterns from size one to size $maxN$ where size is the number of edges (Lines~\ref{alg4:extendstart}-\ref{alg4:extendend} ).}
Before calculating the domains for a pattern $q$ \profwshan{of size $N$}, we first search for its subgraphs $\{q_i\}$ \profwshan{where each $q_i$ has size $N-1$}.
If any of the $q_i$ has an empty domain (i.e., $D_{q_i}(u) = \emptyset$ for any $u \in V_{q_i}$), we can skip processing $q$ according to the anti-monotone property; $q$ also has at least one empty domain (Lines~\ref{alg4:hasemptystart}-\ref{alg4:hasemptyend}).
\profwshan{Otherwise, we start ``walk'' to determine whether $q$ is tangled  (Lines~\ref{alg4:getminimum}-\ref{alg4:getfailure}).
Specifically, for each $u \in V_q$, we first select a set of candidates from the set of domains $\{D_{q_i}(u') \,|\, u' \in V_{q_i} \wedge \, \exists m:\, m(u') = u\}$, which will be used in random walks. Among such domains, we select the minimum-size domain, $\hat{D}_q(u) := D_{\hat{q}}(\hat{u})$ where $(\hat{q}, \hat{u}) = \argmin_{q_i, u' \in V_{q_i} \wedge \, \exists m:\, m(u') = u}\abs{D_{q_i}(u')}$, as the set of candidates since it has the highest pruning power. Then, we can get $\hat{D}_q=\{\hat{D}_{q}(u)\}_{u \in V_q}$ (Line~\ref{alg4:getminimum}). Note that we might refer to different $\hat{q}$ for each $u \in V_q$.
According to the anti-monotone property, $\hat{D}_{q}(u) \supseteq D_q(u)$. 
Then, we perform random walks by running \Alley (Line~\ref{alg4:getfailure}). Specifically, we regard $q$ as a query graph and run Algorithm \ref{alg:main}. We directly use $\hat{D}_q$ \revision{as $D_{q_j}$} instead of running \SearchDomainsRecursive\xspace in Line~\ref{Algo1:searchdomains} of Algorithm~\ref{alg:main}.}

\profwshan{If \Alley ``fails'' (i.e., its failure rate $r$ exceeds a given threshold $\zeta$), we finally ``calculate'' the domains of $q$ (Line~\ref{alg4:calculate}). Here, $r$ is the ratio of the number of calls to \SamplePotentialEmbeddings\xspace that returns $w_1 = 0$ in Line~\ref{Algo1:Sample} of Algorithm \ref{alg:main}. 
Again, when calculating the domains of $q$, we use $\hat{D}_q(u)$ as the candidates for $D_q(u)$ according to the property. If \Alley does not fail, we simply maintain the \emph{PointingTo} edges $(n_q, n_{\hat{q}}, [u, \hat{u}])$ ($\forall u \in V_q$) and regard $D_q$ as $\hat{D}_q$ in Lines~\ref{alg4:pointstart}-\ref{alg4:pointend}.}

\ifFullVersion
\else
\revision{For optimization, we 1) implement a specialized DP-based algorithm for computing domains, 2) reduce the number of patterns by pruning and grouping labels, and 3) stop sampling after the failure rate converges. Details of the optimization techniques are in \cite{AAA}.}
\fi

\begin{figure}[h!]
\centering
\vspace*{-0.2cm}
\includegraphics[width=0.8\columnwidth]{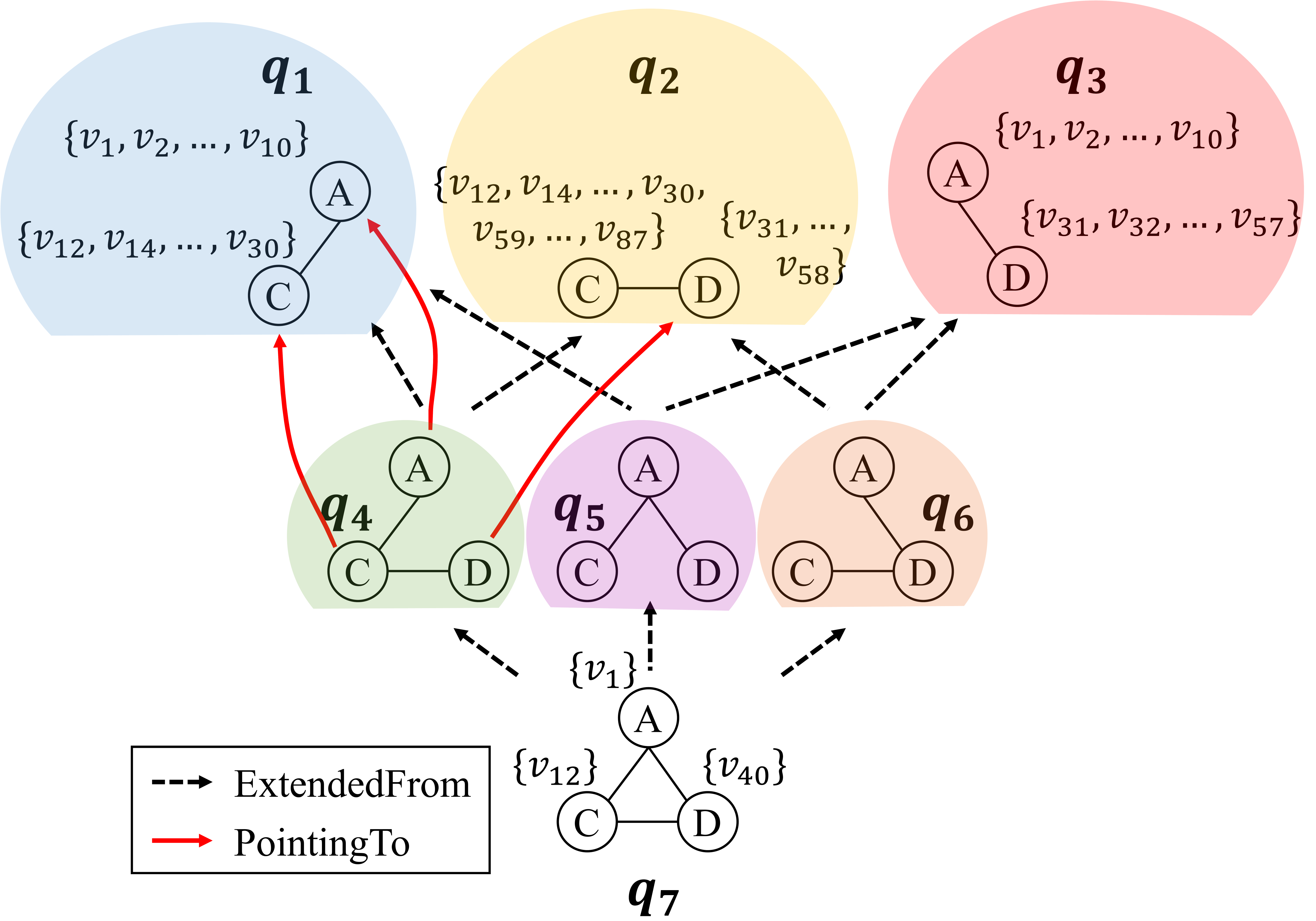}
\vspace*{-0.4cm}
\caption{An example index for $g_2$ in Figure \ref{fig:index-example}.}
\ifFullVersion
\vspace*{-0.2cm}
\else
\vspace*{-0.4cm}
\fi
\label{fig:offline}
\end{figure}

\begin{example}
\vspace*{-0.2cm}
\label{ex:offline}
Figure \ref{fig:offline} shows part of a tangled pattern index for $g_2$ in Figure \ref{fig:index-example}. The 2-edge patterns $q_4$, $q_5$, and $q_6$, are extended from 1-edge patterns, $q_1$, $q_2$, and $q_3$. The $q_7$ is extended from those three 2-edge patterns. Each vertex of $q_4$ points to another vertex in a 1-edge pattern, which represents that $D_q(u)$ has been set to point $\hat{D}_q(u)$ for $u \in V_{q_4}$. This is due to the low failure rate of random walking for $q_4$ using the domains of $q_1$ and $q_2$.
\revision{Note that, since $D_{q_1}(\textcircled{\small{C}})$ is smaller than $D_{q_2}(\textcircled{\small{C}})$, \textcircled{\small{C}} in $q_4$ points to \textcircled{\small{C}} in $q_1$.}
We omit the pointers of $q_5$ and $q_6$. In contrast, random walks for $q_7$ would result in a large failure rate since the domains of $q_7$ are much smaller than those of its subgraphs. Thus, $q_7$ is determined as tangled and $D_{q_7}$ is calculated.
\vspace*{-0.2cm}
\end{example}

\setlength{\textfloatsep}{4pt}
\begin{algorithm} [htb]
\caption{{\TangledPatternMining}($g, maxN, s, \zeta$)} \label{alg:offline}
\small{
    \KwIn{\mbox{A data graph $g$, maximum number of edges of a pattern $maxN$,} \mbox{a sample size $s$, and a threshold $\zeta$}}
    \redcomment{$I_N$ stores N-edge patterns and their domains}
    $I_{N} \leftarrow \emptyset$ for $1 \leq N \leq maxN$ \\ 
    \textbf{populate} $I_1$ with all 1-edge patterns using labels in $g$ \\
    \ForEach{$1 \leq N \leq maxN-1$}
    {
        \ForEach{$q \in I_N$}
        {\label{alg4:extendstart}
            \redcomment{extend $q$ with an edge, generate multiple (N+1)-size patterns using labels in $g$}
            $I_{N+1} \leftarrow I_{N+1} \cup \ExtendPattern(q, L_g)$  \\
        }\label{alg4:extendend}
        \ForEach{$q \in I_{N+1}$}
        {\label{alg4:forloop}
            \If{$\HasUnindexedSubgraph(q, I_N)$}
            {\label{alg4:hasemptystart}
                \textbf{remove} $q$ from $I_{N+1}$; \Continue \\
            }\label{alg4:hasemptyend}
            \redcomment{$\hat{D}_q(u)$ for each $u \in V_q$ points to a precomputed domain $D_{\hat{q}}(\hat{u})$ ($\hat{q} \in \{q_1, q_2, ...\}$)}
            $\hat{D}_q \leftarrow \GetMininumDomainsFromSubgraphs(q, I_N)$ \label{alg4:getminimum} \\
            $r \leftarrow \GetFailureRate(\Alley, q, g, s, \hat{D}_q)$ \label{alg4:getfailure} \\
            \If{$r > \zeta$}
            {\label{alg4:ifzetastart}
                $D_q \leftarrow \CalculateDomains(q, g)$ \label{alg4:calculate} \\
                \If{$\exists u \in V_q: D_q(u) = \emptyset$}
                {
                    \textbf{remove} $q$ from $I_{N+1}$ \\
                    \Continue \\
                }
            }\label{alg4:ifzetaend}
            \Else
            {\label{alg4:pointstart}
                $D_q \leftarrow \hat{D}_q$ \redcomment{just maintain the pointers} 
            }\label{alg4:pointend}
        }
    }
    
    \Return $\{I_2, ..., I_{maxN}\}$ 
}
\end{algorithm}

In conclusion, our approach has five strong points. First, it reduces the number of expensive domain calculations by placing cheap random walks as filters. Second, for tangled patterns $\{q_t\}$, $D_{q_t}(u)$ would be much smaller than $\hat{D}_{q_t}(u)$ since the sampling has failed using $\hat{D}_{q_t}(u)$. This has the effect of bounding the index size similar to using a discriminative ratio as in \cite{DBLP:conf/sigmod/YanYH04}. Third, in order to increase the online effectiveness of \Alley, we allow \Alley itself determine hard cases offline. Fourth, we can control the offline overhead and online effectiveness of our index by changing the threshold $\zeta$. A large $\zeta$ let \Alley reduce the number of domain calculations offline but causes it to use large domains online. Fifth, our index can be generalized to increase the performance of other sampling-based methods, which would be interesting future work.

\noindent \underline{\textbf{Maintenance.}} \revision{
For updates, the index is built by using the RUNSTAT command, as in existing DBMSs. In order to minimize maintenance overhead, we do not update the index whenever there is an update to the graph database. When a considerable amount of updates have occurred, say 5-10\% of data edges, a DBA needs to rebuild the index by issuing RUNSTAT again. We confirmed that the estimation accuracy does not deteriorate much even if 5-10\% of data edges are inserted through experiments
\ifFullVersion
(Figure~\ref{fig:IndexAccuracy-update} in Section~\ref{subsec:exp_index_build}).
\else
(see ~\cite{AAA} for the experimental results).
\fi
We can further optimize performance by incrementally rebuilding the index.}
\ifFullVersion

\noindent \underline{\textbf{Optimizations.}} Even though our filtering approach reduces the overhead of domain computation, problems can still occur if $g$ contains many labels and the number of possible patterns explodes. 
We first reduce the overhead of each call to \CalculateDomains\xspace by implementing a specialized algorithm for computing domains instead of enumerating all embeddings. We utilize dynamic programming to avoid redundant computation, which runs in linear time (i.e., $O(|V_q|\sum_{u \in V_q}{\hat{D}_q(u)})$) for tree queries. To take advantage of dynamic programming, we extend the optimization to cyclic queries allowing non-linear time and approximate computation; domains can contain false positive vertices. 
\ifFullVersion
\todelete{We also apply the techniques to cache intersection results \cite{DBLP:journals/pvldb/MhedhbiS19}, skip particular query and data vertices \cite{DBLP:conf/sigmod/YangFL16}, and exploit neighborhood equivalence class \cite{han2013turbo} to reduce duplicated computation.
\profwshan{We stop extending $q$ if the size of $\hat{D}_q(u)$ is already too small. To avoid intractable computation of some patterns, we set a threshold for the size of Cartesian product.}}

\else 

\fi


Then, we reduce the number of patterns by pruning and grouping labels as follows: First, if $g$ contains edge labels, we use edge labels only without vertex labels. Otherwise, we use vertex labels. If the number of (vertex or edge) labels is large, we group labels and limit the number of groups. 
\ifFullVersion
\todelete{This still might be too large, however, but further decreasing the number of groups may result in too many false positives. In such cases, we use hierarchical grouping, e.g., group labels for path patterns, then make supergroups for general tree patterns and finally, make supergroups of the supergroups for general graph patterns. The motivation is that, while labels tend to determine the tangledness of simple queries (e.g., paths), topologies tend to determine the tangledness of complex queries (e.g., cliques).}
\fi

\revision{Finally, instead of using a predefined sample size, we continuously update the failure rate after calculating each $w_1$ in Algorithm \ref{alg:main} and stop if the rate converges. This prevents an unnecessarily large sample size. Also, we only focus on whether $w_1$ is zero or not, so we stop each call to \SamplePotentialEmbeddings\xspace if we sample an embedding and $w_1$ is guaranteed to be nonzero.}

\else
\fi

\noindent \underline{\textbf{Complexity.}}
\revision{Given a data graph $G=(V_{g}, E_{g}, L_{g})$, there can be $\varepsilon\times|L_{g}|^{(maxN+1)}$ tangled patterns where $\varepsilon$ is the ratio of tangled patterns over all patterns up to $maxN$ edges. The factor $\varepsilon$ can significantly reduce time/space complexity. \CalculateDomains\xspace dominates the other functions which is bounded by $O(|E_{g}|^{maxN})$ time complexity. Therefore, the time complexity of Algorithm \ref{alg:offline} is $O(\varepsilon\times|L_{g}|^{(maxN+1)}\times|E_{g}|^{maxN})$, while the space complexity is $O(\varepsilon\times|L_{g}|^{(maxN+1)}\times|V_{g}|\times(maxN+1))$. The complexities of maintenance are similar to above, but we only need to calculate the domains of affected tangled patterns. Note that these are loose upper bounds since we do not consider 1) the impact of search space reduction by using pointed tangled patterns recursively, 2) the important characteristic of tangled patterns that their domain size is much smaller than $|V_{g}|$, and 3) the optimization techniques. For example, in \YAGO dataset, $\varepsilon$ is $10^{-5}$ and the actual average space per pattern is 4.4K whereas $|V_{g}|\times(maxN+1)$ is 76.8M.}





\ifFullVersion
\else
\vspace*{-0.4cm}
\fi
\subsection{Searching in Index}\label{subsec:online}

In the online phase, we use the built tangled pattern index to reduce the sample space of a given query $q$ (Line~\ref{Algo1:searchdomains} of Algorithm~\ref{alg:main}). The idea is to search for domains of subqueries $\{q_1, q_2, ...\}$ of $q$, and use each domain $D_{q_j}(m^{-1}(o_i))$ to prune candidates $P_{o_i}$ in Line~\ref{alg:intersect} of Algorithm \ref{alg:alley_sample}, where $m$ is the mapping from $V_{q_j}$ to $V_q$. Due to the anti-monotone property, $D_q(o_i) \subseteq D_{q_j}(m^{-1}(o_i))$. Note that $D_{q_j}$ might not be the actual domains of $q_j$, but it points to one of its subgraphs, as described in Section~\ref{subsec:offline}.

Among multiple $D_{q_j}(m^{-1}(o_i))$ for different subqueries $q_j$, we search for the minimum domain $D_{q_j}(m^{-1}(o_i))$ \profwshan{(Lines~\ref{alg5:mindomainstart}-\ref{alg5:mindomainend} in Algorithm~\ref{alg:online})}. \profwshan{Note that the initial \SearchDomainsRecursive\xspace is called right after \ChooseSamplingOrder.} Since enumerating all subqueries of $q$ can be costly for large $q$, we greedily search for maximal subqueries (with size $\leq maxN$) that contain the first $k$ vertices in the given sampling order \profwshan{(Lines~\ref{alg5:greedystart}-\ref{alg5:greedyend})}. After \profwshan{searching for such maximal subqueries, we remove the $k$ vertices} from $q$ \profwshan{(Line~\ref{alg5:remove})} and continue to search from the new first vertex \profwshan{by calling \PriorityFirstSearchFrom\xspace~(Line~\ref{alg5:priorityfirstsearch})}. This is repeated until $q$ is empty. 

\ifFullVersion
\else
\vspace*{-0.3cm}
\fi
\setlength{\textfloatsep}{4pt}
\begin{algorithm} [htb]
\caption{{\SearchDomainsRecursive}($q, o, I$)} \label{alg:online}
\small{
    \KwIn{\mbox{A query graph $q$, a sampling order $o$, and a pre-built index $I$} \mbox{with maximum size $maxN$}}
    \ForEach{maximal connected subquery $\tilde{q}$ of $q$ such that $\abs{E_{\tilde{q}}} \leq maxN$ and $\tilde{q}$ contains the first $k$ vertices in $o$}
    {\label{alg5:greedystart}
        $D_{\tilde{q}} \leftarrow \SearchDomains(\tilde{q}, I)$ \label{alg5:searchdomains}\\
        \ForEach{$u \in V_{\tilde{q}}$}
        {\label{alg5:mindomainstart}
            $D_{q_{min}}(m(u)) \leftarrow \argmin\big(|D_{q_{min}}(m(u))|, |D_{\tilde{q}}(u)|\big)$ \\
        }\label{alg5:mindomainend}
    }\label{alg5:greedyend}
    \textbf{remove} the $k$ vertices from $q$ \label{alg5:remove}\\
    \If{$V_q \neq \emptyset$}
    {
        \redcomment{traverse from $o_{k+1}$ with $o$ as priority ($o_i \prec o_{i+1}$)}
        $o' \leftarrow \PriorityFirstSearchFrom(o_{k+1}, o)$ \label{alg5:priorityfirstsearch}\\
        $\SearchDomainsRecursive(q, o', I)$ \\
    }
        
    
    \Return $D_{q_{min}}$ \\
}
\end{algorithm}
\ifFullVersion
\else
\vspace*{-0.3cm}
\fi

\noindent \underline{\textbf{Time complexity.}} \revision{Given a query $q$, the number of recursive calls is $\ceil{|V_{q}|/2}$ in the worst case. For each recursive call, \PriorityFirstSearchFrom\xspace takes $O(|V_{q}|)$ time (Line~\ref{alg5:priorityfirstsearch}), and the total number of loops (Lines~\ref{alg5:greedystart}-\ref{alg5:greedyend}) is $O(2^{maxN})$. Calculating the BFS code of $\tilde{q}$ dominates \SearchDomains\xspace and takes $O(2^{|V_{\tilde{q}}|})$ where $|V_{\tilde{q}}|$ is always less than $maxN$ (Line~\ref{alg5:searchdomains}). Therefore, the time complexity of Algorithm~\ref{alg:online} is $O(|V_{q}|^2+4^{maxN}|V_{q}|)$, which is quadratic to the query size.}



%% file: 07.proof.tex
\ifFullVersion
\else
\vspace*{-0.1cm}
\fi
\section{Worst-Case Optimal Runtime and Approximation Quality Guarantees}\label{sec:proof}

In this section, we establish the \redtext{runtime and approximation quality} guarantees of \Alley.
First, we formally define the problem of graph pattern cardinality estimation with probabilistic guarantees in worst-case optimal time, which we have discussed in Section \ref{sec:introduction}.

\begin{definition2}
\vspace*{-0.2cm}
\label{def:approximation}
For a given error bound $\epsilon$ and a confidence $\mu \in (0,1)$, if the random variable for the estimated cardinality $Z$ satisfies $\Pr(\abs{Z-\QueryNum} < \epsilon \cdot \QueryNum) > \mu$, then the estimation is a $(1\pm\epsilon)$-approximation of $\QueryNum$.
\end{definition2}


\begin{theorem}
\vspace*{-0.2cm}
\Alley performs $(1\pm\epsilon)$-approximation of $\QueryNum$ in $O(\AGM)$ time.
\label{theorem}
\vspace*{-0.2cm}
\end{theorem}

While the accuracy of sampling-based estimators increases as the sample size increases, the theoretical guarantees of \Alley in Theorem \ref{theorem} indicate that it is enough to set the sample size \profwshan{to} $O(\AGM)$ in order to perform $(1\pm\epsilon)$-approximation of $\QueryNum$, for any $\epsilon$ and $\mu$. Still, smaller $\epsilon$ and larger $\mu$ will increase the constant factor in $O(\cdot)$.
The proof of Theorem \ref{theorem} is based on the following lemmas and propositions, \release{which is analogous to \ST \cite{assadi2018simple}.}
\ifFullVersion
\else
\revision{See our technical paper \cite{AAA} for proofs. Let $Y_1$ denote the random variable \mbox{for $w_1$.}}
\fi


\vspace{-1ex}
\begin{lemma3}
\Alley is consistent, i.e., $\Matches \subseteq \SampleSpace$.
\label{lemma:cons}
\end{lemma3}
\vspace{-2ex}

\ifFullVersion
\begin{proof}
We use proof by contradiction. Assume that there is an embedding $m \in \Matches$ that cannot be sampled by \Alley. If $m = \{o_1 \veryshortarrow v_1, o_2 \veryshortarrow v_2, ..., o_n \veryshortarrow v_n\}$ ($n = |V_q|$), then there exists the smallest $i \leq n$ such that $v_i \not\in P_{o_i}$. Here, $P_{o_i}$ \revision{is calculated from $p = m_i$}
where $m_i$ is the restriction of $m$ on $o_1, o_2, ..., o_{i-1}$.
Since $v \not\in P_{o_i}$, there must be an edge $e^* = (o_k, o_i) \in E_q$ such that \revision{$v_i \not\in adj^{o_i}_{o_k}(p(o_k))$ if $i > k$ and $v_i \not\in V^{o_i}_{o_k}$ if $i < k$.}


i) If $i > k$, then \revision{$adj^{o_i}_{o_k}(p(o_k)) = adj^{o_i}_{o_k}(v_k)$} and $v_i$ must be in this adjacency list, otherwise there is no edge between $v_k$ and $v_i$ that matches $(o_k, o_i)$. Thus, $m$ cannot be an embedding of $q$, which is a contradiction.

ii) If $k > i$, \revision{$v_i \not\in V^{o_i}_{o_k}$} means that $v_i$ has no incident edge $(v, v_i)$ that matches $e^*$. Thus, $m$ cannot be an embedding of $q$, which is a contradiction.
\end{proof}
\fi

\begin{lemma3}
\vspace*{-0.1cm}
\Alley is a stack of simple HT estimators.
\label{lemma:ht}
\vspace*{-0.1cm}
\end{lemma3}

\ifFullVersion
While a simple HT estimator performs $s$ independent sampling and aggregates their weights at once (Algorithm \ref{alg:ht}), \Alley interleaves sampling a data vertex and aggregating weights (Lines~\ref{alg:randomwalk_start}-\ref{alg:randomwalk_end} of Algorithm \ref{alg:alley_sample}).
In that sense, \Alley can be regarded as a stack of simple HT estimators which returns $w_i$ as results.
Then, a natural question arises. What values does $w_i$ estimate for? That is, what is the expectation $\E[w_i]$ of $w_i$ returned by each \SamplePotentialEmbeddings?

To answer the question, we define the following random variables.
Let $a_i$ be the random variable for $\abs{P_{o_i}}$. 
Let $t_i$ be the random variable for branch size, i.e., $t_i = \big\lceil b \cdot \abs{P_{o_i}} \big\rceil$. 
We then recursively define $Y_i|p_i$ as the random variable for $w_i$; $i$ is the recursion depth, and $p_i$ is the current potential mapping that maps $o_1, ..., o_{i-1}$ to data vertices. 

\vspace*{-0.3cm}
\begin{equation}
Y_i|p_i = \left\{
\begin{array}{ll}
      a_i & i = n \\
      a_i \cdot \frac{1}{t_i} \sum_{k=1}^{t_i}{Y^k_{i+1}|p_{i+1}} & i < n \\
\end{array} 
\right. 
\end{equation}

Here, superscript $k$ distinguishes $t_i$ \profwshan{identically distributed} random variables for $Y_{i+1}|p_{i+1}$. Note that these are not independent due to our sampling without replacement policy. 

\begin{lemma3}
$\E[Y_i|p_i] = c(q, g|p_i)$. Here, $c(q, g|p_i)$ denotes the number of embeddings of $q$ in $g$ that contain the vertices and edges specified by $p_i$.
\label{lemma:card}
\end{lemma3}

\fi

\ifFullVersion

\begin{proof}
We use proof by induction.

For base case $i = n$, $\E[Y_n|p_n] = a_n = \abs{P_{o_n}}$. Since each $v \in P_{o_n}$ has an incident edge that matches $e$ for every $e \in E_q, e \ni o_n$, $p_n \cup \{o_n \veryshortarrow v\}$ is an embedding of $q$ in $g$, i.e., $a_n \leq c(q, g|p_n)$. Since \Alley is consistent by Lemma \ref{lemma:cons}, $a_n \geq c(q, g|p_n)$. Combining these two gives $a_n = c(q, g|p_n)$.

For inductive case $i < n$, 

\begin{align*}
    \E[Y_i & \,|\,p_i] \\
    &= \sum_{v \in P_{o_i}}{ \Pr(v \,|\, p_i) \cdot a_i \cdot \frac{1}{t_i} \cdot \sum_{k = 1}^{t_i}{ \E[ Y_{i+1}^k \,|\, p_i \cup \{o_i \veryshortarrow v\}] } } \\
    &\overset{\tiny (1)}{=} \sum_{v \in P_{o_i}}{ \frac{1}{t_i} \cdot \sum_{k = 1}^{t_i}{ \E[ Y_{i+1}^k \,|\, p_i \cup \{o_i \veryshortarrow v\}] } } \\
    &\overset{\tiny (2)}{=} \sum_{v \in P_{o_i}}{ \E[Y_{i+1} \,|\, p_i \cup \{o_i \veryshortarrow v\}] } \\
    &\overset{\tiny (3)}{=} \sum_{v \in P_{o_i}}{ c(q, g \,|\, p_i \cup \{o_i \veryshortarrow v\}) } \\
    &\overset{\tiny (4)}{=} c(q, g \,|\, p_i).
\end{align*}

(1) holds since $\Pr(v|p_i) = 1/a_i$. (2) holds since all $Y_{i+1}^k$'s are identical having the same expectation. (3) holds from the induction hypothesis. (4) holds since summing up $c(q, g|p_i \cup \{o_i \veryshortarrow v\})$ for every possible $v \in P_{o_i}$ results in counting all embeddings specified by $p_i$.

\end{proof}

\else
\fi

\begin{lemma2}
$\E[Y_1] = \QueryNum$.
\label{lemma:1}
\end{lemma2}

\ifFullVersion
Proposition 1 can be readily obtained by Lemma \ref{lemma:card} as $\E[Y_1] = \E[Y_1|p_1] = c(q, g|p_1) = c(q,g) = \QueryNum$.
Therefore, \Alley preserves the unbiasedness of the HT estimator. Furthermore, we bound the estimation variance of \Alley.
\fi

\begin{lemma2}
\vspace*{-0.2cm}
$\Var[Y_1] \leq \frac{b}{1-b} \cdot \big(\frac{1}{b^n} - 1\big) \cdot \QueryNum^2$.
\label{lemma:2}
\end{lemma2}

\vspace{-1ex}

\ifFullVersion
\begin{proof}
Again, we use proof by induction and show that \linebreak $\Var[Y_i|p_i] \leq \frac{b}{1-b} \cdot (\frac{1}{b^{n-i+1}} - 1) \cdot c(q, g|p_i)^2$. If this holds, $\Var[Y_1] = \Var[Y_1|p_1] \leq \frac{b}{1-b} \cdot (\frac{1}{b^n} - 1) \cdot c(q, g|p_1)^2 = \frac{b}{1-b} \cdot (\frac{1}{b^n} - 1) \cdot \QueryNum^2$, completing the proof of the proposition.

For base case ($i = n$), $\Var[Y_n|p_n] = 0$ as $Y_n|p_n$ is constant as $a_n$. Therefore, $\Var[Y_n|p_n] \leq \frac{b}{1-b} \cdot (\frac{1}{b} - 1) \cdot c(q,g|p_n)$.

For inductive case ($i < n$), $\Var[Y_i|p_i] = \E[\Var[Y_i|p_i,v]] + \Var[\E[Y_i|p_i,v]]$ ($v \in P_{o_i}$) by the law of total variance \cite{motwani1995randomized}.
We bound the second term first.

\begin{align*}
    \Var[& \E[Y_i \,|\, p_i, v ]] \\
    &= \Var[\E[Y_i \,|\, p_i \cup \{o_i \veryshortarrow v\} ]] \\
    &\overset{\tiny (1)}{=} \Var[c(q, g \,|\, p_i \cup \{o_i \veryshortarrow v\})] \\
    &\leq \E[c(q, g \,| \,p_i \cup \{o_i \veryshortarrow v\})^2] \\
    &= \sum_{v \in P_{o_i}}{\Pr(v \,|\, p_i) \cdot c(q, g \,|\, p_i \cup \{o_i \veryshortarrow v\})^2} \\
    &\overset{\tiny (2)}{\leq} \sum_{v \in P_{o_i}}{c(q, g \,|\, p_i \cup \{o_i \veryshortarrow v\})^2} \\
    &\leq \Big(\sum_{v \in P_{o_i}}{c(q \,|\, p_i \cup \{o_i \veryshortarrow v\}}\Big)^2 \\
    &= c(q, g \,|\, p_i)^2.
\end{align*}

(1) holds from the inductive proof of Proposition \ref{lemma:1}.
(2) holds since $\Pr(v|p_i) \leq 1$.
Now we bound the first term.

\begin{align*}
    \E[& \Var[Y_i \,|\, p_i, v]] \\
    &= \sum_{v \in P_{o_i}}{ \Pr(v \,|\, p_i) \cdot \Var[Y_i \,|\, p_i, v] } \\
    &= \sum_{v \in P_{o_i}}{ \Pr(v \,|\, p_i) \cdot \Var[Y_i \,|\, p_i \cup \{o_i \veryshortarrow v\}] } \\
    &= \sum_{v \in P_{o_i}}{ \Pr(v\,|\,p_i) \cdot a_i^2 \cdot \frac{1}{t_i^2} \cdot \Var\bigg[ \sum_{k = 1}^{t_i}{  Y_{i+1}^k \,|\, p_i \cup \{o_i \veryshortarrow v\} } \bigg] } \\
    &= \sum_{v \in P_{o_i}}{ \Pr(v\,|\,p_i) \cdot a_i^2 \cdot \frac{1}{t_i^2} \cdot \bigg( \sum_{k = 1}^{t_i}{\Var[ Y_{i+1}^k \,|\, p_i \cup \{o_i \veryshortarrow v\} ]} } \\
\end{align*}    
\begin{align*}
    &\qquad + \sum_{k = 1}^{t_i}{ \sum_{l = 1, l \ne k}^{t_i}{ \Cov\Big(Y_{i+1}^k \,|\, p_i \cup \{o_i \veryshortarrow v\}, Y_{i+1}^l \,|\, p_i \cup \{o_i \veryshortarrow v\}\Big) } } \bigg) \\
    &\overset{\tiny (1)}{\leq} \sum_{v \in P_{o_i}}{ \Pr(v \,|\, p_i) \cdot a_i^2 \cdot \frac{1}{t_i^2} \cdot \sum_{k = 1}^{t_i}{\Var[ Y_{i+1}^k \,|\, p_i \cup \{o_i \veryshortarrow v\} ]} } \\
    &\overset{\tiny (2)}{=} \sum_{v \in P_{o_i}}{ a_i \cdot \frac{1}{t_i} \cdot \Var[  Y_{i+1} \,|\, p_i \cup \{o_i \veryshortarrow v\} ] } \\
    &\overset{\tiny (3)}{\leq} \sum_{v \in P_{o_i}}{ \frac{1}{b} \cdot \Var[ Y_{i+1} \,|\, p_i \cup \{o_i \veryshortarrow v\} ] } \\
    &\overset{\tiny (4)}{\leq} \sum_{v \in P_{o_i}}{ \frac{1}{b} \cdot \frac{b}{1-b} \cdot \Big(\frac{1}{b^{n-i}}-1\Big) \cdot c(q, g \,|\, p_i \cup \{o_i \veryshortarrow v\})^2 } \\
    &= \frac{1}{b} \cdot \frac{b}{1-b} \cdot \Big(\frac{1}{b^{n-i}}-1\Big) \cdot \sum_{v \in P_{o_i}}{c(q, g \,|\, p_i \cup \{o_i \veryshortarrow v\})^2} \\
    &\leq \frac{1}{b} \cdot \frac{b}{1-b} \cdot \Big(\frac{1}{b^{n-i}}-1\Big) \cdot \Big(\sum_{v \in P_{o_i}}{c(q, g \,|\, p_i \cup \{o_i \veryshortarrow v\})}\Big)^2 \\
    &= \frac{1}{1-b} \cdot \Big(\frac{1}{b^{n-i}}-1\Big) \cdot c(q, g \,|\, p_i)^2.
\end{align*}

(1) holds since the covariance between $Y_{i+1}^k$ and $Y_{i+1}^l$ is negative due to sampling without replacement \cite{banerjee2012simple}. (2) holds since $\Pr(v|p_i) = 1/a_i$, and all $Y_{i+1}^k$'s are identical having the same variance. (3) holds since $t_i = \ceil{b \cdot a_i} \geq b \cdot a_i$. (4) holds from the induction hypothesis. 
Finally, we add the two terms and complete the proof.

\begin{align*}
    \Var[Y_i& \,|\, p_i] \\
    &= \E[\Var[Y_i \,|\, p_i,v]] + \Var[\E[Y_i \,|\, p_i,v]] \\
    &\leq \frac{1}{1-b} \cdot \Big(\frac{1}{b^{n-i}}-1\Big) \cdot c(q, g \,|\, p_i)^2 + c(q, g \,|\, p_i)^2 \\
    &= \bigg(\frac{1}{1-b} \cdot \Big(\frac{1}{b^{n-i}}-1\Big) + 1\bigg) \cdot c(q, g \,|\, p_i)^2 \\
    &= \frac{1}{1-b} \cdot \Big(\frac{1}{b^{n-i}}-b\Big) \cdot c(q, g \,|\, p_i)^2 \\
    &= \frac{b}{1-b} \cdot \Big(\frac{1}{b^{n-i+1}}-1\Big) \cdot c(q, g \,|\, p_i)^2.
\end{align*}

\end{proof}
\fi

Propositions 1 and 2 explain the accuracy part of Theorem \ref{theorem}. Proposition 3 explains the remaining efficiency part. 
\ifFullVersion
The proof can be found at the end of Section \ref{subsec:branching}.
\else
\fi

\begin{lemma2}
\vspace*{-0.2cm}
\profwshan{Realizing $Y_1$ can be done in  $O(\AGM)$ time.}
\label{lemma:3}
\end{lemma2}


\ifFullVersion
\textbf{\textsc{Proof of Theorem 1.}}
From the propositions, if we repeat Lines~\ref{Algo1:whilestart}-\ref{Algo1:whileend} in Algorithm \ref{alg:main} by \profwshan{$h$} times and take the average of $Y_1$'s as $Z$ (the final estimate), the following statements hold ($h=\frac{ \frac{b}{1-b} \cdot (\frac{1}{b^n} - 1)  }{\epsilon^2 \cdot (1-\mu)}$).

1) $\E[Z] = \E[Y_1] = \QueryNum$ since the $Y_1$'s are independent of each other and $\E[Y_1] = \QueryNum$ from Proposition \ref{lemma:1}.

2) $\Var[Z] = \frac{\Var[Y_1]}{h} \leq \frac{1}{h} \cdot \frac{b}{1-b} \cdot (\frac{1}{b^n} - 1) \cdot \QueryNum^2$ from Proposition \ref{lemma:2}, and the rightmost term is equal to $\epsilon^2 \cdot (1-\mu) \cdot \QueryNum^2$ from the definition of $h$.

3) From the Chebyshev inequality, $\Pr(Z - \E[Z] \geq \epsilon \cdot \E[Z])$ $\leq \frac{ \Var[Z] }{ \epsilon^2 \cdot \E[Z]^2 }$ where we know $\frac{ \Var[Z] }{ \epsilon^2 \cdot \E[Z]^2 } \leq \frac{ \epsilon^2 \cdot (1-\mu) \cdot \QueryNum^2 }{ \epsilon^2 \cdot \QueryNum^2 } = 1-\mu$ from the above statements.

Regarding $b$ and $n$ as constants, $h$ is also a constant. Acquiring $Y_1$ by $h$ times can be done in $O(\AGM)$ time, and we can obtain $Z$ in $O(\AGM)$ time that satisfies $\Pr(Z - \E[Z] < \epsilon \cdot \E[Z]) > \mu$.
\endproof
\fi

%% file: 08.experiments.tex
\ifFullVersion
\else
\vspace*{-0.3cm}
\fi
\section{Experiments}\label{sec:exp}

We now evaluate the performance of \Alley to answer the following research questions. 

\begin{itemize}
\vspace*{-0.1cm}
  \item \textbf{Q1.} Compared to existing estimators, how well does \Alley perform cardinality estimation on various datasets and queries, in terms of accuracy and efficiency? (Section \ref{subsec:exp_overall}) 
  \item \textbf{Q2.} How much does the tangled pattern index improve the accuracy of \Alley? (Section \ref{subsec:exp_overall}) 
  \item \textbf{Q3.} Does \Alley effectively reduce sampling failures, even for small sampling ratios? (Section \ref{subsec:exp_sampling_failure}) 
  \item \textbf{Q4.} How does our novel mining approach (i.e., walk-fail-then-calculate) improve indexing performance compared to a naive pattern mining approach? (Section \ref{subsec:exp_index_build}) %
  \vspace*{-0.1cm}
\end{itemize}

\input{experiment_subchapters/1.setup}

\input{experiment_subchapters/2.overall}

\input{experiment_subchapters/3.sampling_failure}


\input{experiment_subchapters/5.index_build}

%% file: experiment_subchapters/1.setup.tex
\vspace{-0.3cm}
\subsection{Experimental Setup}\label{subsec:exp_setup}

\noindent \underline{\textbf{Datasets and query sets.}} We use six real-world and synthetic datasets and the corresponding query sets shown in Table \ref{table:dataset}. \LUBM \cite{guo2005lubm}, \YAGO    \cite{suchanek2008yago}, \AIDS \cite{AIDS}, and \Human are used in \cite{park2020g}, while \HPRD and \Youtube are used in \cite{sun2020memory}. We use the same queries used in \cite{park2020g} and \cite{sun2020memory} except for \Youtube. For \Youtube, we use 763 queries with less than 10B embeddings out of 1,800 queries in the original query set, since calculating the true cardinality $\QueryNum$ requires tremendous computation.
\revision{For \LUBM, we use scale factor 80 by default as in \cite{park2020g}, and additionally use larger datasets with scale factors 160, 320, 480, 640, and 800; $|E_g|$ and $|V_g|$ increase linearly to the scale factor.}
{For \HPRD and \Youtube, query size denotes $|V_q|$ as in \cite{sun2020memory}, while it denotes the number of RDF triples for other datasets as in \cite{park2020g}.}


\begin{table}[h]
\renewcommand{\tabcolsep}{1mm}
\scriptsize
\vspace*{-0.3cm}
\caption{Statistics of datasets.}
\vspace*{-0.4cm}
\label{table:dataset}
\begin{tabular}{c|c|c|c|c|c|c}
\hline
\textbf{Dataset} & \textbf{\LUBM} & \textbf{\YAGO} & \textbf{\AIDS} & \textbf{\Human} & \textbf{\HPRD} & \textbf{\Youtube} \\ \hline
\textbf{\# vertices} & 2.6M & 12.8M & 254K & 4.7K & 9.5K & 1.1M \\
\textbf{\# edges} & 12.3M & 15.8M & 548K & 86K & 70K & 6.0M \\
\textbf{\# vertex labels} & 35 & 188K & 50 & 89 & 307 & 25 \\
\textbf{\# edge labels} & 35 & 91 & 4 & 0 & 0 & 0 \\
\textbf{\# queries used} & 6 & 1,366 & 780 & 49 & 1,800 & 763 \\
\textbf{Query size} & 4 to 6 & 3 to 12 & 3 to 12 & 3 & 4 to 32 & 4 to 32 \\
\textbf{Cardinality} & 15 to 22K & 1 to 28.7M & 1 to 952K & 1 to 9.6K & 1 to 3.2B & 3 to 10B \\ \hline
\end{tabular}
\vspace*{-0.3cm}
\end{table}

\noindent \underline{\textbf{Measure.}} We measure  accuracy and efficiency using  $q$-error \cite{moerkotte2009preventing} and  elapsed time, respectively. 
The $q$-error quantifies the ratio between the actual and the estimated cardinality. The $q$-error is always greater than one, and the smaller the $q$-error, the more accurate the estimation is. Formally, $\textrm{$q$-error} = \max\bigg(\frac{\max(1,Z)}{\max(1,\QueryNum)}, \frac{\max(1,\QueryNum)}{\max(1,Z)}\bigg)$.

For sampling-based methods, we run 30 times for each query while we run one time for deterministic summary-based methods, as in \cite{park2020g}. For \LUBM, we report the mean and the standard deviation of the $q$-error for each query. For the other datasets having numerous queries, we grouped queries by topology, query size, and true cardinality and then, we report the quartiles (i.e., the 25\%, 50\%, and 75\%-tiles) and the whiskers of the $q$-errors for each group.

\noindent \underline{\textbf{Running Environment.}} 
We conducted experiments on a Linux machine with 16 Intel Xeon E5-2450 2.10 GHz CPUs and 32 GB RAM \revision{by default, and a machine with 512 GB RAM for scalability test,} using a single thread for all experiments. We set one minute as the timeout threshold. For \spellcheck{a} fair comparison, if a method raises a timeout at least once in a particular query group, we exclude the results of the method for that group.
Following \cite{park2020g}, we use sample size $s = N \cdot r$ where $N$ is the number of triples in $g$ that match a triple in $q$, and $r \in (0, 1)$ is the given sampling ratio. We set $r = 0.1\%$ as default and use 1\%, 0.1\%, and 0.01\% for sensitivity analysis. Note that during query optimization a cardinality estimator might be invoked more than a thousand times for a single query. Thus, the estimator should be able to estimate in a few milliseconds. In this respect, we choose the default sampling ratio where sampling-based methods can complete estimation in about one millisecond on most datasets. We report the performance of both using only the sampling technique without any synopsis (denoted by \Alley in this section) and with the tangled pattern index (denoted by \AlleyPM). We use $b = 1/32$ as default. When building the tangled pattern index, we use $\zeta = 0.9$, $maxN = 5$ when using edge labels and $maxN = 4$ when using vertex labels. We set the maximum number of label groups in the index to 32.

\noindent \underline{\textbf{Competitors.}} We include all seven methods in~\cite{park2020g} as our competitors from both graph and relational domains.
\newredtext{We additionally considered two more recent competitors, \IBJS~\cite{leis2017cardinality} and subgraph catalog, a summary-based method used in~\cite{DBLP:journals/pvldb/MhedhbiS19}. However, subgraph catalog was originally proposed for graphs with few or zero labels and does not scale well for heterogeneous graphs. It took more than a day to build a summary for almost all datasets we used. Therefore, we exclude subgraph catalog from our experiments.}
For methods evaluated in~\cite{park2020g}, \CSET~\cite{neumann2011characteristic} and \SumRDF~\cite{stefanoni2018estimating} are summary-based methods for graphs, while \WanderJoin~\cite{li2016wander} (aka \WJ) and \Correlated~\cite{vengerov2015join} (aka \CS) are sampling-based methods for relations. 
\IMPR~\cite{DBLP:conf/icdm/ChenL16} is a sampling-based method that estimates the cardinality of small graphs with three to five vertices.
\BSK~\cite{cai2019pessimistic} is a summary-based method for relations that estimates the upper bound of $\QueryNum$. 
\JSUB in ~\cite{park2020g} also estimates the upper bound of $\QueryNum$ using the sampling strategy of~\cite{zhao2018random}.
We use the public implementation of \GCARE~\footnote{https://github.com/yspark-dblab/gcare} for the seven estimators. We additionally implement \Alley and \IBJS on top of \GCARE. Since \IBJS was originally proposed to obtain good-quality samples that aid cost-based query optimizers, it does not choose any sampling order but obtains samples for all subqueries and injects the samples into query optimizers. This is inappropriate for our experiments with fixed sampling ratio; thus, we applied the sampling order of \WJ, which is known to be the most accurate and efficient, to \IBJS.
\IBJS works similarly to \WJ, but it samples a batch of edges for each walk step instead of sampling an edge as in \WJ.
\revision{For \WJ, we additionally implemented two optimization techniques as described in Section 3.6 of \cite{li2017wander} on top of the \GCARE implementation.}

%% file: experiment_subchapters/2.overall.tex
\vspace*{-0.2cm} \subsection{Overall Performance}\label{subsec:exp_overall}



\begin{figure}[h!]
\begin{subfigure}{\columnwidth}
\ifFullVersion
\else
\vspace*{-0.2cm}
\fi
\centering
\includegraphics[width=0.90\columnwidth]{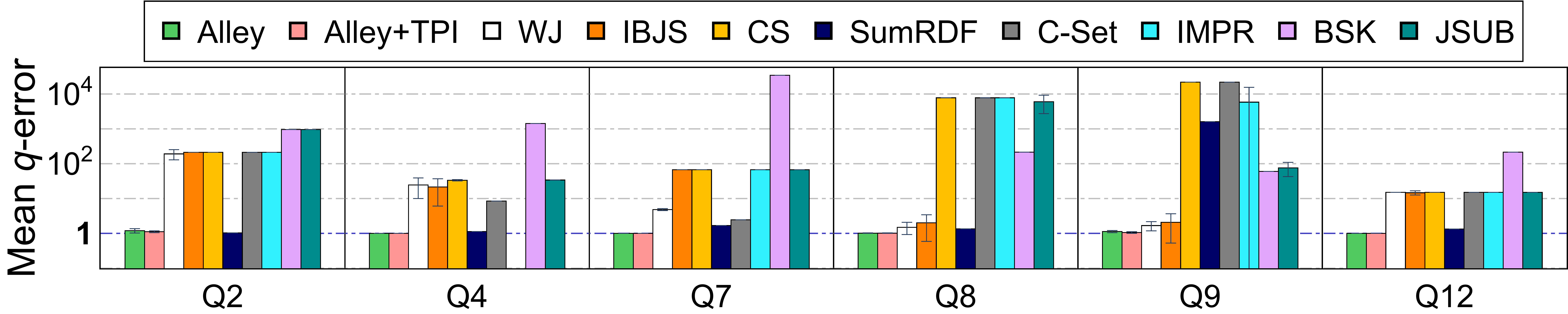}
\vspace*{-0.2cm}
\caption{\revision{Varying query (\LUBM).}}
\label{subfig:SmallQuery-LUBM}
\end{subfigure}
\begin{subfigure}{\columnwidth}
\centering
\includegraphics[width=0.90\columnwidth]{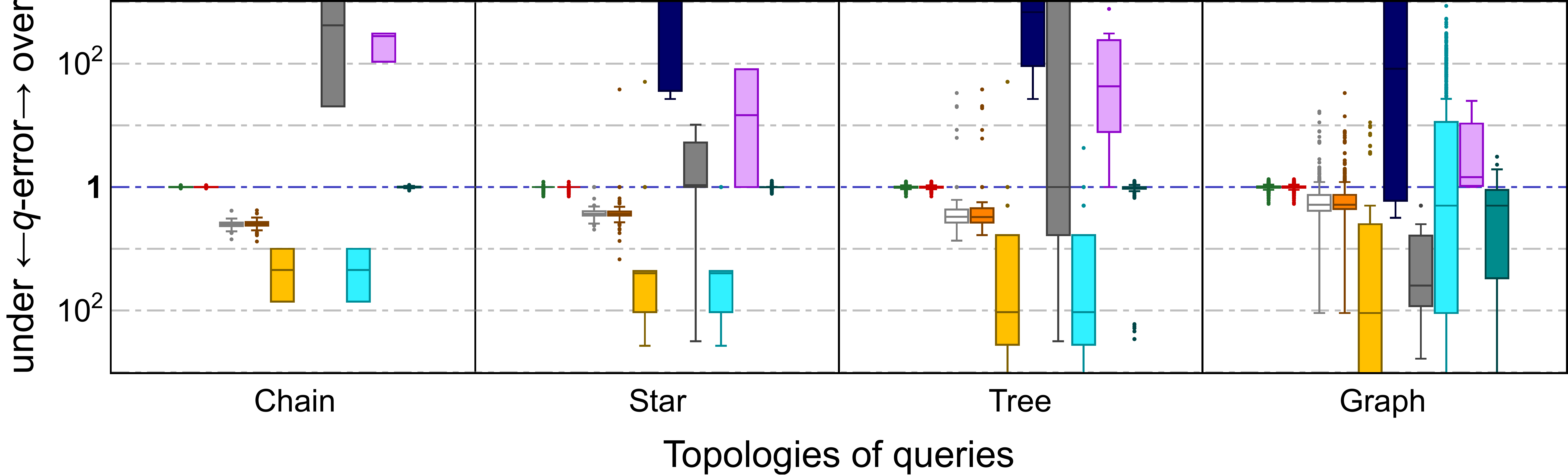}
\vspace*{-0.2cm}
\caption{\revision{Varying query topology (\Human).}}
\label{subfig:SmallQuery-Human}
\end{subfigure}
\vspace*{-0.4cm}
\caption{\revision{Accuracy using small queries.}}
\label{fig:SmallQuery}
\ifFullVersion
\else
\vspace*{-0.2cm}
\fi
\end{figure}

\vspace*{-0.5ex} \noindent \underline{\textbf{Using Small Queries.}}
For \LUBM and \Human, \Alley performs a near perfect estimation, achieving almost one $q$-error (Figure~\ref{fig:SmallQuery}). 
This is due to our novel sampling strategy, random walk with intersection that reduces the sample space. However, other sampling-based methods, i.e., \WJ, \IBJS, \IMPR, \CS, and \JSUB, result in large $q$-error (up to four orders of magnitude higher than \Alley's) due to sampling failures even for these small queries.
\SumRDF is relatively accurate on \LUBM but significantly over-estimates on \Human, since \SumRDF was originally proposed for RDF graphs, such as \LUBM, while \Human is a non-RDF graph. These results indicate that summary-based methods do not guarantee consistent performance over various graphs. 

\begin{figure}[h!]
\centering
\includegraphics[width=0.95\columnwidth]{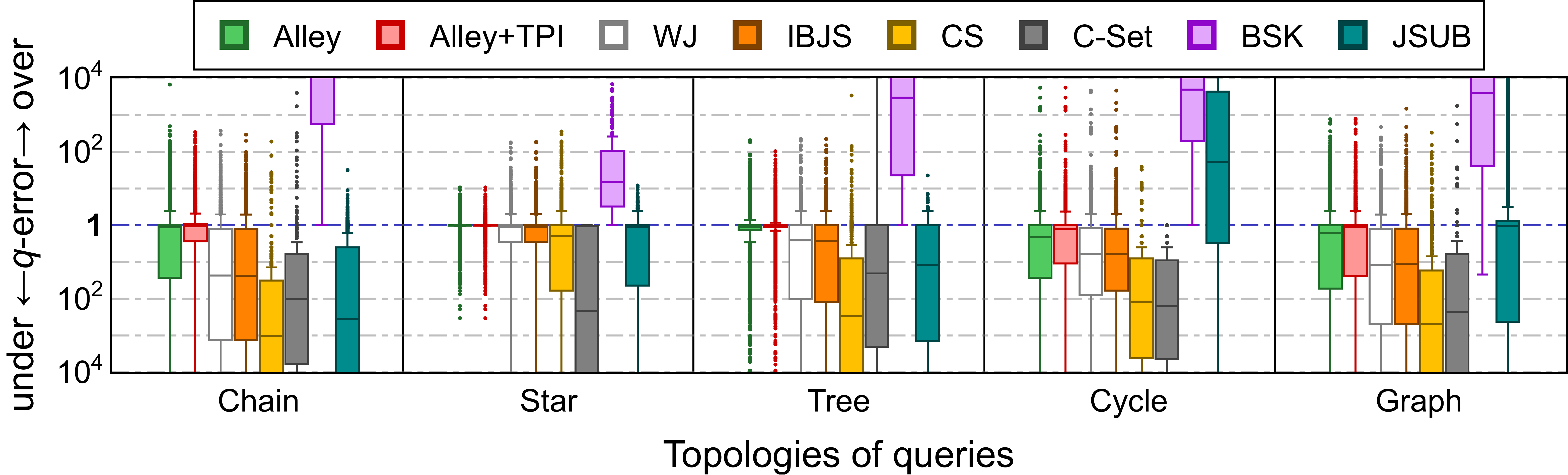}
\ifFullVersion
\else
\vspace*{-0.4cm}
\fi
\caption{\revision{Varying query topology (\YAGO).}}
\label{fig:MediumQuery}
\ifFullVersion
\else
\vspace*{-0.4cm}
\fi
\end{figure}

\vspace*{-0.5ex} \noindent \underline{\textbf{Using Medium-size Queries.}} Figure \ref{fig:MediumQuery} shows the results on \YAGO for various query topologies. \SumRDF is excluded in this experiment since at least one query of any topology timed out. 
Overall, more complex queries involving long chains and cycles lead to larger $q$-error. Nevertheless, \Alley consistently and significantly outperforms the others for all topologies. For stars, obviously, \Alley shows extremely high accuracy since they can be covered by a 1-hop intersection. For trees, \Alley still achieves superior accuracy by effectively reducing the sample space. For long chains and cyclic queries (i.e., Cycle and Graph), all sampling-based methods often under-estimate due to the highly selective structures in queries. \Alley clearly outperforms all the others for these queries as well.

\begin{figure}[h!]
\centering
\includegraphics[width=0.95\columnwidth]{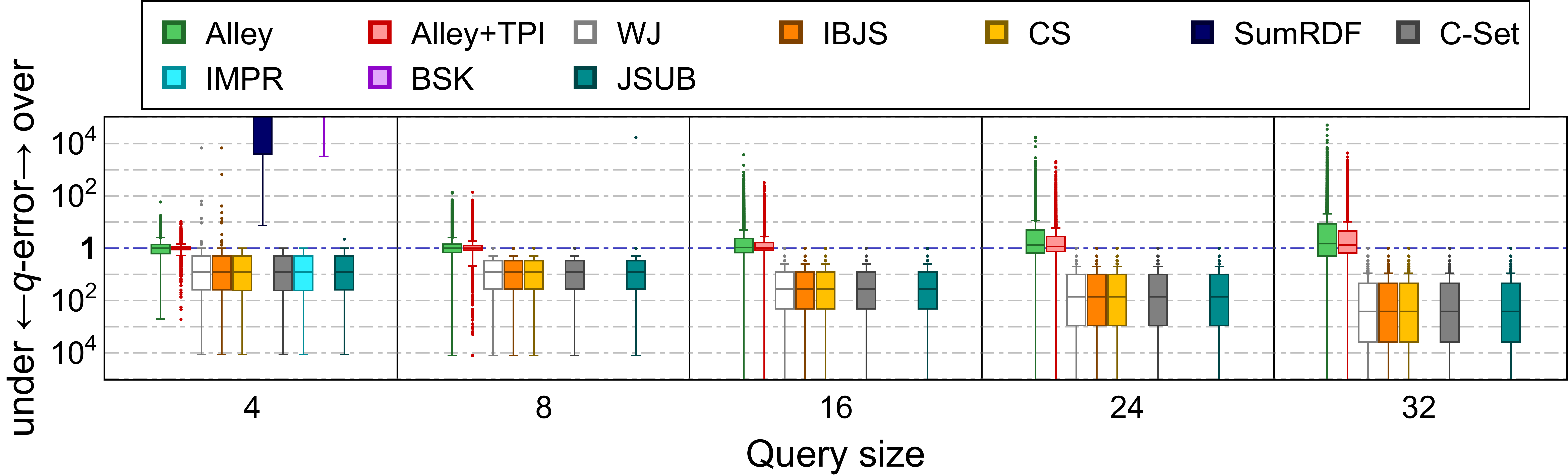}
\vspace*{-0.4cm}
\caption{\revision{Varying query size (\HPRD).}}
\ifFullVersion
\else
\vspace*{-0.2cm}
\fi
\label{fig:LargeQuery}
\end{figure}

\noindent \underline{\textbf{Using Large Queries.}}
For larger queries with up to 32 vertices on \HPRD, \Alley significantly outperforms the others (Figure \ref{fig:LargeQuery}). All competitors suffer from significant under-estimation or time-out (\SumRDF and \BSK). 
Specifically, all the sampling-based methods except \Alley have similar $q$-error since, for nearly all queries, they fail to sample any embedding and report zero. \Alley, however, shows robust performance with $q$-error less than 10 for more than half of the trials. The results show that reducing the sample space is a "must" for large and complex queries.
These experiments show the great advantage of interleaving intersections with random walks.
\ifFullVersion
For other datasets, we observed similar trends (Figures~\ref{fig:VaryingQuerySize-AIDS}-\ref{fig:VaryingQuerySize-YAGO}).
\else
We observed similar trends on other datasets (see~\cite{AAA} for details).
\fi

\ifFullVersion
\begin{figure}[h!]
\centering
\includegraphics[width=0.95\columnwidth]{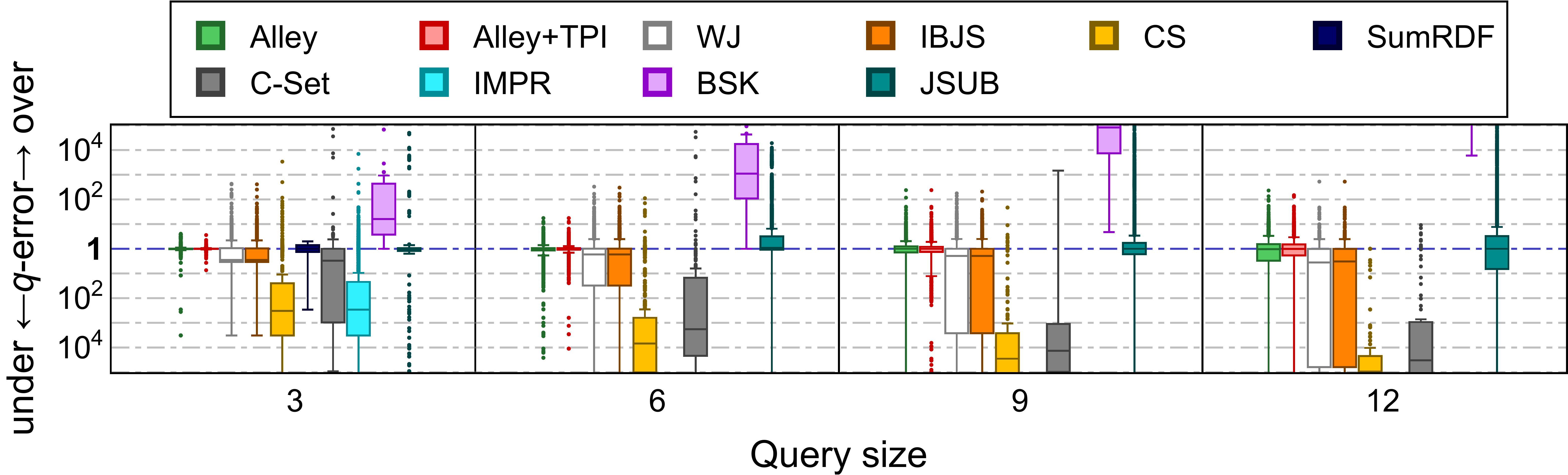}
\vspace*{-0.4cm}
\caption{Varying query size (\AIDS).}
\label{fig:VaryingQuerySize-AIDS}
\end{figure}

\begin{figure}[h!]
\centering
\includegraphics[width=0.95\columnwidth]{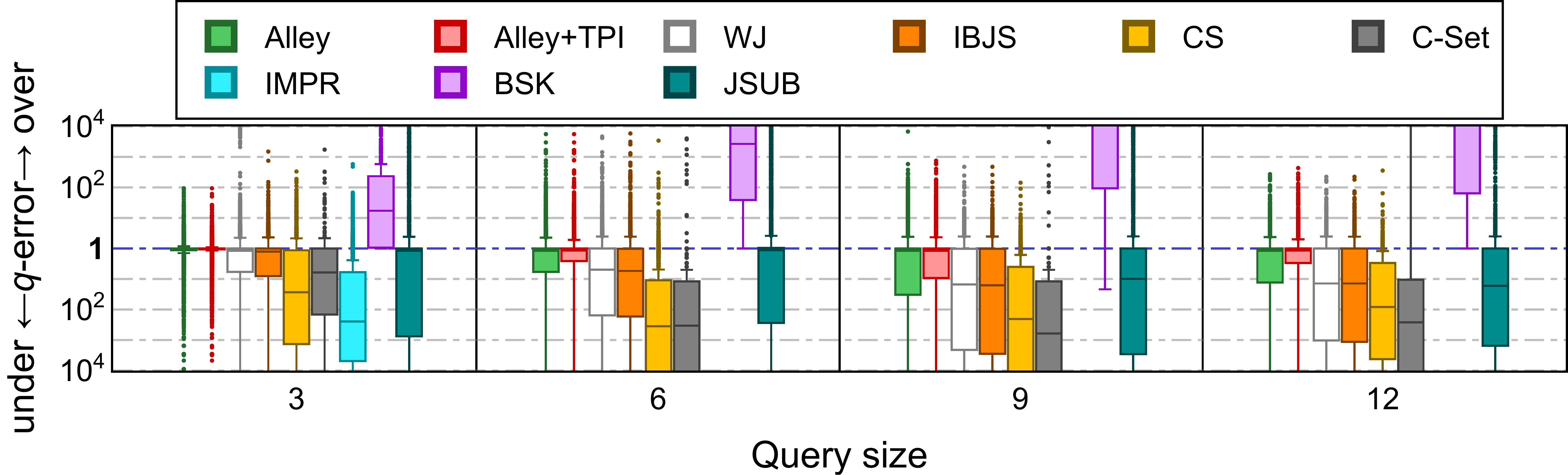}
\vspace*{-0.4cm}
\caption{Varying query size (\YAGO).}
\label{fig:VaryingQuerySize-YAGO}
\end{figure}

\fi

\begin{figure}[h!]
\centering
\ifFullVersion
\else
\vspace*{-0.2cm}
\fi
\includegraphics[width=0.95\columnwidth]{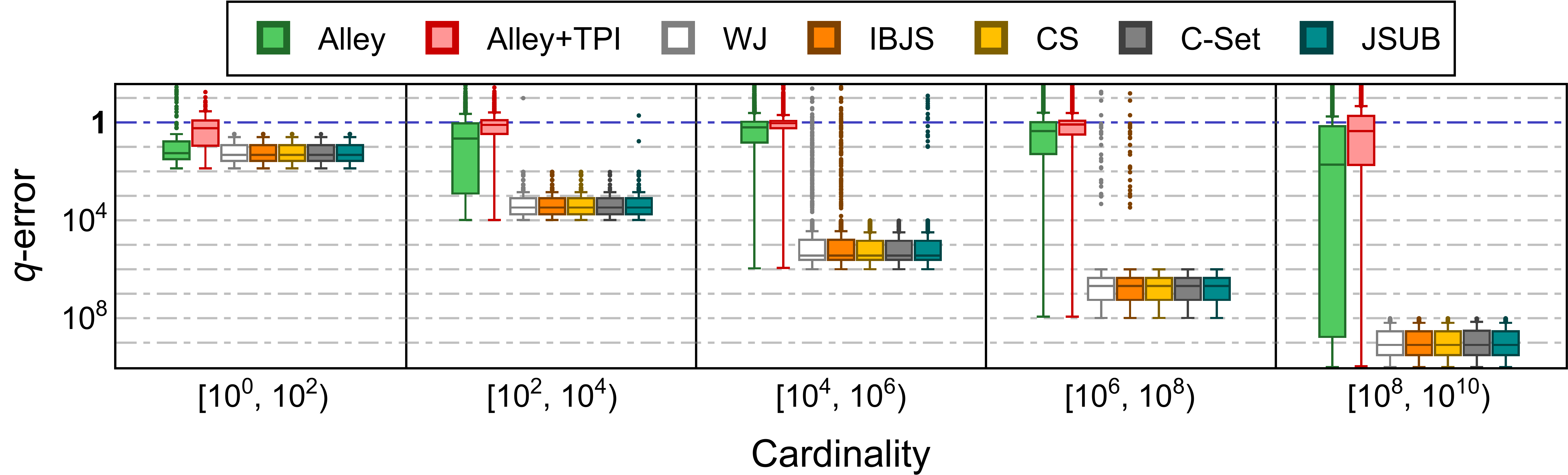}
\vspace*{-0.4cm}
\caption{\revision{Varying true cardinality (\Youtube).}}
\label{fig:TrueCard}
\ifFullVersion
\else
\vspace*{-0.2cm}
\fi
\end{figure}

\noindent \underline{\textbf{Varying True Cardinality.}}
We now show the trend by varying $\QueryNum$ on \Youtube (Figure \ref{fig:TrueCard}), where \Youtube queries have larger $\QueryNum$ than \HPRD queries.
The $q$-error of \Alley tends to increase as $\QueryNum$ increases, but suffers less from severe under-estimation, which occurs in all of the other estimators.
\ifFullVersion
This trend is also shown in Figures~\ref{fig:TrueCarde-Human}-\ref{fig:TrueCarde-Yago}.
\fi

\ifFullVersion

\begin{figure}[h!]
\centering
\includegraphics[width=0.95\columnwidth]{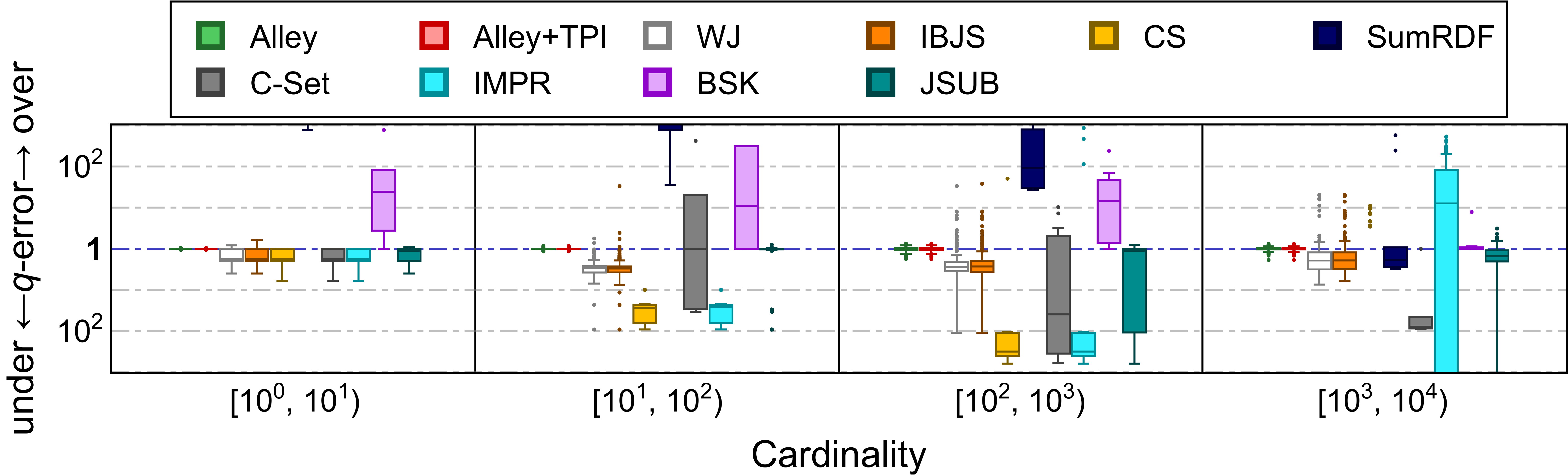}
\vspace*{-0.4cm}
\caption{\revision{Varying true cardinality (\Human).}}
\label{fig:TrueCarde-Human}
\end{figure}

\begin{figure}[h!]
\centering
\includegraphics[width=0.95\columnwidth]{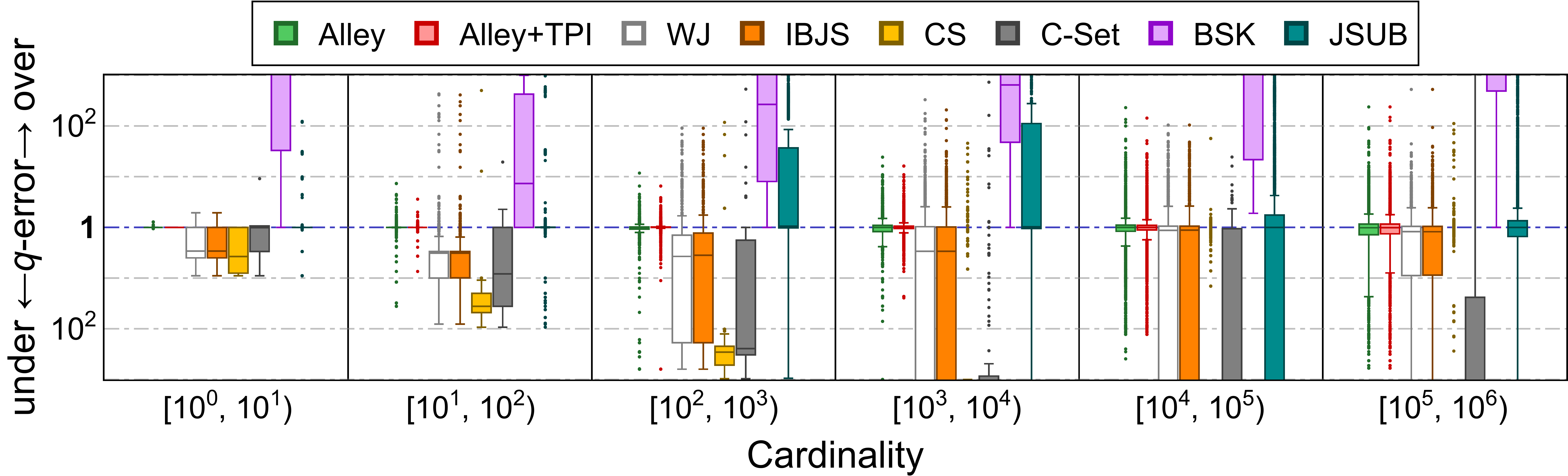}
\vspace*{-0.4cm}
\caption{Varying true cardinality (\AIDS).}
\label{fig:TrueCarde-AIDS}
\end{figure}

\begin{figure}[h!]
\centering
\includegraphics[width=0.95\columnwidth]{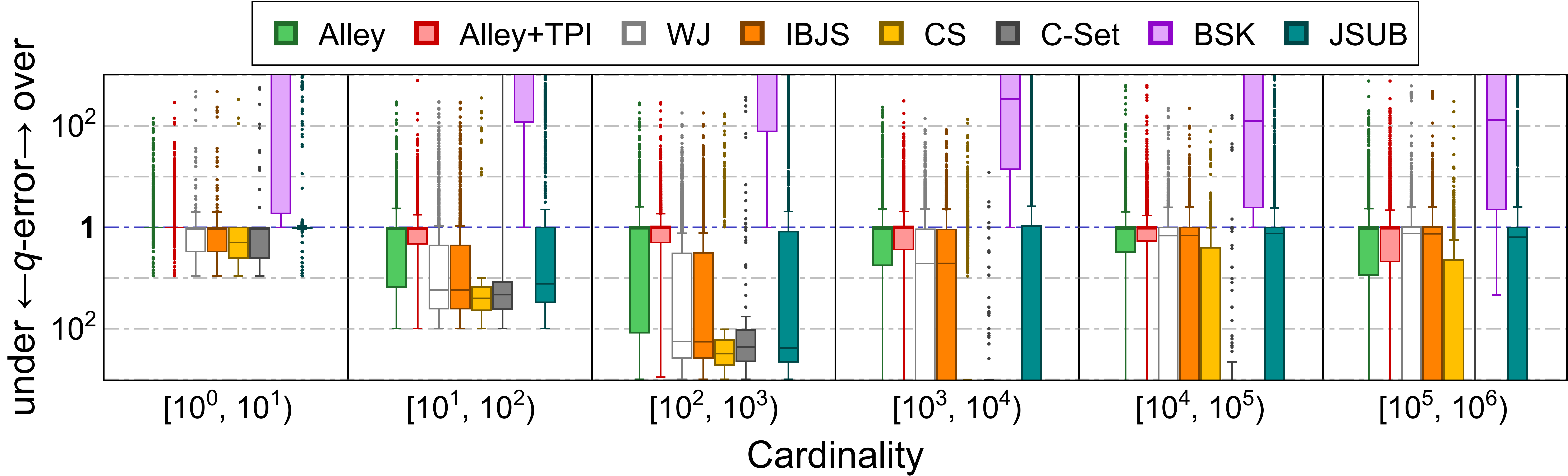}
\vspace*{-0.4cm}
\caption{\revision{Varying true cardinality (\YAGO).}}
\label{fig:TrueCarde-Yago}
\end{figure}
\fi

\begin{figure}[h!]
\ifFullVersion
\else
\vspace*{-0.2cm}
\fi
\begin{subfigure}{\columnwidth}
\centering
\includegraphics[width=0.90\columnwidth]{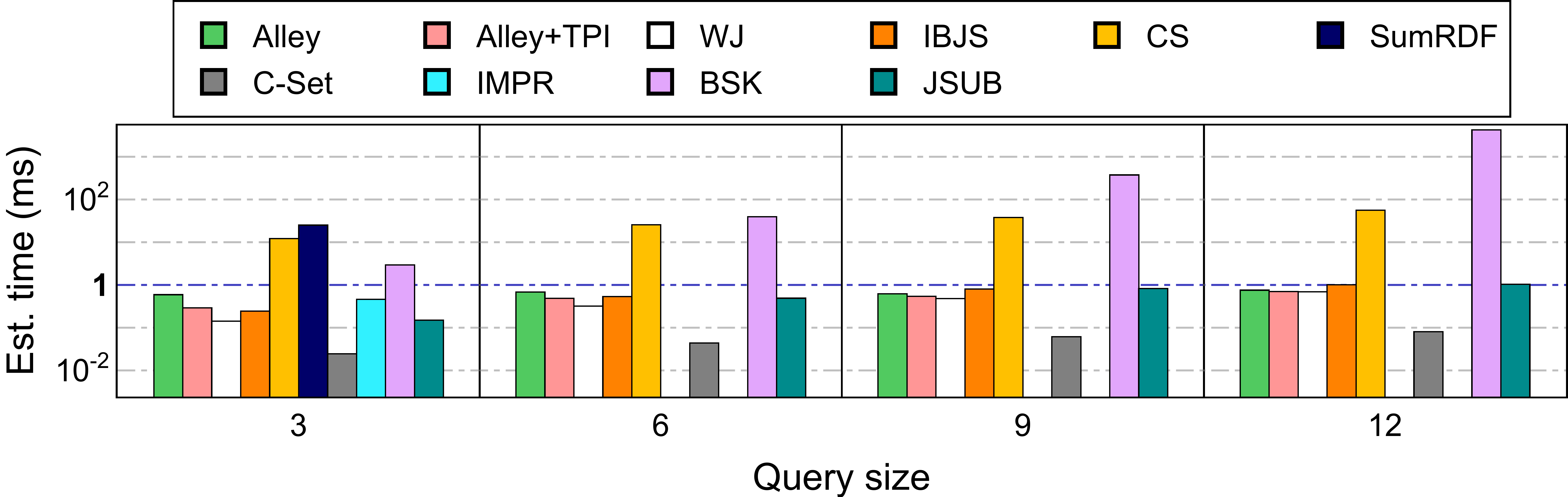}
\vspace*{-0.2cm}
\caption{\revision{\AIDS.}}
\label{subfig:etime-AIDS}
\end{subfigure}
\begin{subfigure}{\columnwidth}
\centering
\includegraphics[width=0.90\columnwidth]{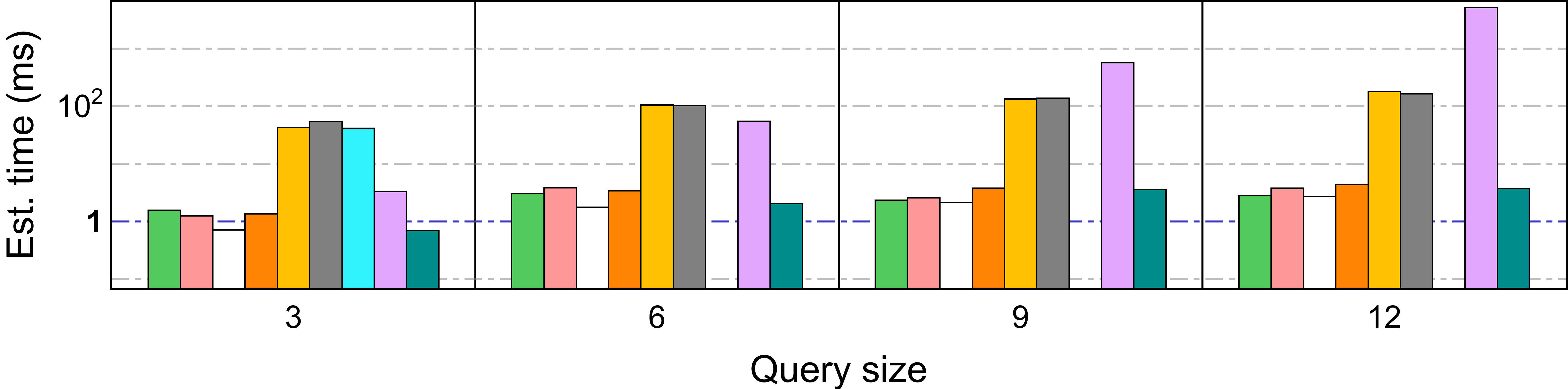}
\vspace*{-0.2cm}
\caption{\revision{\YAGO.}}
\label{subfig:etime-YAGO}
\end{subfigure}
\vspace*{-0.4cm}
\caption{\revision{Efficiency by varying query size.}}
\ifFullVersion
\else
\vspace*{-0.4cm}
\fi
\label{fig:etime}
\end{figure}

\noindent \underline{\textbf{Measuring Efficiency.}}
We measure the elapsed time on \AIDS and \YAGO (Figure~\ref{fig:etime}). 
When query graphs are small, \Alley is slightly slower than \WJ or \IBJS, which are the most efficient ones among sampling-based methods. However, as query size grows, \Alley achieves  efficiency similar to theirs. This is due to the early stopping effect of intersections,  explained in Section~\ref{subsec:multiway}. Note that, in addition to reasonable accuracy, \Alley achieves high efficiency with less than a millisecond latency. 
While \CSET is faster than \Alley on \AIDS, it does not scale well on \YAGO since \spellcheck{\YAGO has many labels. \CSET generates and maintains} a large number of entries in its summary, thereby increasing the search time.
\ifFullVersion
We observed the same phenomena in the other datasets (Figures~\ref{fig:Efficiency-HPRD}-\ref{fig:Efficiency-Youtube}).
\fi


\ifFullVersion

\begin{figure}[h!]
\centering
\includegraphics[width=0.95\columnwidth]{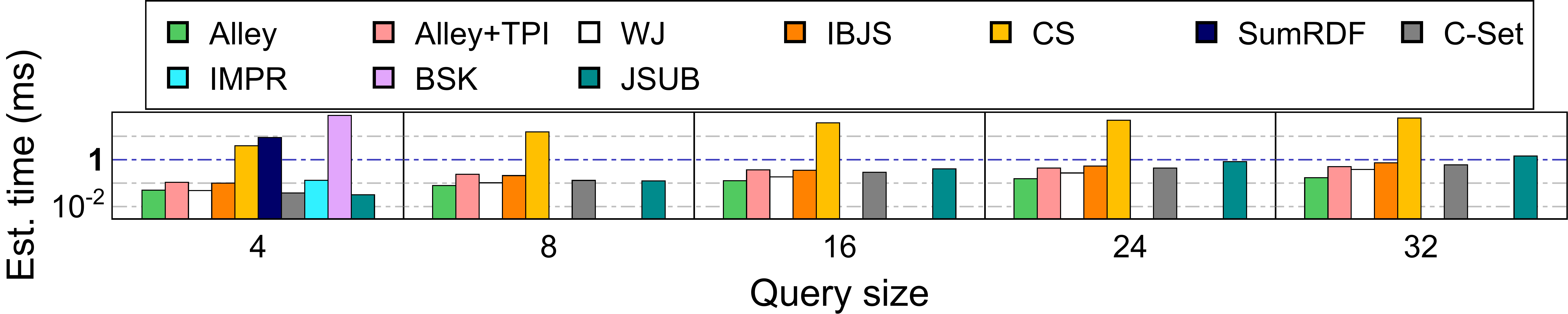}
\vspace*{-0.3cm}
\caption{Efficiency by varying query size (\HPRD).}
\label{fig:Efficiency-HPRD}
\end{figure}

\begin{figure}[h!]
\centering
\includegraphics[width=0.95\columnwidth]{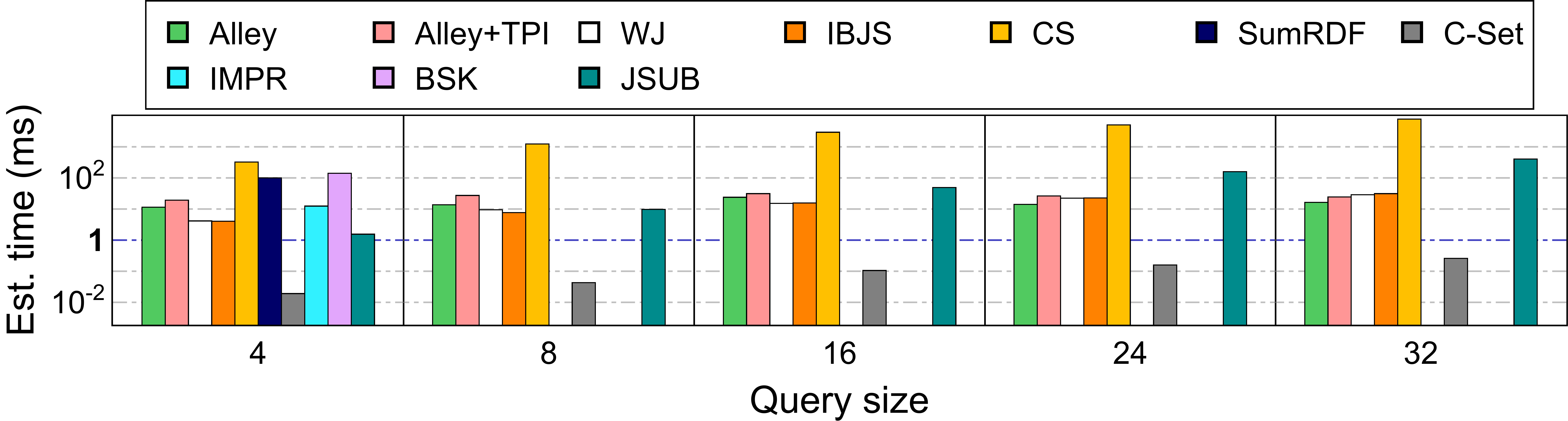}
\vspace*{-0.2cm}
\caption{Efficiency by varying query size (\Youtube).}
\label{fig:Efficiency-Youtube}
\end{figure}

\fi

\noindent \underline{\textbf{Combining Sampling with Synopsis.}}
Compared to \Alley, \AlleyPM always has better accuracy while having almost the same (sometimes even better) efficiency. This improvement becomes more apparent for large and complex queries, as we claimed earlier. The efficiency of \AlleyPM comes from reducing sample space and intersection overhead by using the index, thereby reducing the estimation time. Due to this advantage along with the superiority of our mining technique that effectively reduces index size, the computational overhead of \spellcheck{the} index search is almost hidden.

%% file: experiment_subchapters/3.sampling_failure.tex
\subsection{Sampling Failures}\label{subsec:exp_sampling_failure}

In this section, we examine the failures of sampling-based methods.
We count extreme failure cases where each method outputs zero due to no sampling success. We compare \Alley and \AlleyPM with the two best-performing sampling-based competitors, \WJ and \IBJS.
Table~\ref{tab:failure} shows the ratio of extreme failure cases to total trials for all queries.
\Alley and \AlleyPM significantly reduce the failure rates compared to the other methods for all datasets. In particular, when decreasing the sampling ratio on \LUBM, the failure rates of the other methods increase greatly by up to \revision{70.6\%}, while \Alley fails only \revision{1.1\%} of the total trials.
\revision{This is due to the small sample space of \Alley, for instance, the average size of $|P_{o_i}|$ for \Alley is 107 on \LUBM ($r=0.01\%$), which is less than 0.05\% of \WJ's.}
This results in extremely high accuracy and robustness, as shown in Figure \ref{subfig:SmallQuery-LUBM}.

\begin{table}[h]
\scriptsize
\vspace*{-0.2cm}
\caption{\revision{The ratio of zero-estimation cases.}}
\vspace*{-0.4cm}
\label{tab:failure}
\begin{tabular}{c|ccc|c|c|c|c}
\hline
\textbf{Dataset} & \multicolumn{3}{c|}{\textbf{\LUBM}} & \textbf{\YAGO} & \textbf{\AIDS} & \textbf{\HPRD} & \textbf{\Youtube} \\ \hline
\textbf{\begin{tabular}[c]{@{}c@{}}Sampling \\ Ratio ($r$)\end{tabular}} & \textbf{1\%} & \textbf{0.1\%} & \textbf{0.01\%} & \textbf{0.1\%} & \textbf{0.1\%} & \textbf{0.1\%} & \textbf{0.1\%} \\ \hline
\textbf{\Alley} & {\ul \textbf{0.0\%}} & {\ul \textbf{0.0\%}} & {\ul \textbf{1.1\%}} & 23.8\% & 5.2\% & 15.9\% & 17.4\% \\
\textbf{\AlleyPM} & {\ul \textbf{0.0\%}} & {\ul \textbf{0.0\%}} & {\ul \textbf{1.1\%}} & {\ul \textbf{16.6\%}} & {\ul \textbf{1.4\%}} & {\ul \textbf{8.2\%}} & {\ul \textbf{6.3\%}} \\
\textbf{\WJ} & 20.6\% & 43.3\% & 70.6\% & 49.9\% & 23.8\% & 100\% & 99.3\% \\
\textbf{\IBJS} & 21.7\% & 41.1\% & 67.2\% & 49.8\% & 23.8\% & 100\% & 99.3\% \\ \hline
\end{tabular}
\end{table}

\begin{figure}[h!]
\begin{subfigure}{\columnwidth}
\ifFullVersion
\else
\vspace*{-0.2cm}
\fi
\centering
\includegraphics[width=0.95\columnwidth]{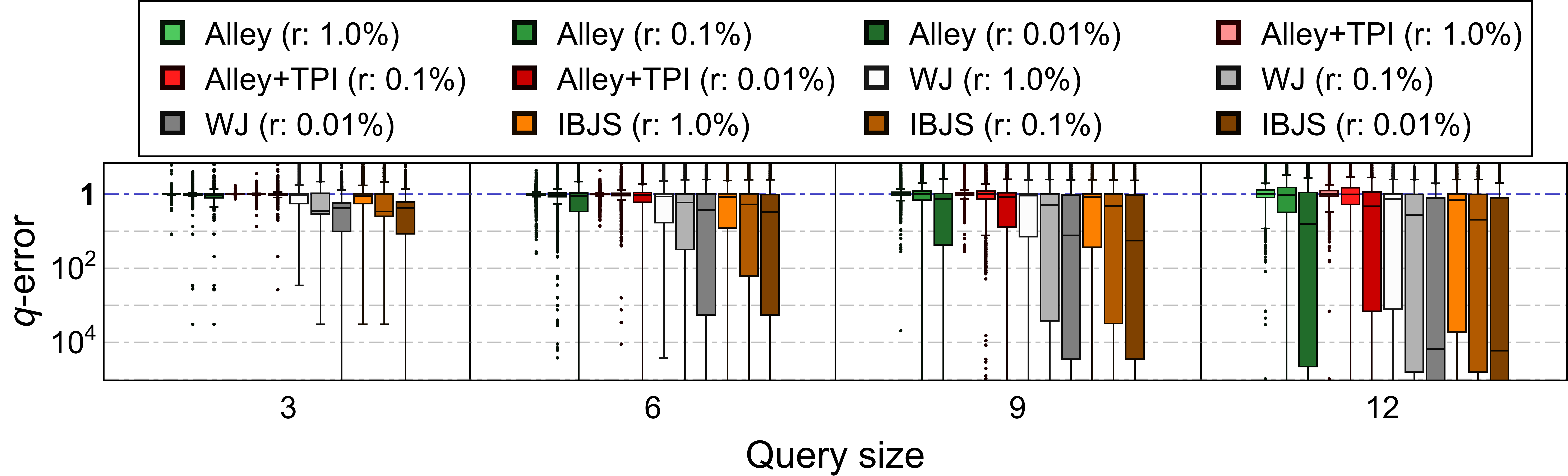}
\vspace*{-0.2cm}
\caption{\revision{\AIDS}}
\end{subfigure}
\begin{subfigure}{\columnwidth}
\centering
\includegraphics[width=0.95\columnwidth]{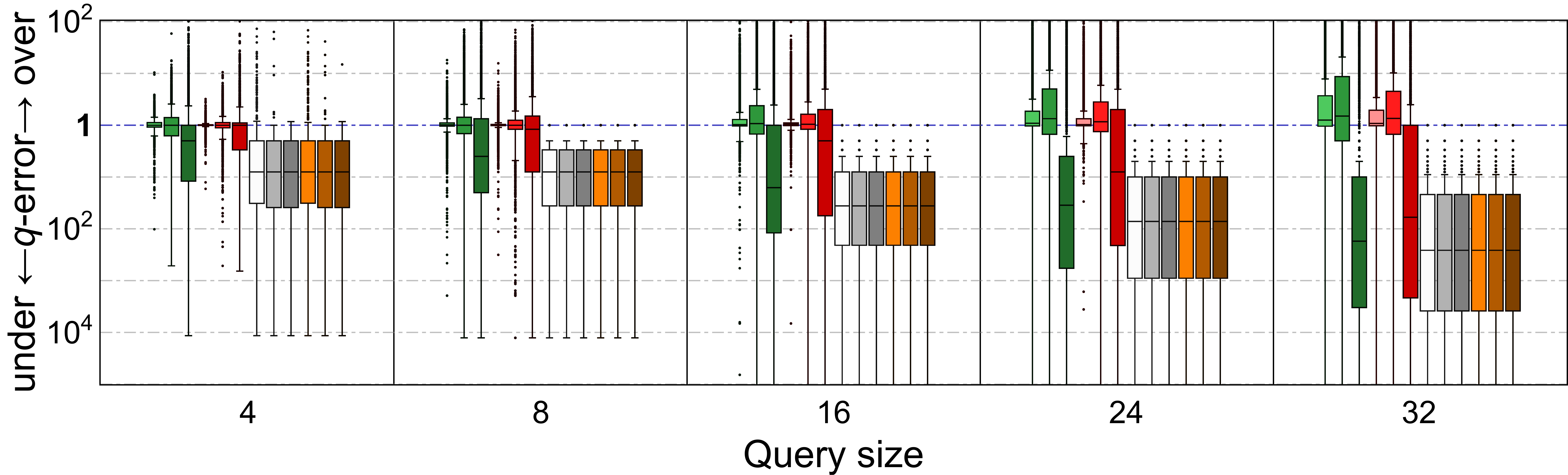}
\vspace*{-0.2cm}
\caption{\revision{\HPRD}}
\label{subfig:VaryingSamplingRatio-HPRD}
\end{subfigure}
\vspace*{-0.4cm}
\caption{\revision{Varying sampling ratio $r$.}}
\ifFullVersion
\else
\vspace*{-0.4cm}
\fi
\label{fig:VaryingSamplingRatio}
\end{figure}

\noindent \underline{\textbf{Varying Sampling Ratio.}}
In order to further investigate the robustness of \Alley, we vary the sampling ratio $r$ from 1\% (easy case) to 0.01\% (extremely hard case) on
\ifFullVersion
\AIDS, \HPRD, and \YAGO (Figures~\ref{fig:VaryingSamplingRatio}-\ref{subfig:VaryingSamplingRatio-Yago}).
\else
\AIDS and \HPRD (Figure~\ref{fig:VaryingSamplingRatio}).
\fi
We plot notches for better visibility. 
As the sampling ratio increases, the estimation variance of \Alley decreases and \Alley becomes more robust, as explained in Section \ref{sec:ht} and \ref{sec:proof}. Again, due to the superiority of our sampling strategy, \Alley and \AlleyPM outperform the others by orders of magnitudes for all ratios. Moreover, with $r=0.01\%$ accuracy, \Alley is comparable or better than the other methods with only $r=1\%$ accuracy.

As shown in Figure~\ref{subfig:VaryingSamplingRatio-HPRD}, for \HPRD, \WJ and \IBJS fail 100\% even with $r=1\%$, while \Alley and \AlleyPM show reasonable accuracy with $r=1\%$ and $0.1\%$.
However, when $r=0.01\%$ and the query size is larger than 10, \Alley also fails often. \AlleyPM is better than \Alley but still has \spellcheck{a} large variance. Depending on the complexity of the target data and query, one needs to adjust or adaptively determine the sampling ratio in order to perform an accurate estimation. Note that among all methods compared in the experiments, only \Alley is able to achieve this functionality in a reasonable time. Others suffer from 1) irrecoverable under-estimation (for sampling-based), 2) irrecoverable and large errors due to loss of information (for summary-based), or even 3) large estimation time.

\ifFullVersion

\begin{figure}[h!]
\centering
\includegraphics[width=0.95\columnwidth]{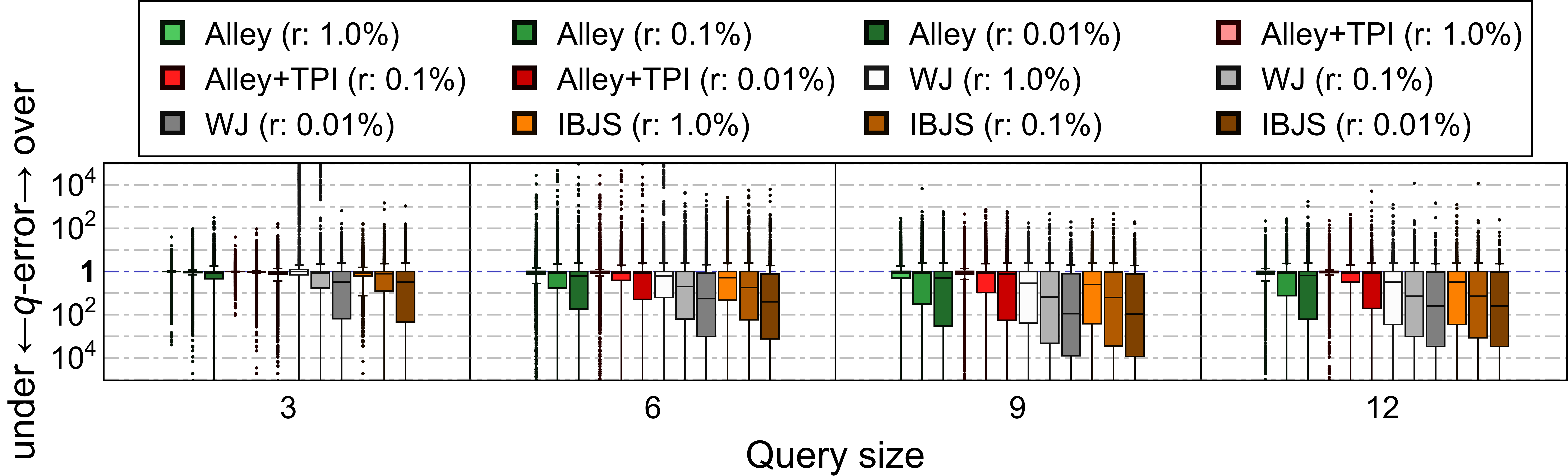}
\vspace*{-0.2cm}
\caption{Varying sampling ratio (\YAGO).}
\label{subfig:VaryingSamplingRatio-Yago}
\end{figure}

\fi

%% file: experiment_subchapters/5.index_build.tex
\subsection{Mining Performance}\label{subsec:exp_index_build}

We now show the impact of our novel mining approach on the performance of indexing and online estimation. We first evaluate how much our "walk-fail-then-calculate" approach improves the performance of building an index in terms of build time and index size, compared to a naive mechanism that stores domains for all small patterns. \revision{We also show the scalability of indexing over varying data sizes.} Then, we evaluate whether the tangled pattern index really retains important information. In other words, we evaluate whether \AlleyPM shows comparable performance with \Alley using the naive index (denoted as \AlleyNaive). We build the naive index using the same code as \AlleyPM with the only difference being setting $\zeta$ to zero. Note that we use only one thread for building those indices, as mentioned in Section~\ref{subsec:exp_setup}, although it can easily be parallelized by calculating multiple patterns simultaneously.

The result in Table \ref{tab:IndexPerformance} shows that our ``walk-fail-then-calculate'' approach effectively reduces both index size and time by up to \revision{56} and \revision{9.7} times, respectively. For $maxN=5$ in \YAGO, even the naive indexing approach fails since materializing the domains of all patterns requires a tremendous amount of memory. 
While increasing $maxN$ exponentially increases the size of the naive index and time as in typical frequent pattern mining methods, our approach can mitigate the overhead, resulting in a much lighter and more efficient index than using a naive approach.

\begin{table}[h]
\scriptsize
\ifFullVersion
\else
\vspace*{-0.2cm}
\fi
\caption{\revision{Ablation results of mining approaches.}}
\vspace*{-0.4cm}
\label{tab:IndexPerformance}
\begin{tabular}{ccccc}
\hline
\multicolumn{1}{c|}{} & \multicolumn{2}{c|}{\textbf{Max pattern size: 5}} & \multicolumn{2}{c}{\textbf{Max pattern size: 4}} \\ \hline
\multicolumn{1}{c|}{} & \begin{tabular}[c]{@{}c@{}}Walk-fail-\\ then-calculate\end{tabular} & \multicolumn{1}{c|}{\begin{tabular}[c]{@{}c@{}}Calculate all \\ ($\zeta=0$)\end{tabular}} & \begin{tabular}[c]{@{}c@{}}Walk-fail-\\ then-calculate\end{tabular} & \begin{tabular}[c]{@{}c@{}}Calculate all\\ ($\zeta=0$)\end{tabular} \\ \hline
\multicolumn{1}{l}{} & \multicolumn{4}{c}{\textbf{Dataset: \AIDS}} \\ \hline
\multicolumn{1}{c|}{\textbf{\begin{tabular}[c]{@{}c@{}}Index size in MB\\ (Relative size \\ to input data)\end{tabular}}} & \textbf{\begin{tabular}[c]{@{}c@{}}88.3\\ (6.17)\end{tabular}} & \multicolumn{1}{c|}{\begin{tabular}[c]{@{}c@{}}4,950\\ (346)\end{tabular}} & \textbf{\begin{tabular}[c]{@{}c@{}}16.2\\ (1.13)\end{tabular}} & \begin{tabular}[c]{@{}c@{}}640\\ (44.8)\end{tabular} \\ \hline
\multicolumn{1}{c|}{\textbf{Index time (sec)}} & \textbf{39.0} & \multicolumn{1}{c|}{316} & \textbf{4.0} & 38.8 \\ \hline
\multicolumn{1}{l}{} & \multicolumn{4}{c}{\textbf{Dataset: \YAGO}} \\ \hline
\multicolumn{1}{c|}{\textbf{\begin{tabular}[c]{@{}c@{}}Index size in MB\\ (Relative size \\ to input data)\end{tabular}}} & \textbf{\begin{tabular}[c]{@{}c@{}}2,218\\ (3.73)\end{tabular}} & \multicolumn{1}{c|}{\begin{tabular}[c]{@{}c@{}}(Out-of-\\memory)\end{tabular}} & \textbf{\begin{tabular}[c]{@{}c@{}}297\\ (0.50)\end{tabular}} & \begin{tabular}[c]{@{}c@{}}3,244\\ (5.46)\end{tabular} \\ \hline
\multicolumn{1}{c|}{\textbf{Index time (sec)}} & \textbf{1,800} & \multicolumn{1}{c|}{-} & \textbf{121} & 358 \\ \hline
\end{tabular}
\ifFullVersion
\else
\vspace*{-0.2cm}
\fi
\end{table}

\begin{figure}[h!]
\centering
\vspace*{-0.2cm}
\includegraphics[width=0.74\columnwidth]{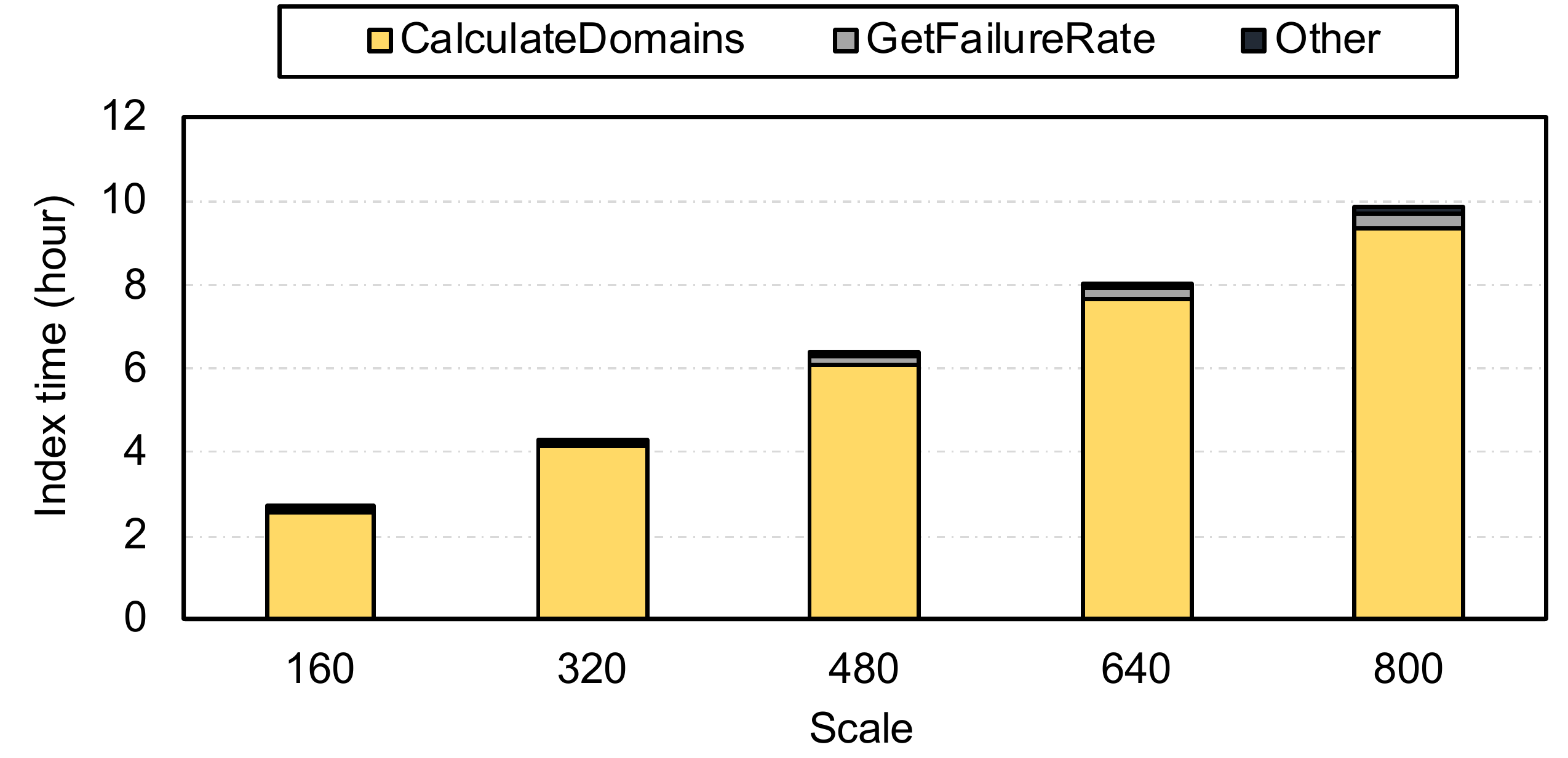}
\vspace*{-0.4cm}
\caption{\revision{Indexing time by varying data size.}}
\label{fig:Scalability}
\vspace*{-0.2cm}
\end{figure}

\revision{Figure \ref{fig:Scalability} shows the indexing performance over large \LUBM datasets with scale factors from 160 to 800. The indexing time increases linearly to the scale, and \CalculateDomains\xspace dominates other functions, occupying 95\% of the total time. \GetFailureRate\xspace accounts for only 3\%. The numbers of calls for \CalculateDomains\xspace and \GetFailureRate\xspace are similar for all data sizes, which are 9K and 150K, respectively; each call to \GetFailureRate\xspace is about 500 times faster than \CalculateDomains\xspace. Therefore, our ``walk-fail-then-calculate'' approach does not incur a significant overhead yet effectively prunes the patterns to index. 
}

\begin{figure}[h!]
\vspace*{-0.2cm}
\begin{subfigure}{\columnwidth}
\centering
\includegraphics[width=0.95\columnwidth]{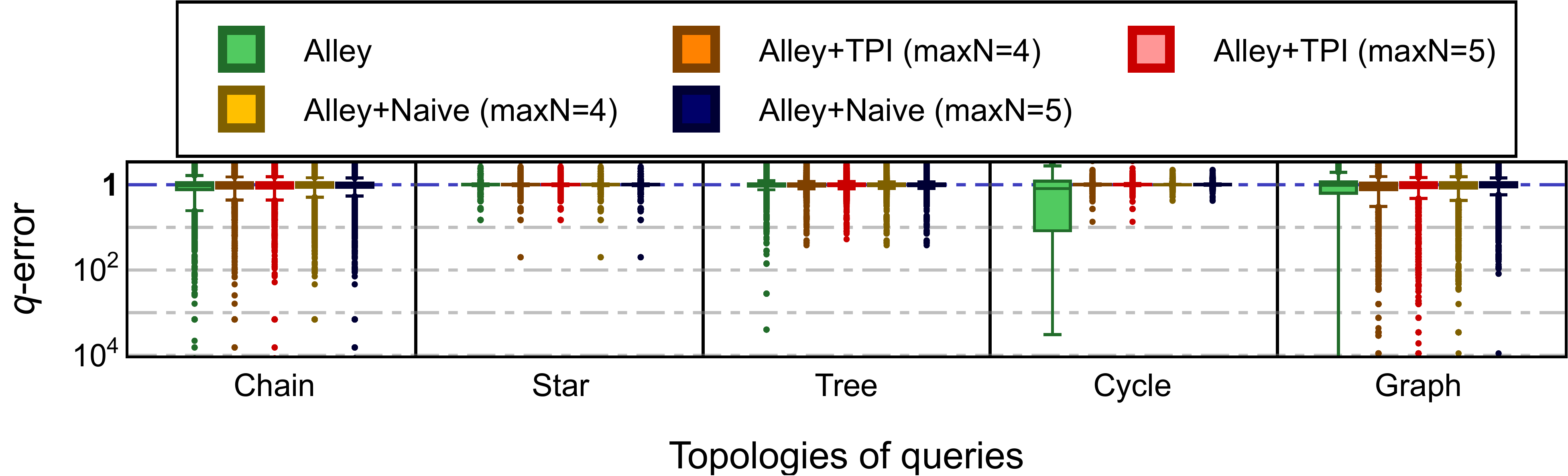}
\vspace*{-0.2cm}
\caption{\revision{\AIDS.}}
\label{subfig:qerror-AIDS}
\end{subfigure}
\begin{subfigure}{\columnwidth}
\centering
\includegraphics[width=0.93\columnwidth]{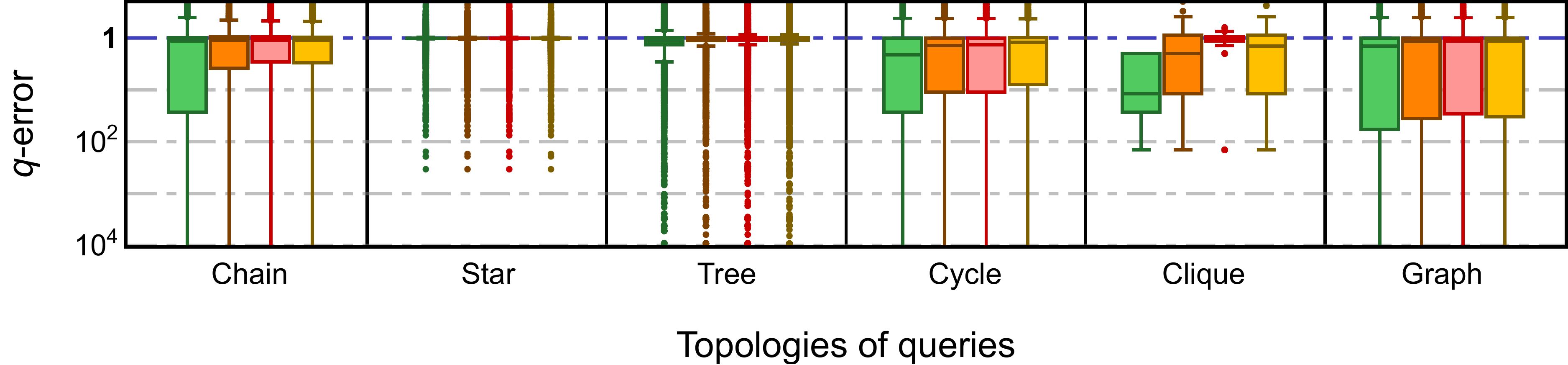}
\vspace*{-0.2cm}
\caption{\revision{\YAGO.}}
\label{subfig:qerror-YAGO}
\end{subfigure}
\vspace*{-0.4cm}
\caption{\revision{Accuracy by varying indexing approach.}}
\label{fig:IndexAccuracy}
\ifFullVersion
\else
\fi
\end{figure}

Figure \ref{fig:IndexAccuracy} shows that the tangled pattern index is as effective as the naive index. Both \AlleyPM and \AlleyNaive show higher accuracy than \Alley without using an index. These experimental results illustrate that 1) combining with synopses can increase the performance of sampling in cardinality estimation and 2) \redtext{filtering out domain calculation by random walks} during mining can increase scalability while preserving effectiveness.

\ifFullVersion
We additionally investigate the robustness of \AlleyPM regarding data updates. Edge insertion can introduce additional embeddings, which causes false negatives of the pre-built index. Therefore, \AlleyPM would increasingly underestimates as the number of inserted edges increases. In order to simulate evolving graphs, we build an index by using 90\% of data edges, and run cardinality estimation on 90\%, 92\%, 94\%, 96\%, 98\%, and 100\% of data edges. We tested with \AIDS and \YAGO. Figure~\ref{fig:IndexAccuracy-update} shows the accuracy of each amount of data edges inserted after the index build. The results show that the estimation accuracy does not deteriorate much even if 5-10\% of data edges are inserted. From the result, we can see that we do not need to rebuild a tangled pattern index until there are significant updates to the data graph.

\begin{figure}[h!]
\begin{subfigure}{\columnwidth}
\centering
\includegraphics[width=0.95\columnwidth]{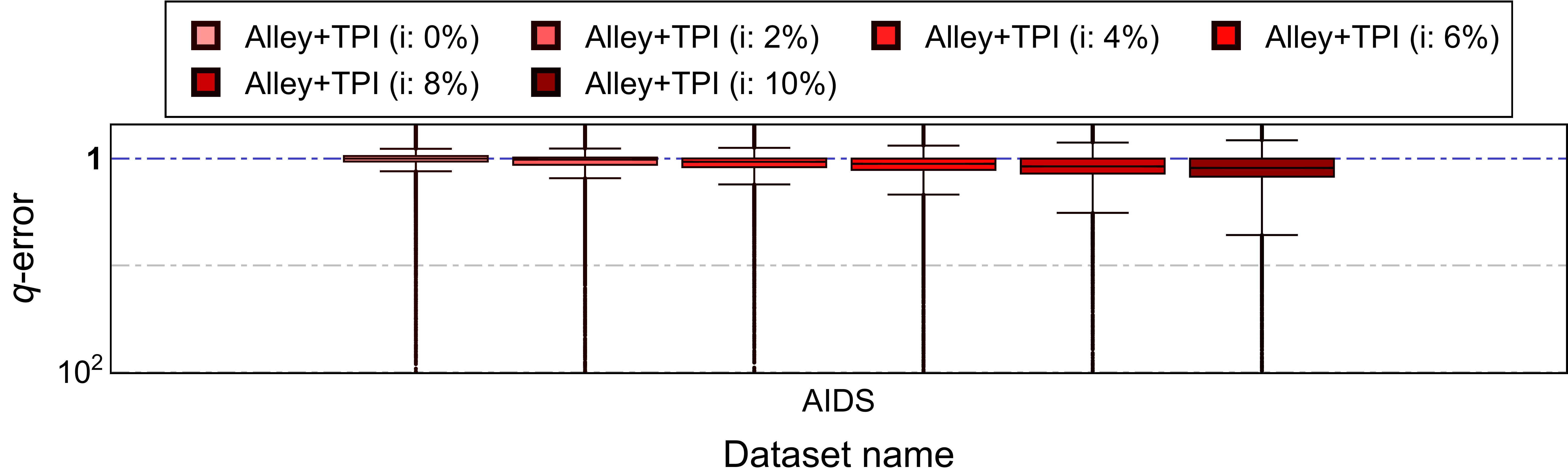}
\end{subfigure}
\begin{subfigure}{\columnwidth}
\centering
\includegraphics[width=0.93\columnwidth]{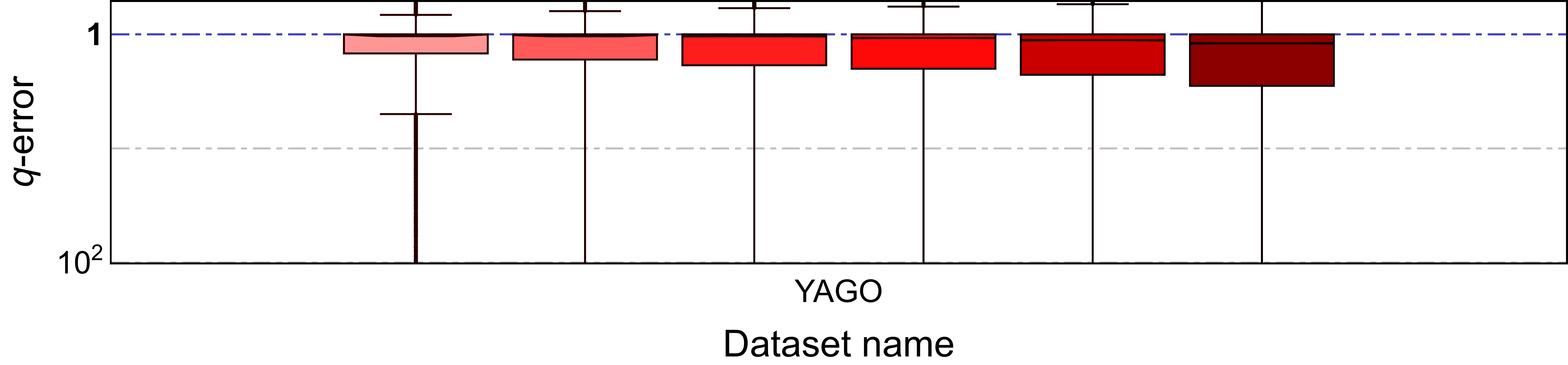}
\end{subfigure}
\vspace*{-0.4cm}
\caption{\revision{Accuracy of \AlleyTPI by varying amount of inserted edges.}}
\label{fig:IndexAccuracy-update}
\end{figure}
\fi

%% file: 09.related.tex
\ifFullVersion
\else
\fi
\section{Related Work} \label{sec:related}

\noindent \underline{\textbf{Graph Pattern Cardinality Estimation.}}
\ifFullVersion
Table \ref{table:summary} shows a summary of graph pattern cardinality estimators with four comparison aspects. The fourth aspect is based on our empirical study.

\begin{table}[htp]
\caption{\revision{Comparison aspects of estimators. G, R, S, and A refer to \underline{G}raph, \underline{R}elation, \underline{S}ynopsis, and s\underline{A}mpling. \faCheck, \faTimes, and \faMinus\xspace represent  positive, negative, and omission in our experiments, respectively.}}
\label{table:summary}
\small
\begin{tabular}{|c|c|c|c|c|}
\hline
Estinator               & Domain & Method & Unbiased?               & Accurate?                     \\ \hline
\CSET \cite{neumann2011characteristic}   & \textbf{G}      & \textbf{S}      & \faTimes & \faTimes       \\ \hline
\SumRDF \cite{stefanoni2018estimating}  & \textbf{G}      & \textbf{S}      & \faTimes & \faTimes       \\ \hline
\Catalog \cite{DBLP:journals/pvldb/MhedhbiS19} & \textbf{G}      & \textbf{S}      & \faTimes & \faMinus                             \\ \hline
\IMPR \cite{DBLP:conf/icdm/ChenL16}  & \textbf{G}      & \textbf{A}      & \faCheck & \faTimes       \\ \hline
\ST \cite{assadi2018simple}     & \textbf{G}      & \textbf{A}      & \faCheck & \faMinus                             \\ \hline
\WJ \cite{li2016wander}    & \textbf{R}      & \textbf{A}      & \faCheck & \faCheck       \\ \hline
\IBJS \cite{leis2017cardinality}   & \textbf{R}      & \textbf{A}      & \faCheck & \faCheck       \\ \hline
\CS \cite{vengerov2015join}     & \textbf{R}      & \textbf{A}      & \faCheck & \faTimes       \\ \hline
\JSUB \cite{zhao2018random, park2020g}   & \textbf{R}      & \textbf{A}      & \faTimes & \faTimes       \\ \hline
\BSK \cite{cai2019pessimistic}   & \textbf{R}      & \textbf{S}      & \faTimes & \faTimes       \\ \hline
\Alley (ours)   & \textbf{G}      & \textbf{S}, \textbf{A}   & \faCheck & \faCheck\faCheck \\ \hline
\end{tabular}
\end{table}

Summary-based approaches have been widely adapted for graph pattern cardinality estimation~\cite{aboulnaga2001estimating, maduko2008graph, neumann2011characteristic, stefanoni2018estimating}.
\else
Summary-based appro-aches have been widely adapted for graph pattern cardinality estimation~\cite{aboulnaga2001estimating, maduko2008graph, neumann2011characteristic, stefanoni2018estimating}.
\fi
Early work~\cite{aboulnaga2001estimating,maduko2008graph} stores the exact cardinality of small acyclic patterns (paths or trees) to estimate larger subgraphs in a data graph. More recent work includes \CSET~\cite{neumann2011characteristic}, \SumRDF~\cite{stefanoni2018estimating}, and subgraph catalog~\cite{DBLP:journals/pvldb/MhedhbiS19}. \CSET~\cite{neumann2011characteristic} decomposes $q$ into independent subqueries. \SumRDF~\cite{stefanoni2018estimating} merges the vertices and edges of $g$ into a smaller graph $S$, and extends the embeddings calculated in $S$ to the embeddings in $g$ under a uniformity assumption. However, these ad-hoc assumptions have been avoided in sampling-based work as they can significantly degrade estimation accuracy~\cite{park2020g}. Subgraph catalog  \cite{DBLP:journals/pvldb/MhedhbiS19} precomputes the average expansion ratio from each \emph{k}-vertex pattern to a (\emph{k}+1)-vertex pattern and multiplies these ratios under the uniformity assumption. 
However, as mentioned in Section~\ref{subsec:exp_setup}, storing all small  patterns can lead to scalability issues, especially for heterogeneous graphs.

\newredtext{Join cardinality estimation in relational databases, which is closely related to graph pattern cardinality estimation, often adopts sampling approaches~\cite{li2016wander,leis2017cardinality}.
However, we have shown that these sampling-based methods have the serious under-estimation problem due to sampling failures in large sample spaces.
\revision{While sampling failures have been addressed in \cite{li2016wander}, it briefly mentions that too many failed random walks will slow down the convergence of estimation, and failures occur more for cyclic queries. However, resolving these is not the main contribution of \cite{li2016wander}. Instead, \cite{li2016wander} focuses on selecting a good walk order using a round-robin approach and incorporating trial and failed walks in estimation.}
In contrast, we propose random walk with intersection, which can reduce sampling failures by cutting down the sample space.
We further reduce sampling failures by combining sampling and synopses.}


Theoretical bounds on estimation variance or runtime have been established by {\cite{eden2017approximately,aliakbarpour2018sublinear,assadi2018simple}. For the most part, these works solve very specific problems, e.g., estimating the number of triangles \cite{eden2017approximately}, stars \cite{aliakbarpour2018sublinear}, or cliques \cite{eden2018approximating}. The most recent work, \ST~\cite{assadi2018simple}, generalizes the previous theoretical estimators to arbitrary-shaped queries.}
\release{However, \ST lacks practicality since it does not perform random walks but independently samples edges as $T_1$ and $T_2$ in Section \ref{sec:ht}.}


\noindent \underline{\textbf{Frequent Pattern Mining.}} \newredtext{Frequent pattern mining finds all patterns (up to a limited size) that appear frequently in graph data. Here, data can be a set of multiple small graphs~\cite{DBLP:conf/sigmod/YanYH04} or a single large graph~\cite{abdelhamid2016scalemine, elseidy2014grami}. Then, a domain of pattern $p$ represents either the set of small graphs that contain $p$ or the set of vertices that participate in an \revision{embedding} of $p$. Other metrics that satisfy the anti-monotone property are also used \cite{DBLP:conf/icdm/FiedlerB07, DBLP:journals/datamine/KuramochiK05}.
The anti-monotone property permits the mining algorithms to search domains of larger patterns filtered by the domains of smaller sub-patterns.
Despite these optimizations, scalability has been a serious concern~\cite{jiang2013survey, abdelhamid2016scalemine}. Recent methods try to solve this problem \spellcheck{by} using distributed \mbox{computing.}}


\newredtext{Several methods use sampling for frequent pattern mining~\cite{zou2010frequent, abdelhamid2016scalemine}. These methods employ a sampling-based estimator to detect a set of patterns that are frequent with high probability and prune out infrequent patterns~\cite{abdelhamid2016scalemine}, or they sample the data graph to approximate the frequencies of patterns~\cite{zou2010frequent}. In contrast, we propose a novel indexing algorithm to store the domains of tangled patterns level-by-level by performing sampling on the previously computed domains of tangled sub-patterns.}

%% file: 10.conclusion.tex
\ifFullVersion
\else
\vspace*{-0.4cm} 
\fi
\revision{\section{Discussion \& Conclusion}\label{sec:con}}

\newredtext{We presented \Alley, an accurate and efficient graph pattern cardinality estimator. \Alley is a hybrid method that combines sampling and synopses, which includes 1) a new sampling strategy based on random walk with intersection and branching, and 2) an efficient mining algorithm and index for tangled patterns. Combining these two novel ideas, \Alley achieves high accuracy within a reasonable latency. \Alley also guarantees worst-case optimal time complexity for any given error bound and confidence level.}


\revision{\Alley can be used in the following scenarios. Given a pattern matching query $q$, a cost-based optimizer enumerates valid subgraph patterns of $q$ using dynamic programming and computes the cost for each pattern. The cost model calculates the cost based on the estimated cardinality. Here, \Alley computes such estimated cardinality. Furthermore, from the anti-monotone property, domains of a subquery $q^{\prime}$ of $q$ can be used as the matching candidates for $q$. Since the tangled patterns have strong structural (e.g., cycles) or label correlations, the search space can be effectively reduced.}


\revision{While \Alley can be integrated into any in-memory graph database system in its query optimizer or for answering approximate count queries, it would be beneficial if the system has data structures optimized for intersection, which is the main bottleneck in \Alley. For example, in order to efficiently process intersections, the edges in a data graph should be stored in sorted adjacency lists. A recent graph database system, \EmptyHeaded \cite{aberger2016emptyheaded}, can further benefit from its hybrid representation using bitmaps and SIMD operations, enabling faster intersections. Recent databases with GPU support might further benefit from GPU-based \mbox{intersection \cite{DBLP:conf/icde/ZengZOHZ20}.}}

\revision{For exploiting parallelism, index construction can be easily parallelized in a multi-core, distributed setting as in previous pattern mining studies. Our sampling-based estimation can also be easily parallelized in a multi-core setting by calling Line \ref{alg4:forloop} of Algorithm \ref{alg:offline} in parallel. However, it is non-trivial to implement an efficient online estimator in a distributed setting due to the communication overhead. We leave it as future work to extend the sampling-based estimation to a distributed setting.}